%% file: 2020-TR-Stallmann.tex
\documentclass[twoside,twocolumn,leqno]{article}
\usepackage{W-paper}
\usepackage{balance}

\begin{document}

\title{
Graph Profiling for Vertex Cover:\\
Targeted Reductions in a Branch and Reduce
Solver
}


\author{\OMIT{Matthias F. Stallmann\thanks{North Carolina State University}}
  \and \OMIT{Yang Ho\thanks{Sandia National Laboratory}}
  \and \OMIT{Timothy D. Goodrich\thanks{North Carolina State University}}
}

\date{\today}

\begin{titlepage}

\pagestyle{empty}

\maketitle

\fancyfoot[R]{\scriptsize{Copyright \textcopyright\ 2020\\
Copyright for this paper is retained by authors}}

\begin{abstract} \small\baselineskip=9pt
\input{0-abstract}
\end{abstract}

\end{titlepage}

\fancyfoot[CO,CE]{\thepage}
\setcounter{page}{1}

\section{Introduction}

\input{1-introduction}

\section{Background and Motivation} \label{sec:background}

\input{2-background}

\section{Reduction Configurations} \label{sec:configs}

\input{2-configs}

\section{Problem Instances} \label{sec:instances}

\input{3-problem-instances}

\section{Experimental Results} \label{sec:results}

\input{4-results}

\section{Special Cases}  \label{sec:special}

\input{4-special_cases}



\section{Conclusions and Future Work} \label{sec:conclusions}

\input{5-conclusions}

\smallskip
\noindent
\textbf{Acknowledgement.}
The authors thank Yoichi Iwata for his help navigating some of the details of \VCS, and for providing insights about the \lpRed reduction that allowed us to create a stand-alone implementation.

\clearpage

\onecolumn

\bibliographystyle{siam}

\bibliography{Z-combined_references}

\appendix

\twocolumn

\section{Branch and reduce algorithm} \label{app:branch_reduce}

\input{A-branch_reduce}

\onecolumn

\section{Options and Statistics Provided by \VCSP} \label{app:vcsp}

\input{A-vcsp}

\clearpage

\section{DIMACS challenge instances} \label{app:dimacs}

\input{A-dimacs}

\clearpage

\section{OCT instances} \label{app:oct}

\input{A-oct}

\clearpage

\section{Exceptions to the main hypotheses}
\label{app:exceptions}

\input{A-exceptions} \label{app:exceptions}

\clearpage

\section{Runtime tables} \label{app:runtime_tables}

\input{A-runtime_tables}

\clearpage

\end{document}

%% file: 0-abstract.tex
Akiba and Iwata [TCS, 2016] demonstrated that a branch and reduce (\BR) solver
for the vertex cover problem can compete favorably with integer linear
programming solvers (e.g., CPLEX).
Given a well-engineered \BR\ solver taking a reduction routine configuration as
input, our research question is \emph{are there graph characteristics that
determine which reductions will be most effective?}
Not only is the answer affirmative, but
the relevant characteristics are easy to identify and compute.

In order to explore our ideas rigorously, we provide an
enhanced implementation of the Akiba-Iwata solver so that it can
(a)~be configured with any subset of reductions and any applicable lower bounds;
(b)~print statistics such as time taken and number of vertices reduced by each
reduction type; and (c)~print trace information with additional details.

Based on extensive experiments with both benchmark and random instances we
demonstrate that (i)~doing more reductions does not necessarily lead to better
runtimes (in fact, sometimes the best strategy is to use no reductions at all);
(ii)~in most cases, the subset of reductions leading to the best (or nearly the
best) runtime can be predicted based on measurable characteristics of a graph,
such as density of the graph and degree distribution; and
(iii)~the exceptions have structural characteristics
that may be known in advance; examples include large sparse graphs, geometric
graphs, and planar graphs.

Our primary contributions are
\begin{enumerate}
\item A thorough examination reduction routine performance in the context of graph
  characteristics.
\item Three primary hypotheses suggesting simple suites of reductions as the
  most efficient options.
\item Experiments with a large corpus of data to validate our hypotheses.
  \item Measures that quantify a problem instance on two key dimensions to
    make our hypotheses concrete.
\item An enhanced open-source version of the Akiba-Iwata solver that enables
  our investigations and creates opportunities for future exploration.
\end{enumerate}

Our main objective is to provide guidance to a user so that, faced with a given problem instance or set of instances, they may most effectively use the available reductions. Ultimately these efforts can lead to an automated process.

\bigskip
\noindent
\textbf{Categories and Subject Descriptors:} G.2.2 [Discrete Mathematics]:
Graph Theory -- \emph{Graph Algorithms}.

\bigskip
\noindent
\textbf{General Terms:} Algorithms, Experimentation, Graphs, Performance.

\bigskip
\noindent
\textbf{Additional Key Words and Phrases:}
vertex cover,
branch and reduce,
dominance,
folding.


%% file: 1-introduction.tex
Originally one of Karp's 21 NP-complete problems \cite{karp1972reducibility}, Vertex Cover has a rich history in combinatorial optimization.
While there are simple 2-approximation algorithms,
the problem cannot be approximated below 1.36~\cite{dinur2005hardness}, or below~2 when assuming the Unique Games Conjecture \cite{khot2008vertex}.
When solving exactly, Vertex Cover is fixed-parameter tractable in both the natural parameter (solution size) and also in structural graph properties such as treewidth \cite{cygan2015parameterized}.
Such exact solutions may be of interest in their own right \cite{abu2004kernelization,wernicke2014algorithmic} or as a structural property utilized by an algorithm \cite{fellows2008graph}.
Additionally, Vertex Cover relates to several other optimization problems such as Independent Set, Maximum Clique, and Odd Cycle Transversal \cite{BB:Akiba16}.
In some cases, reformulating these optimization problems and solving them with a Vertex Cover solver leads to faster algorithms~\cite{lokshtanov2014faster}.

In recent years researchers have developed vertex cover solvers scalable to
large data. Minimum vertex cover (\mvc) is easily converted to an Integer Linear Programming (ILP) instance, enabling the use of industrial-quality solvers such as CPLEX and Gurobi.
Akiba and Iwata~\cite{BB:Akiba16} showed that a branch-and-reduce (\br) framework populated with several reduction routines, lower bound estimators, and branching rules could compete with both CPLEX and a Maximum Clique solver on social-network--sized datasets.
In this paper we report how this \br solver can be tuned for faster performance based on the provided instance.
Specifically, we study how the choice of reduction configuration affects total
performance.

In order to carry out these explorations, we provide an enhanced version of the
Akiba-Iwata solver (\VCSP)\footnote{When the distinction is necessary, we
  refer to the original solver as \VCS and our enhanced version as \VCSP.}
where (a)~any subset of the provided reductions and
lower bound methods can be selected; and (b)~the output includes detailed
statistics on the number of vertices reduced and the runtime spent for each
reduction. Statistics related to the effectiveness of verious lower bounds
are also provided.
In our analysis we find that
\begin{compactitems}
  \item The set of reductions chosen for a particular problem instance
    matters, sometimes a great deal.
  \item Easily computed characteristics of a problem instance,
    based on concrete measures we define, can pinpoint
    the set of reductions that are likely to be most effective.
  \item In some cases, structural properties of the instance, such as whether
    it is planar, based on geometry, or simply very large and sparse, are the primary
    factor influencing the most efficient set of reductions.
\end{compactitems}

\noindent
To make the relationships between graph characteristics and performance of
reductions more precise,
we propose and experimentally validate five hypotheses.
Outlined here, these are described and quantified in Section~\ref{sec:results}.

\begin{enumerate}
\item
  If the graph has low degree variation and density is large, the most
  promising option is to do no reductions at all.
\item
  If degree variation and density are large, the \domRed reduction
  is effective.
\item
  Simple reductions, such as \degRed and \foldRed, are sufficient for most
  sparse graphs.
\item
  The effectiveness of the \lpRed reduction is related to how close the graph
  is to being bipartite.
\item
  Sparse graphs with low degree variation are much harder to solve using \br
  than others with the same number of vertices.
\end{enumerate}

Our preliminary experiments
involved over 40,000 trials on roughly 10,000 instances. The results
reported here focus on a carefully selected set of roughly one thousand instances, with
'interesting' runtimes, and at least six trials (using a different set of
reductions for each trial) per instance.\footnote{We also did multiple trials
with permuted inputs on some instance/reduction set combinations to verify
that runtime variance was not significant.}

The remainder of the paper is organized as follows.
Section~\ref{sec:background} provides definitions and outlines the concepts
relevant to \VCS;
Section~\ref{sec:configs} motivates our study and distinguishes our
experimental approach from that of Akiba and Iwata;
Section~\ref{sec:instances} lists the problem instances in our corpus;
Section~\ref{sec:results} presents our main experimental results;
Section~\ref{sec:special} discusses special cases;
and Section~\ref{sec:conclusions} summarizes our findings and suggests future
work.

Appendix~\ref{app:branch_reduce} outlines the \br algorithm implemented by
\VCS;
Appendix~\ref{app:vcsp} shows the options and statistics provided by \VCSP.
Appendices~\ref{app:dimacs} and~\ref{app:oct} describe problem instances from
other sources in detail;
Appendix~\ref{app:exceptions} describes the exceptions to the main
hypotheses;
and Appendix~\ref{app:runtime_tables} has detailed tables that complement the
figures in the main text.

Code and scripts for \VCSP are available at
\textsf{https://github.com/mfms-ncsu/VC-BR} and a CPLEX driver at
\textsf{https://github.com/mfms-ncsu/CPX-ILP}.


%% file: 2-background.tex
A \emph{graph} $G = (V, E)$ has a vertex set $V$ and edge set $E \subseteq V \times V$.
In this paper we assume that all graphs are simple and undirected.
When clear from context, we denote $n = |V|$ and $m = |E|$.
The neighborhood of $v$, set of adjacent vertices, is denoted \neighbors{v}.
If $v$ is included we use \Neighbors{v}.

A \emph{vertex cover} of $G$ is a set of vertices $S \subseteq V$ such that
$G \setminus S$ is edgeless. Formulated as an optimization problem, the
objective of Vertex Cover is to find a minimum sized vertex cover.
We refer to this as the \mvc problem.
A \mvc instance can be converted into an Integer Linear Programming (ILP)
instance by minimizing $\sum_{v \in V} x_v$ subject to $x_u + x_v \geq 1$ for each edge $uv$, and $x_v \in \{0, 1\}$ for each vertex $v$.
In other words, minimize the number of vertices in the cover, such that every edge is covered and each vertex is either in the cover or not.
We implicitly use this conversion when solving \mvc instances with CPLEX.

A related problem is Odd Cycle Transversal (OCT), the problem of computing a minimum set of vertices whose removal renders a graph odd cycle-free.
Using a canonical transformation \cite{BB:Akiba16}, an OCT instance can be transformed into a \mvc instance with only constant blowup; in fact, there is precedent to think that OCT instances are best solved with \mvc solvers \cite{lokshtanov2014faster}.

In the remainder of this section we overview branching techniques such as branch-and-bound and branch-and-reduce, summarize the reduction rules provided in \cite{BB:Akiba16}, and provide motivation for the current study.

\subsection{Branch-and-reduce}

Similar to branch-and-bound, the branch-and-reduce (\br) paradigm iterates
between branching, checking bounds, and running an ensemble of reduction
routines to simplify the problem instance.
A graph instance is reduced until the reductions no longer apply, then two
branches are created (by choosing a vertex and either including it in the
cover or excluding it), and each resulting sub-instance is reduced.
At each step the partial solution and a lower bound on the current instance is compared against an upper bound, allowing the algorithm to quit a branch early
if the lower bound is greater than or equal to the known upper bound.

Unlike preprocessing routines, only run once at the beginning of execution, the reduction routines are executed throughout the course of the full algorithm.
Whereas ineffective preprocessing routines will incur wasted time for the
initial instance only, ineffective reduction routines could waste time at a
potentially exponential number of subinstances.
Therefore it is of interest to identify which reduction routines will prove effective on the structures and substructures found over the full course of an execution.

Appendix~\ref{app:branch_reduce} gives details of the general branch and
reduce algorithm and Iwata's implementation, \mbox{\VCS}.
Our enhanced version, \VCSP, makes
no changes beyond enabling an unconstrained choice of reductions and lower
bounds.
In addition, \VCSP outputs the actual solution obtained (as a bit vector), to
allow verification and potential post-processing (when the solution is not
optimal);
and it outputs a large variety of statistics relevant to our study -- see
Appendix~\ref{app:vcsp}.

\subsection{Lower Bounds}

The three nontrivial lower bounds in \VCS are
\begin{compactlist}
\item \textbf{clique.}
  If $C$ is a $k$-clique, then any vertex cover must include at least $k-1$
  vertices of $C$. \textsf{VCSolver} employs a simple greedy strategy that
  collects vertices into cliques, largest first.
\item \textbf{LP.}
  The linear programming relaxation of the ILP for \mvc is a lower bound.
  This bound is applied when an \lpRed{} reduction, see below, is done.
\item \textbf{cycle.}
  If $C$ is a cycle with $k$ vertices, any vertex cover
  must include at least $\ceil{k/2}$ vertices of $C$.
  \textsf{VCSolver} applies the cycle lower bound \emph{only if} an \lpRed{}
  reduction has taken place and information about odd cycles is readily
  available.\footnote{In a yet to be released C++ solver, we have had some
    success using a variant of breadth-first to search for small odd cycles.}
\end{compactlist}
Options provided by \VCS are: $-l0$ (trivial lower bound only),
$-l1$ (clique lower bound), $-l2$ (LP lower bound),
$-l3$ (LP and cycle lower bounds),
$-l4$ (all lower bounds).

\subsection{Reduction rules}

The reduction rules used in \VCS are as follows.
We refer the reader to Akiba and Iwata~\cite{BB:Akiba16}, Xiao and Nagamochi~\cite{BB:Xiao13,BB:Xiao17} and
\OMIT{Ho~\cite{2018-MSThesis-Ho}} for more details and proofs.

\begin{compactlist}
  \item \textbf{\degRed.} A degree-one vertex can be removed and its neighbor added to
    the cover.
  \item \textbf{\domRed.} If $vw$ is an edge and
    $\Neighbors{v} \subseteq \Neighbors{w}$,
    then $w$ \emph{dominates} $v$ and we can add $w$ to the cover.
  \item \textbf{\foldRed.} If $\degree{v} = 2$ and its neighbors $u$ and $w$ are not
    adjacent, then we contract $u,v,w$ into a new vertex $z$ to form $G'$;
    if $C'$, a minimum cover of $G'$, includes $z$ then
    $C'-z \cup \{u,w\}$ is a minimum cover of $G$; otherwise $C' \cup v$ is a
    minimum cover.\footnote{The \VCS implementation, when doing \foldRed
      reductions, also performs simple dominance reductions where the neighbors
    of a degree-2 vertex \emph{are} adjacent.}
  \item \textbf{\twinRed.} If $\neighbors{v} = \neighbors{w}$ and
    $\degree{v} = \degree{w} = 3$ then $v$ and $w$ are twins;
    we contract $\neighbors{v}$ into a single vertex $z$ to form $G'$;
    if $C'$, a minimum cover of $G'$, contains $z$, then
    $C'-z \cup \neighbors{v}$ is a minimum cover of $G$; otherwise
    $C' \cup v$ is.

    The \foldRed and \twinRed reductions are special cases of a
    $k$-independent-set reduction -- see~\cite{BB:Xiao13,BB:Xiao17}.

  \item \textbf{\lpRed.} If $S$ is a solution to the LP-relaxation of the ILP
    for \mvc,
    then there exists a minimum vertex cover that includes
    every vertex $v$ with $x_v = 1$ and excludes every $v$
    with $x_v = 0$. Iwata et
    al.~\cite{LP:Iwata14} give an algorithm that minimizes the number of
    variables with half-integer values.
  \item \textbf{\unconfinedRed.}~\cite{BB:Xiao13} A vertex is \emph{unconfined} if the following algorithm returns true.
    \begin{enumerate}
    \item $S = \{v\}$.
    \item Find $u \in N(S)$ with $|N(u) \cap S| = 1$ and having minimum value of $|N(u) \setminus N[S]|$.
    \item If no such $u$ exists or $|N(u) \setminus N[S]| > 1$, then return \False. \label{step:confined}
    \item If $N(u) \setminus N[S] = \emptyset$, then return \True. \label{step:empty}
    \item Otherwise, $S = S \cup (N(u) \setminus N[S])$ and start again from step 2.
    \end{enumerate}
    Unconfined reductions generalize \domRed{} -- if the algorithm returns \True
    during step~\ref{step:empty} in the first iteration, then $v$ dominates $u$.

  \item \textbf{\funnelRed.} If, for some $w \in \neighbors{v}$, $\Neighbors{v} - w$
    is a clique, we create $G'$ by removing $v$, $w$, and
    $\neighbors{v} \cap \neighbors{w}$, and adding an edge $st$ between each
    $s \in \neighbors{v} \setminus \Neighbors{w}$
    and each $t \in \neighbors{w} \setminus \Neighbors{v}$.
    Let $C'$ be a \mvc of $G'$. If
    $\neighbors{v} \setminus \neighbors{w} \subseteq C'$ then
    $C = C' \cup w \cup (\neighbors{v} \cap \neighbors{w})$ is a minimum cover of $G$;
    otherwise $\neighbors{w} \setminus \neighbors{v} \subseteq C'$ and
    $C = C' \cup v \cup (\neighbors{v} \cap \neighbors{w})$ form a minimum cover of $G$.

  \item \textbf{\deskRed.} If $v_1,v_2,v_3,v_4$ is a chordless four-cycle,
      $\degree{v_i} \geq 3$ for $i=1,\ldots,4$, and
      $\neighbors{v_1,v_3} \cap \neighbors{v_2,v_4} = \emptyset$, then
      create $G'$ by removing $v_1,v_2,v_3,v_4$ and adding an edge between
      each
      $s \in \neighbors{v_1,v_3}$ and each
      $t \in \neighbors{v_2,v_4}$.
      Let $C'$ by a minimum cover of $G'$.
      If $\neighbors{v_1,v_3} \subseteq C'$ then
      $C = C' \cup \{v_2,v_4\}$ is a minimum cover of $G$;
      else $\neighbors{v_2,v_4} \subseteq C'$ and
      $C = C' \cup \{v_1,v_3\}$ is a minimum cover of $G$.
\end{compactlist}

\noindent
The \funnelRed and \deskRed reductions are special cases of
\emph{alternative} reductions -- see~\cite{BB:Xiao13}.

\subsection{Related work}

Some recent research addresses the relationship between graph
characteristics and vertex cover complexity.
For example, Bl\"asius et al.~\cite{2019-arXiv-Blasius}
show that \mvc on \emph{hyperbolic random graphs} can be solved in polynomial
time.
These graphs model the degree distribution and clustering of many real-world
graphs.
Though we have not experimented specifically with these graphs,
we note that \VCSP, with appropriate choice of reductions, is efficient on
graphs with similar characteristics.
Recent work also addresses use of various combinations of reductions to
obtain algorithms that are efficient in practice.
Hespe et al.~\cite{2018-ALENEX-Hespe} make effective use of parallelism.
Chang et al.~\cite{2017-Sigmod-Chang} use reductions in a linear-time
heuristic that generates high quality solutions in practice (for the related
maximum independent set problem).
Other researchers have demonstrated that the combination of \degRed and
\foldRed, used in \emph{preprocessing}, is effective for many real-world
instances -- see, e.g.,
Strash~\cite{2016-COCOON-Strash}.
We have found that this combination is effective when used throughout
execution
for most benchmark and
randomly generated instances, including ones from the recent PACE-2019 vertex
cover challenge~\cite{pace2019data}.


%% file: 2-configs.tex
One of our main contributions is the analysis of individual
reductions and combinations of reductions, particularly those not offered by
the original \VCS.

The reduction options offered by \VCS are limited and cumulative, such that \reduction{0}
$\subseteq$ \reduction{1} $\subseteq$ \reduction{2} $\subseteq$
\reduction{3}.
The \reduction{0} level includes \degRed, \domRed, and \foldRed;
level \reduction{1} adds \lpRed{}; level \reduction{2} adds \twinRed,
\deskRed, \unconfinedRed, and \funnelRed; and level \reduction{3} adds
\packingRed.
The \degRed, \foldRed, \twinRed, \deskRed, and \packingRed reductions all
take linear time, while \lpRed is linear after an initial $O(m\sqrt{n})$ to
set up an auxiliary graph and compute a matching.
Dominance and \unconfinedRed{} are
$\Theta(d^2n)$ and $\Theta(d^3n)$, respectively, where $d$ is the maximum degree.

Based on a limited set of problem instances, Akiba and Iwata concluded that
using all reductions and all lower bounds (their \mbox{\textsf{-r3 -l4}} options) is (almost) always best.
We discovered, however, that for many of the problem instances from the
full benchmark corpus, the fastest runtime was achieved \emph{without using
  any reductions at all} and using only the lower bound based on a clique
cover.
For these instances at least, we show, effectively, that \VCS is a
well-engineered branch-and-bound solver.

Table~\ref{tab:none_intro} is a stark illustration.
These particular
instances were selected from the thousands in our experiments based purely on
runtime, with the following characteristics: (i)~runtime differences are
significant, often differing by an order of magnitude; and (ii)~CPLEX is not
a good option (runtime yet another order of magnitude worse).
One characteristic they share is that they are extremely dense;
so it is no surprise that reductions designed for sparse instances are
ineffective, whereas a clique lower bound is.
In our experiments
we explore a middle ground between choosing no reductions and
choosing all of them.

\input{Y-none_intro}

\input{Y-reduction_effectiveness_efficiency}

To that end, we analyzed the effectiveness and efficiency of various
reductions over our corpus -- see
Table~\ref{tab:reduction_effectiveness_efficiency}.
A disadvantage of \VCS, already noted, is that the choice of reductions is limited.
In particular, the smallest subset includes \domRed, overall the most
time-consuming and least effective.
To emphasize the flexibility of \VCSP,
we use the term \emph{reduction configuration} (or simply \emph{\config})
to refer to a set of reductions and lower bounds
performed by a trial of \VCSP.
In preliminary experiments (before we compiled the results in
Table~\ref{tab:reduction_effectiveness_efficiency})
we discovered that, on most instances where
reductions paid off at all, the \config that includes most of the linear time ones,
\cheap (\degRed, \foldRed, \deskRed, \twinRed), gave the best runtimes.

\input{Y-configs}

Table~\ref{tab:configs} lists the important \configs
used in our trials.
We used the first group in preliminary experiments on most instances
to establish the effectiveness of key reductions.
In later experiments, focused on
\emph{goldilocks} instances, ones
where at least one of the initial \configs took $> 2$~seconds and at least
one took $< 600$, we added \foldTwo and \dft (the two most efficient reductions).
For comparison, we also added \configs offered by the original \VCS:
\rOlI, \rtlf, and \rthreelf.
Finally, we added \rOlIU, the \textsf{-r0 -l1} option of \VCS with
\unconfinedRed added (the most frequent after \foldRed).
So five additional \configs.

When doing experiments with special cases -- large sparse, geometric, and
planar graphs, we added \unconfinedRed and/or \lpRed to the \cheap config and
to the \dft config, as suggested by
Table~\ref{tab:reduction_effectiveness_efficiency}.

Focusing on runtime, a reduction configuration is said to be
\emph{competitive} for a given problem instance $x$ if its runtime for $x$ is
within a factor of~2 (binary order of magnitude) of the minimum.
The definition is robust in the sense that it appears to be
invariant for trials on radically different machine architectures (cache size
being the major factor: 512~KB, 4~MB, 12~MB, 20~MB, or 30~MB) and multiple trials with permuted
inputs for the same instance.

A collection $\mathcal{C}$ of reduction configurations is \emph{globally
competitive} over a set of problem instances $\mathcal{I}$ if, for every
instance $x \in \mathcal{I}$, at least one reduction set in $\mathcal{C}$ is
competitive for $x$.
In our preliminary experiments the collection $\{\none, \cheap, \all\}$ was
globally competitive for all instances.
In the more comprehensive experiments on goldilocks instances the collection
$\{\none, \dft, \rOlI\}$ was globally competitive for all but a few
exceptions.


%% file: Y-none_intro.tex
\begin{table}
  \caption{Selected instances where the best runtimes are achieved with no reductions.}
  \label{tab:none_intro}

  \medskip
  \small
  \begin{center}
  \begin{tabular}{|l||r|r|r|r|r|r|r|r|}
    \hline
    Instance & \none & \textsf{-r2 -l4}
    & \textsf{-r3 -l4} & \textsf{CPLEX}\\ \hline
    san1000 & 30.93 & 305.46 & 762.71 & $> 900$\\ \hline
    DSJC1000.9 & 27.90 & 232.12 & 699.34 & $> 900$ \\ \hline
    p\_hat700-1 & 12.47 & 97.47 & 218.92 & $> 900$ \\ \hline
    c-fat500-1 & 2.26 & 7.29 & 7.32 & 71.92 \\ \hline
  \end{tabular}

  \medskip
  (a) runtimes: \none means clique lower bound only; the other columns represent
  \VCS options.
  
  \medskip
  \begin{tabular}{|l||r|r|r|r|r|r|r|r|}
    \hline
    Instance & vertices & min & avg
    & max \\ \hline
    san1000 & 1,000 & 449 & 498.0 & 554 \\ \hline
    DSJC1000.9 & 1,000 & 870 & 898.9 & 924 \\ \hline
    p\_hat700-1 & 700 & 413 & 524.7 & 624 \\ \hline
    c-fat500-1 & 500 & 479 & 481.2 & 482 \\ \hline
  \end{tabular}

  \medskip
  (b) degree statistics
  \end{center}

\end{table}


%% file: Y-reduction_effectiveness_efficiency.tex
\begin{table}
  \caption{The median effectiveness and efficiency of various reductions.}
  \label{tab:reduction_effectiveness_efficiency}
  \centering

  \medskip
    \begin{tabular}{|l||r|r||r|r|}
      \hline
      reduction
      & \multicolumn{2}{c||}{$\mu$sec/vertex}
      & \multicolumn{2}{c|}{\% reduced} \\
      \cline{2-5}
      & \multicolumn{1}{c|}{med} & \multicolumn{1}{c||}{geo}
      & \multicolumn{1}{c|}{med} & \multicolumn{1}{c|}{geo}
      \\
      \hline\hline
      fold2 & 1.2 & 1.3 & 71.2 & 62.4 \\ \hline
      unconf. & 24.6 & 21.6 & 9.1 & 9.9 \\ \hline
      lp & 51.5 & 70.0 & 4.5 & 3.5 \\ \hline
      pack. & 59.4 & 69.4 & 3.5 & 2.4 \\ \hline
      deg1 & 6.1 & 8.1 & 2.8 & 1.9 \\ \hline
      fun. & 81.8 & 136.3 & 2.0 & 1.2 \\ \hline
      desk & 108.5 & 129.7 & 0.8 & 0.8 \\ \hline
      twin & 96.3 & 117.2 & 0.5 & 0.5 \\ \hline
      dom. & 236.8 & 331.7 & 0.3 & 0.3 \\ \hline
    \end{tabular}









  \medskip
  \begin{minipage}{\columnwidth}
    \small
    Efficiency, for a reduction type,
    is measured as the time spent per vertex reduced, in microseconds.
    Effectiveness is percentage of vertices reduced
    with respect to number of vertices reduced overall.
    Measurements are based on the \mbox{\textsf{-r3~-l4}} options, which include all
    reductions and all lower bounds.
    Given the large variation among problem instances in the corpus
    (goldilocks instances), we report both the medians and the geometric
    means.
    The reductions are sorted by decreasing frequency (last two columns).



  \end{minipage}

\end{table}


%% file: Y-configs.tex
\begin{table*}
  \caption{Configurations used in our experiments}
  \label{tab:configs}

  \medskip
  \begin{minipage}{\textwidth}
    \small
    \begin{center}
    \begin{tabular}{|l||l|l|}
      \hline
      \multicolumn{1}{|c||}{\textbf{Name}}
      & \multicolumn{1}{c|}{\textbf{Reductions}}
      & \multicolumn{1}{c|}{\textbf{Lower Bounds}}
      \\
      \hline
      \multicolumn{3}{|c|}{\emph{Used in preliminary experiments}} \\
      \hline
\textsf{None} &   & clique \\ \hline
\textsf{Deg1} & deg1 & clique \\ \hline
\textsf{DD} & deg1 dom & clique \\ \hline
\textsf{Cheap}
& deg1 fold2 desk twin & clique \\ \hline
\textsf{All} & deg1 dom fold2 LP unconfined funnel desk twin & clique lp \\
\hline
\multicolumn{3}{|c|}{\emph{Used in comprehensive experiments}}
\\
\hline
\textsf{Fold2} & fold2 & clique \\ \hline
\textsf{DF2}
& deg1 fold2 & clique \\ \hline
r0\_l1 & deg1 dom fold2 & clique \\ \hline
\textsf{r0\_l1+U} & deg1 dom fold2 unconfined & clique \\ \hline
r2\_l4 & deg1 dom fold2 LP unconfined twin funnel desk & clique lp cycle
\\ \hline
r3\_l4 & deg1 dom fold2 LP unconfined twin funnel desk packing & clique lp cycle \\ \hline

\multicolumn{3}{|c|}{\emph{Used in experiments with large sparse graphs}}
\\ \hline
\textsf{Cheap+U}
& deg1 fold2 unconfined desk twin & clique \\ \hline
\textsf{Cheap+LP}
& deg1 fold2 LP desk twin & clique lp \\ \hline
\textsf{Cheap+LPU}
& deg1 fold2 LP unconfined desk twin & clique lp \\ \hline
\textsf{DF2+U}
& deg1 fold2 unconfined & clique \\ \hline
\textsf{DF2+LP}
& deg1 fold2 LP & clique lp \\ \hline
\textsf{DF2+LPU}
& deg1 fold2 LP unconfined & clique lp \\ \hline


    \end{tabular}
    \end{center}
  \end{minipage}
\end{table*}


%% file: 3-problem-instances.tex
Before introducing problem instances we define the measures that characterize
them and the resulting two-dimensional landscape.
We take special care to introduce random instances that populate as much of
the landscape as possible.

\subsection{Instance measures} \label{sec:measures}

The most obvious of measure is density -- many of the reductions are designed
specifically for sparse graphs while the densest graphs have large cliques,
making the clique lower bound effective without reductions.
We use \emph{normalized average degree} (\nad), a hybrid of traditional
density and actual average degree.
Specifically, if average degree is $> 20$, we normalize it using a factor of
$200/n$ (most of our randomly generated graphs have roughly 200 vertices), so
a complete graph has \mbox{$\nad = 199$}.
When $\leq 20$ we use the actual average degree; the threshold is arbitrary,
but, at the low end, reductions and branching lead to trivial subinstances at
a rate determined by actual average degree.

The \emph{degree spread}, or simply \emph{spread}, captures the fact that
the when \degree{v} is small and $v$ has high degree neighbors,
max-degree branching will quickly make $v$ eligible for \degRed or \foldRed reductions.
We define spread to be $t/b$, where $t$ is the degree at the 95-th percentile
and $b$ at the 5-th. This is an arbitrary choice, but the landscape does not change much
if we use, for example, the 90-th and the 10-th.
In the extreme, when both spread and \nad are high, the probability of
dominance becomes non-negligible: $\frac{k!}{(k-j)!}\cdot\frac{(n-k)!}{n!}$
for two vertices of degree $k$ and $j\ll k$.

To effectively visualize trials on a corpus of hundreds of graphs we use a
\emph{landscape} plot (log-log) with \spread on the
x-axis and \nad on the y-axis. Fig.~\ref{fig:instance_landscape} shows our randomly generated
instances on the landscape.

\input{Y-instance_landscape}

\subsection{Random instances}

Our random instances fall into several categories. The first of these, \blg,
is of our own design, keeping the landscape in mind. The rest are based on
standard techniques.

\medskip\noindent
\textbf{blg.}
The generator for these, \emph{bucket list generator} (\blg) was created
specifically for our preliminary experiments to saturate the landscape.
The \blg takes as input the number of vertices, average degree, and a \degVar parameter.
The generator guarantees connectivity by initially creating a random spanning
tree; then edges are added until the average degree is achieved.

Vertices are maintained in `buckets' based on their degree. The
parameter \degVar determines the choice of endpoints for each edge:
$\degVar = 0$ means lowest degree vertices are chosen -- this leads to
regular (or nearly regular) graphs; if $\degVar = 1$, endpoints are chosen
uniform randomly; if $\degVar < 1$ bias is toward endpoints of low degree, if
$\degVar > 1$ then bias is toward those of high degree -- this leads to degree
distributions that are approximately normal or approximately exponential,
respectively.

In addition to the mandatory parameters, the generator has optional parameters
to control (as much as possible) \mindeg and \maxdeg.
Vertices whose degree is less than \mindeg are chosen unconditionally, while
vertices with degree greater than \maxdeg are no longer
considered.

Most of our \blg graphs have 200 vertices. In some cases, where this led to
instances that were too easy or too hard, we generated corresponding ones
with 250 or 150 vertices, respectively.
Naming convention is \mbox{\textsf{blg-}$n$\_$a$\_$dv$\_$m$\textsf{d}$M$},
where $n$ is the number of vertices, $a$ is the average degree,
and $m$ and $M$ are the desired minimum and maximum degree,
respectively.

\medskip\noindent
\textbf{ba.} (Barabasi-Albert)
We used the preferential attachment generator provided by the python
\textsf{networkx} package, function \textsf{barabasi\_albert\_graph}.
The only parameters are number of vertices and number of edges to be added by
each new vertex. These graphs are connected.
In our corpus most of the \textsf{ba} graphs have 256 vertices, but when
these turned out to be too easy, we added ones with 512 or 999 vertices.
Number of attachment edges ranged from 8 to 90.
Naming convention is \textsf{ba\_}$n$\_$k$\_$s$, where $n$, $k$, and $s$ are number of
vertices, number of edges to be added, and the random seed, respectively.

\medskip\noindent
\textbf{cl and clspk.} (Chung-Lu)
Here we used the \textsf{expected\_degree\_graph} function from the
\textsf{networkx} package. This requires a complete list of vertex degrees
and the result is at best an approximation.
Our \textsf{cl} graphs start with a uniform distribution of degrees from a
specified min to a specified max. The result is closer to a normal
distribution.
Our \textsf{clspk} graphs start with one min and one max degree vertex and
put other degrees somewhere in between.
The result is a normal distribution with low standard deviation (but
nontrivial spread for the sparser ones).
Naming conventions are
\mbox{\textsf{cl\_}$n$\_\textsl{min}\_\textsl{max}\_$s$}
and \mbox{\textsf{clspk}\_$n$\_\textsl{min}\_\textsl{max}\_\textsl{avg}\_$s$},
where $n$ is number of vertices, \textsl{min} and \textsl{max} are minimum
and maximum degree, respective, $s$ is the random seed, and in the case of
\textsf{clspk}, \textsl{avg} is the desired average degree (leading to a
skewed distribution in some cases).

\medskip\noindent
\textbf{geo.} (Geometric)
These are classic two-dimensional geometric graphs using our own generator.
Given a desired number of vertices $n$ and edges $m$, the generator estimates a
distance $d$ such that, when randomly placed points (vertices) within distance $d$ of each
other are connected, the number of edges will be roughly $m$.
We guarantee connectivity in a post-processing phase that constructs a
spanning tree on the connected components.
The resulting number of edges tends to be larger than desired for sparser
graphs, smaller for denser graphs.
Our \textsf{geo} graphs have 512~vertices with 1024, 2048, 4096, 8192, 16384,
and 32768 desired edges; the actual graphs have roughly 1200, 2000, 3700,
7100, 13500, and 25000 edges, respectively.

\medskip\noindent
\textbf{planar.}
These are actually two (extreme) special cases of planar graphs, both based
on Delaunay triangulations.
The first set, \textsf{tri-inf} are triangulations with the infinite face
also triangulated -- they have exactly 1000 vertices and 2994 edges.
Since most of the reductions preserve planarity, the subinstances created by
\VCS are more general planar graphs.

The other set is the \textsf{dual} graphs -- duals of infinite-face
triangulations and therefore guaranteed to be 3-regular.
We used these to compare with random 3-regular graphs and discover the
extent to which planarity matters;
they have to have 1024 vertices and 1536 edges.
Our initial experiments showed that, with appropriate choice of reductions,
these dual graphs are 
much easier to solve than their
general 3-regular cousins.
Hence the larger size.
The same holds for the triangulations and graphs similar profile
(average degree~6 and \spread close to~2).

\subsection{Instances from other sources}

The following benchmark instances come from other collections.

\medskip\noindent
\textbf{DIMACS graph coloring challenge.} We included graphs from the 1993
DIMACS graph coloring implementation challenge~\cite{DIMACS} as \mvc{} instances.
These include several varieties and appear scattered in our landscape. Some
have well-defined structures. Details are given in Appendix~\ref{app:dimacs}.

\medskip\noindent
\textbf{Odd cycle transversal instances.}
We included graph instances from experiments reported by Goodrich et
al.~\cite{goodrich2018practical} and provided on their web site~\cite{goodrich2018code}.
Specifically, the classic \emph{Minimum Site Removal} dataset from Wernicke
\cite{wernicke2014algorithmic}, used in the Akiba-Iwata experiments is included, along with graphs of interest in quantum computing, originally provided in Beasley's OR library~\cite{beasley} and the GKA dataset~\cite{glover1998adaptive}.
Of the random graph instances provided in this repository, we sub-selected from these graphs with the (arbitrary) $\texttt{seed} = 7$.
For figures showing where the OCT instances fall on
our landscape, see Appendix~\ref{app:oct}.

\medskip\noindent
\textbf{PACE 2019 challenge instances.}
The 2019 Parameterized Algorithms and Computational Experiments (PACE)
challenge includes a track on Vertex Cover. Results for both the public and
the contest instances~\cite{pace2019data} are included in our experiments.
The results confirm our hypotheses on instances that \VCS is able to solve at
all (within four hours on a powerful server). 
See Section~\ref{sec:pace} for more details.


%% file: Y-instance_landscape.tex
\begin{figure}
  \centering

  \includegraphics[width=\columnwidth]{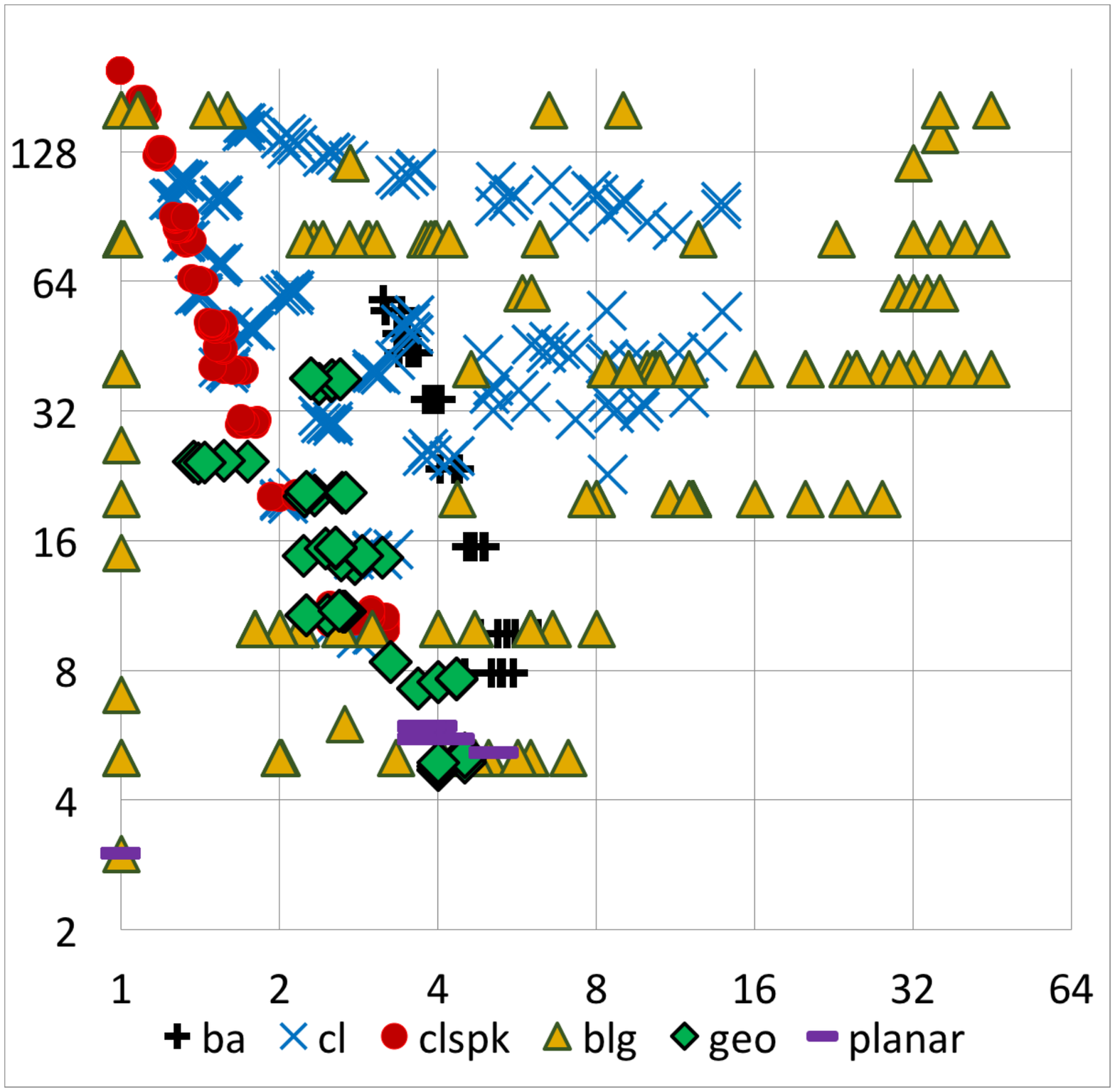}

   \small
  \medskip
  \begin{minipage}{\columnwidth}
    Randomly generated instances of various flavors: Barabasi-Albert, two
    types of Chung-Lu (uniform/normal distribution and clspk is 'spiked' -- a
    few vertices at extreme degrees and almost all in the middle), geometric,
    planar, and blg.
    The blg (bucket list generator) produces connected graphs and gives full control of min, max, and
    average degree, thus allowing fuller coverage of the landscape.
  \end{minipage}
  
  \caption{The landscape of random instances.}
  \label{fig:instance_landscape}
\end{figure}


%% file: 4-results.tex
Recall that we call a \config \emph{competitive} for an instance $\mathcal{I}$ if its runtime on $\mathcal{I}$ is within a factor of two of the minimum over the configs used in our comprehensive experiments.
In this section we evaluate the competitiveness of the configs
(described in Table~\ref{tab:configs}), formalize our observations as hypotheses,
and discuss how the hypotheses apply to various subsets of the overall
corpus, including the
the recent PACE 2019 instances.
Special cases such as geometric, planar, and large sparse graphs are treated
separately in Section~\ref{sec:special}.

We performed our experiments on a
server with dual Intel E5645 (2.4GHz, 12MB cache) processors and 4GB DDR3 RAM,
running Red Hat 4.8.5-16 Linux.
The \VCSP solver was compiled and run using Java, version~1.8.
We ran CPLEX in default mode.

To allow time for trials on several thousand instances with at least six reduction
configurations and CPLEX, we set a time limit of 900~seconds.
Larger instances were given timeouts of 4 hours or 24 hours and were run on
platforms with more memory.

\subsection{Main hypotheses}

Three main hypotheses emerged from our preliminary experiments and \OMIT{Ho's
thesis~\cite{2018-MSThesis-Ho}}.

\begin{hyp} \label{hyp:none}
  If \spread is small ($\leq 4$) and \nad is large ($> 20$), the \none
  config is competitive.
\end{hyp}

Hypothesis~\ref{hyp:none} is easily explained by the presence
of larger cliques.

\begin{hyp} \label{hyp:dom}
  If both \spread and \nad are large \mbox{($\geq 16$)}, a config that includes
  \domRed, e.g., \rOlI, is competitive.
\end{hyp}

As observed in Section~\ref{sec:measures}, large \spread and \nad make \domRed
reductions more likely.

\begin{hyp} \label{hyp:dft}
  If \nad is small,
  the \dft config is competitive.
\end{hyp}

Here branching is likelier to lead to degree-1 and degree-2 vertices than in
the situations covered by the previous hypotheses.
An important point is that, even when a degree-2 vertex is not a candidate
for a \foldRed reduction, \VCS does the obvious \domRed reduction: if the two
neighbors of the degree-2 vertex are adjancent, they both dominate it.

\input{Y-none_df2_r0_l1}

\input{Y-borders}

Fig.~\ref{fig:none_df2_r0_l1} shows competitive configs for all
626~goldilocks instances in our general corpus.
Each data point shows the first config in the list \none, \dft, \rOlI that is
competitive with respect to those in the first two categories of
Table~\ref{tab:configs}.
This may not be the only competitive one nor the one with
minimum runtime.
With very few exceptions, all three hypotheses hold.
In fact, Table~\ref{tab:borders} shows that, in the few cases where instances
in the chart do not quite meet the numerical thresholds,
both of the relevant configs are competitive.
There are 12~instances that completely fail to validate the three hypotheses.
We address these in detail in Appendix~\ref{app:exceptions}.

The companion (to Fig.~\ref{fig:none_df2_r0_l1}) tables in Appendix~\ref{app:runtime_tables}
show that, not only are the
relevant configs competitive, but their runtimes are often close to minimum.
In situations where \none is competitive (as shown in the chart),
it is often better to add
\degRed and/or \mbox{\foldRed{}} -- these have minimal overhead.
Where \dft is competitive, sometimes \foldRed by itself works as well or
better.
A perusal of the \rtlf columns reveals that the \rtlf config is almost never
competitive; the same holds for \rthreelf, not shown.

\input{Y-other}

The reader may wonder if the \rOlI and \rtlf \configs are as good as, or
at least competitive, where Hypotheses~\ref{hyp:none} and \ref{hyp:dft} apply.
Fig.~\ref{fig:none_df2_r0_l1} shows that,
while \rOlI is competitive over much of the landscape, there are still many
instances where it is not, specifically in regions where \none and \dft
\emph{are}.
And lest the reader believe that \rtlf is competitive in any of the regions
specified by the three hypotheses,
Fig.~\ref{fig:r2_l4} shows that \rtlf is a poor choice for most instances.

\subsection{OCT and LP reductions}

\input{Y-to_instances}

\input{Y-oct_vs_lp}

Ho's thesis~\cite{2018-MSThesis-Ho} proposed another hypothesis.

\begin{hyp} \label{hyp:OCT}
  If (estimated) OCT is small ($< 20\%$ of vertices), a config that includes
  \lpRed, e.g., \rIlf, is competitive.
\end{hyp}

Since computing OCT directly is an NP-hard
problem, we rely on estimates provided by heuristics from~\cite{goodrich2018code}.
Table~\ref{tab:to_instances} shows data for instances in our goldilocks
corpus that do not necessarily fit the first three hypotheses,
but for which
\lpRed reductions lead to significantly lower runtimes.
CPLEX is competitive with \br{} in all but two instances (instance names in bold);
there it spends extra time solving the instance algebraically at the root
and its overall runtime is not significant.
In one instance, CPLEX is far superior (runtime in bold italic). In all cases, CPLEX
does significantly less branching, but more algebraic processing and
more cuts.\footnote{Our CPLEX driver is instrumented to report a variety of
  information about, for example, simplex iterations and cuts.}
Also, \lpRed{} by itself with only an LP lower bound is as
good as the \textsf{r1\_l4} config for the instances in the table.

The ratio between runtimes for r0\_l1 and r1\_l4 generally decreases with
increasing OCT percentage -- estimated OCT as a percentage of vertices;
the influence of adding LP reductions and corresponding lower bounds
becomes less pronounced as OCT increases. Fig.~\ref{fig:oct_vs_lp} shows this
relationship for the whole corpus.

Interestingly, all of the instances in Table~\ref{tab:to_instances} are tunable
OCT instances (see Section~\ref{sec:instances}), i.e., randomly generated
to mimic instances from the OCT corpus.
Except for \textsf{aa41}, the originals were all too easy to be
goldilocks.

\subsection{Degree of difficulty}

\input{Y-hardness}

It is also important to identify instances where \br is likely to encounter
difficulty.
Fig.~\ref{fig:hardness} gives some guidance that, while only small instances
are shown, turns out to scale to much larger ones.
The main message is that the hardest instances tend to have low to medium average
degree and small \spread.
Conversely, instances with
high average degree and/or large \spread are easy.
None of the instances we report have average degree $<3$; such instances,
when small, turn
out to be trivial -- we address \emph{large} sparse instances in
Section~\ref{sec:real_world}. This leads to another hypothesis, harder to quantify
than the others.

\begin{hyp} \label{hyp:hardness}
Sparse instances, those with
average degree ranging from 5 to 20 (roughly) and small spread, are significantly harder to solve
than others.
\end{hyp}

\subsection{Evaluating Hypotheses on PACE Data} \label{sec:pace}

\input{Y-pace_2019}

Recently we tested our hypotheses on all 200~public instances from the
PACE-2019 vertex cover challenge.
We used a more powerful server\footnote{Quad-Core AMD Opteron(tm) 8374 HE (2.2GHz, 512KB cache)
  processor and 128GB DDR3 RAM, running Red Hat Linux.} and set a
four-hour timeout.
For the sake of thoroughness we chose configs \none, \dft, \rOlI, \cheap,
\cheapU, \cheapLP, \cheapLPU, \rOlIU, \rtlf, \rthreelf, and \all.

The instances that \VCS was able to solve confirmed Hypothesis~\ref{hyp:dft}.
All were in the region where \dft should be competitive and all but a few had competitive runtimes for
\dft, usually minimum ones.
The exceptions are large and, where they are sparse with low spread,
required \unconfinedRed reductions to reduce them effectively.
Table~\ref{tab:pace_table} in Appendix~\ref{app:runtime_tables} gives results
for instances with minimum runtime $>5$ seconds. The instances where
\dft was not competitive are in bold.
Table~\ref{tab:pace_2019-degree_stats} gives degree statistics for the same
instances.
Here, the instances where \rthreelf has minimum or within 1.5 of minimum
runtime are highlighted in bold-italic or bold, respectively.

CPLEX has minimum runtimes on almost all instances where the best runtime for
the \configs under consideration was more than a second.
On the instances with small runtimes, CPLEX preprocessing time dominated the
actual branching -- they were solved at the root.
No \config of \VCS{} solved any instances beyond
\textsf{vc-exact\_100} within our time limit, and only 78 of the first 100
were solved.
In contrast, CPLEX was able to solve 142 out of the 200 total instances and
72 of the 100 contest instances.



%% file: Y-none_df2_r0_l1.tex
\begin{figure}
  \centering
  \includegraphics[width=\columnwidth]{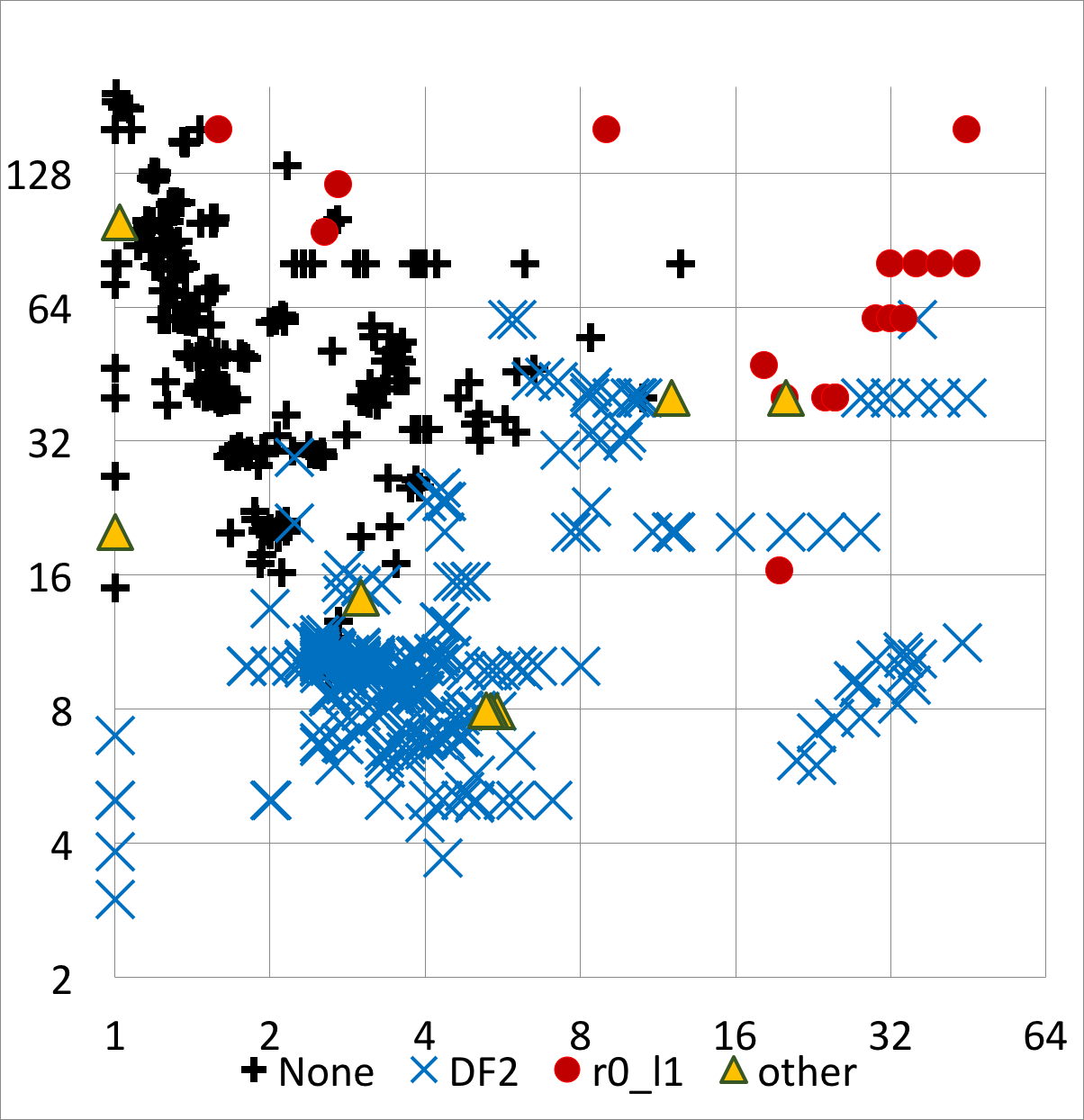}

  \caption{Where \none, \dft, and \rOlI are competitive.}
  \label{fig:none_df2_r0_l1}
\end{figure}


%% file: Y-borders.tex
\begin{table*}
  \caption{Runtime data for instances on the boundaries of our hypotheses.}
  \label{tab:borders}
  \small

  \medskip
  \centering
  \begin{tabular}{|l||r|r||r|r|r|r|r|r||l||}
    \hline
Instance & spread & nad & None & DF2 & r0\_l1 & r2\_l4 & min & CPLEX &
competitive configs \\ \hline
\multicolumn{10}{|c|}{shown as \none in chart but
  Hypothesis~\ref{hyp:none_dft} says \dft competitive} \\ \hline
cl\_200\_010\_080\_4 & 4.9 & 43.3 & \textbf{2.32} & \textbf{1.41} &
\textbf{1.53} & 5.86 & 1.21 & 45.44 & \none, \dft, \rOlI \\ \hline
blg-200\_040\_01\_05d060 & 4.6 & 40 & \textbf{7.74} & \textbf{\emph{4.89}} & \textbf{6.48} & 19.44 & 4.66 & 238.31 & \none, \dft, \rOlI \\ \hline
\hline
\multicolumn{10}{|c|}{shown as \dft in chart but
  Hypothesis~\ref{hyp:dom} says \domRed competitive} \\ \hline
blg-200\_040\_16\_05d140 & 28 & 40 & 99.28 & \textbf{0.42} & \textbf{0.51} & 0.89 & 0.37 & 1.00 & \dft, \rOlI \\ \hline
blg-250\_050\_16\_05d150 & 30 & 40 & 816.18 & \textbf{0.51} & \textbf{\emph{0.39}} & 0.85 & 0.39 & 2.48 & \dft, \rOlI \\ \hline
blg-200\_040\_16\_05d160 & 32 & 40 & 244.71 & \textbf{\emph{1.62}} & \textbf{\emph{1.57}} & 3.92 & 1.57 & 3.29 & \dft, \rOlI \\ \hline
blg-200\_040\_16\_05d180 & 36 & 40 & 607.35 & \textbf{1.89} & \textbf{1.46} & 3.34 & 1.13 & 4.60 & \dft, \rOlI \\ \hline
blg-250\_050\_16\_05d200 & 40 & 40 & t/o & \textbf{\emph{3.41}} & \textbf{\emph{3.42}} & \textbf{5.83} & 3.41 & 15.02 & \dft, \rOlI, \rtlf \\ \hline
blg-250\_050\_16\_05d225 & 45 & 40 & t/o & \textbf{7.09} & \textbf{7.71} & \textbf{10.52} & 6.21 & 17.41 & \dft, \rOlI, \rtlf \\ \hline
blg-200\_020\_16\_05d080 & 16 & 20 & 225.73 & \textbf{\emph{0.53}} & \textbf{0.66} & 1.59 & 0.53 & 1.07 & \dft, \rOlI \\ \hline
blg-200\_020\_16\_05d100 & 20 & 20 & 780.26 & \textbf{\emph{1.66}} & \textbf{\emph{1.8}} & 5.95 & 1.66 & 3.74 & \dft, \rOlI \\ \hline
blg-200\_020\_16\_05d120 & 24 & 20 & t/o & \textbf{5.32} & \textbf{5.05} & 11.01 & 4.32 & 11.86 & \dft, \rOlI \\ \hline
blg-200\_020\_16\_05d140 & 28 & 20 & t/o & \textbf{7.27} & \textbf{8.52} & 14.80 & 6.45 & 12.93 & \dft, \rOlI \\ \hline
  \end{tabular}

\end{table*}


%% file: Y-other.tex
\begin{figure}
  \centering
 \includegraphics[width=\columnwidth]{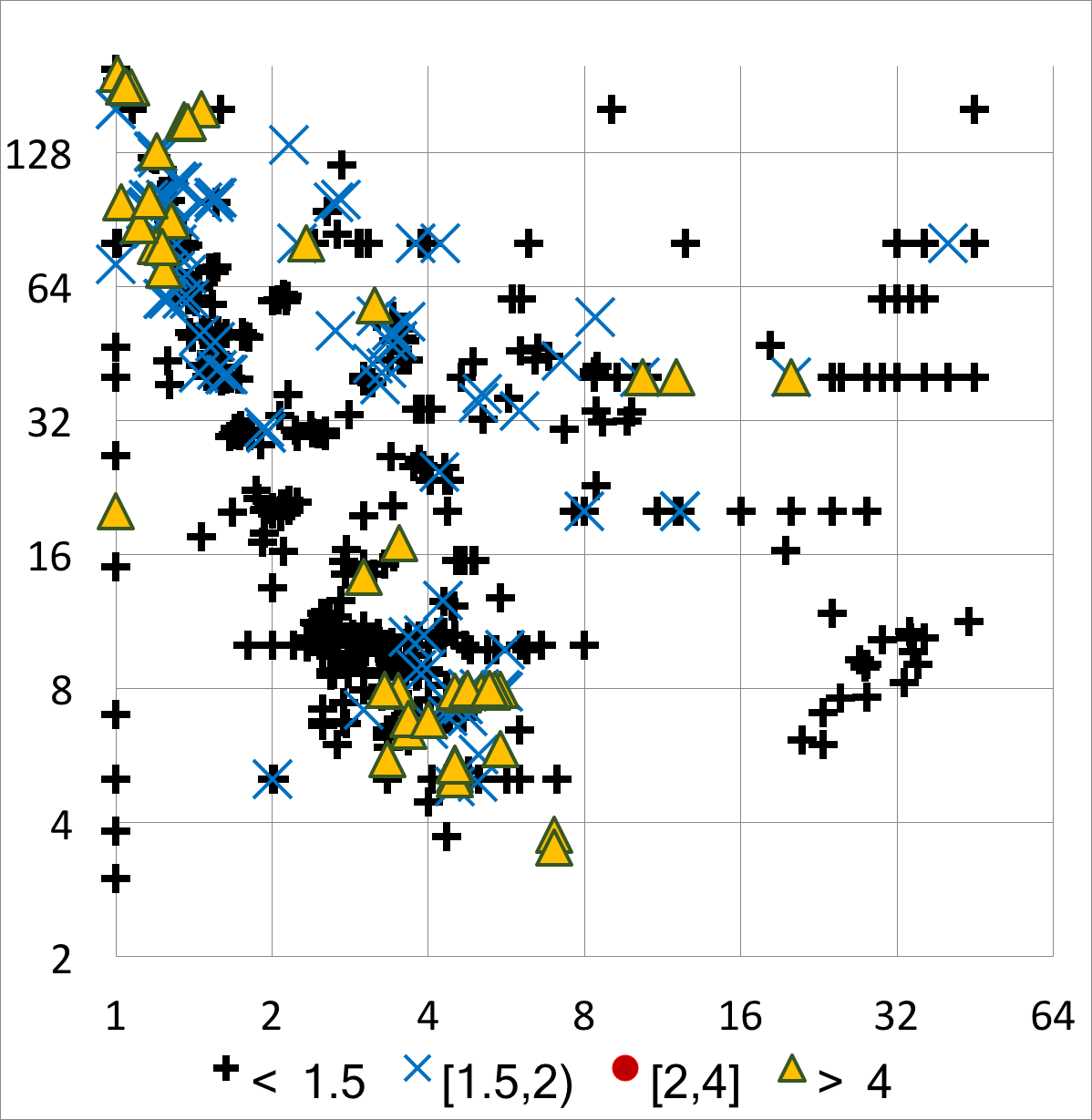}
  \caption{Ratio of \rOlI runtime to the minimum.}
  \label{fig:r0_l1}
\end{figure}

\begin{figure}
  \centering
  \includegraphics[width=\columnwidth]{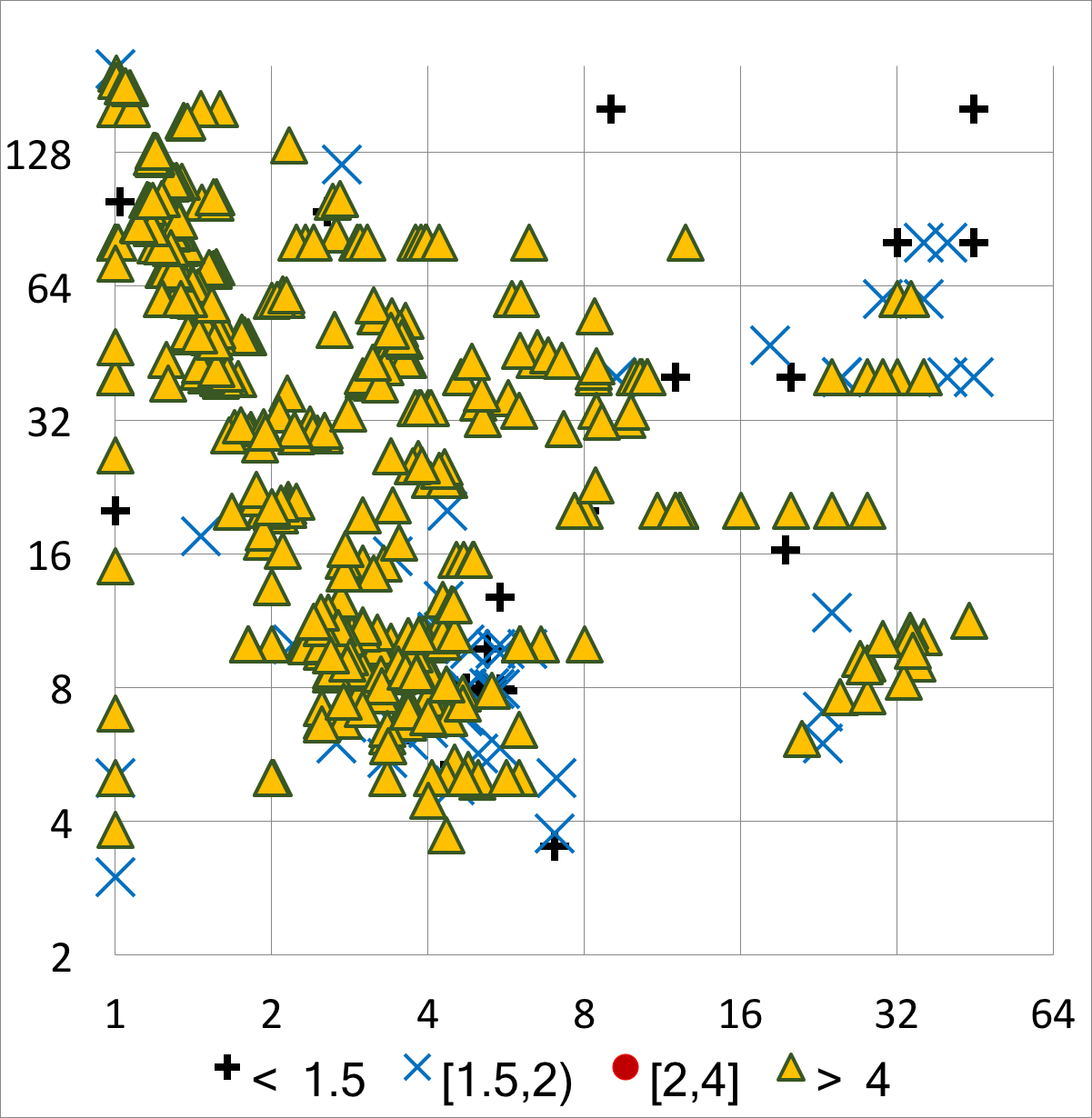}

  \caption{Ratio of \rtlf runtime to the minimum.}
  \label{fig:r2_l4}
\end{figure}


%% file: Y-to_instances.tex
\begin{table*}
  \centering
  \caption{OCT instances where LP reductions are effective.}
  \label{tab:to_instances}

  \medskip
  \small
  \begin{tabular}{|l|r|r|r||r|r|r|r||r|r|r|r|}
    \hline
    \multicolumn{4}{|c||}{} & \multicolumn{4}{c||}{runtime} & \multicolumn{4}{c|}{branches}
    \\
    \hline
00-Instance & spread & nad & oct \% & DF2 & r0\_l1 & r1\_l4 & CPLEX & DF2 & r0\_l1 & r1\_l4 & CPLEX \\ \hline
aa41-to-7 & 24.0 & 11.8 & 13.2 & t/o & t/o & 216.27 & \textbf{\emph{4.90}} & t/o & t/o & 365,790 & \textbf{\emph{6}} \\ \hline
aa42-to-7 & 3.7 & 6.8 & 10.8 & 67.31 & 89.82 & 7.19 & 3.13 & 496,208 & 496,173 & 12,428 & 26 \\ \hline
aa32-to-7 & 4.7 & 7.8 & 18.9 & \textbf{12.48} & 16.80 & 6.67 & 3.38 & 125,716 & 125,270 & 21,628 & 148 \\ \hline
aa29-to-7 & 4.5 & 5.4 & 9.0 & 51.04 & 65.76 & 4.18 & 2.97 & 396,403 & 389,063 & 5,292 & 4 \\ \hline
aa28-to-7 & 3.5 & 7.9 & 15.9 & 11.56 & 14.66 & 3.99 & 2.67 & 103,365 & 103,199 & 8,875 & 50 \\ \hline
aa17-to-7 & 3.7 & 6.4 & 14.7 & \textbf{6.97} & 9.06 & 3.84 & 3.59 & 57,942 & 57,876 & 6,606 & 166 \\ \hline
aa24-to-7 & 5.5 & 5.9 & 7.5 & 30.32 & 41.20 & 3.23 & 2.70 & 248,317 & 248,250 & 4,324 & 14 \\ \hline
aa20-to-7 & 4.5 & 5.0 & 7.8 & 15.82 & 18.81 & 2.71 & 2.83 & 132,851 & 132,628 & 3,036 & 29 \\ \hline
aa40-to-7 & 4.4 & 6.9 & 15.4 & \textbf{4.45} & \textbf{5.26} & 2.65 & 2.15 & 23,170 & 23,132 & 4,161 & 32 \\ \hline
aa19-to-7 & 4.5 & 5.1 & 8.6 & 6.11 & 6.49 & 1.60 & 2.36 & 39,821 & 39,753 & 2,654 & 1 \\ \hline
aa22-to-7 & 4.5 & 5.4 & 7.8 & 1.46 & 1.93 & 0.41 & 0.32 & 5,880 & 5,874 & 506 & 1 \\ \hline
\textbf{aa46-to-7} & 4.5 & 4.8 & 8.9 & 0.81 & \textbf{0.67} & 0.38 & 1.93 & 2,345 & 2,325 & 345 & 1 \\ \hline
\textbf{aa34-to-7} & 3.3 & 5.6 & 9.3 & 0.75 & 0.90 & 0.32 & 0.86 & 3,355 & 3,334 & 385 & 1 \\ \hline
aa33-to-7 & 7.0 & 3.8 & 1.5 & 0.17 & 0.20 & 0.05 & 0.06 & 344 & 344 & 30 & 1 \\ \hline
j20-to-7 & 7.0 & 3.5 & 0.0 & 0.13 & 0.17 & 0.01 & 0.02 & 199 & 199 & 1 & 1 \\ \hline

  \end{tabular}

  \medskip
  Instances are sorted by decreasing r1\_l4 runtime.

  


\end{table*}


%% file: Y-oct_vs_lp.tex
\begin{figure}
  \centering
  \includegraphics[width=\columnwidth]{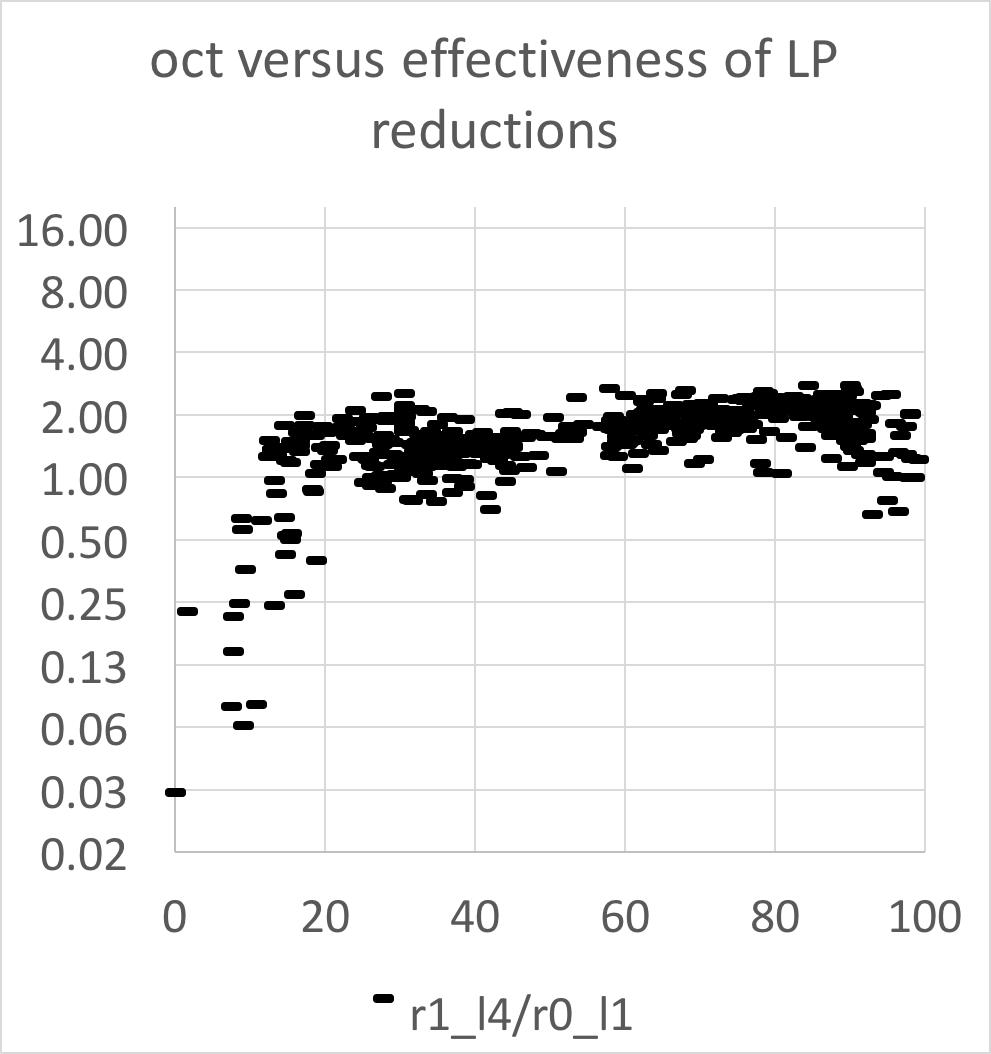}

  \medskip
  \begin{minipage}{\columnwidth}
    The x-axis is a good oct percentage upper bound; the y-axis represents
    the ratio between runtime of a config without \lpRed{} and one with
    \lpRed{} added.
    Up to 20\%, adding \lpRed{} is a good idea. Beyond that,
    \textsf{r0\_l1} is always within at least a factor of 2.
  \end{minipage}
  
  \caption{Effectiveness of LP reductions as a function of estimated oct.}
  \label{fig:oct_vs_lp}
\end{figure}


%% file: Y-hardness.tex
\begin{figure}
  \centering
  \includegraphics[width=\columnwidth]{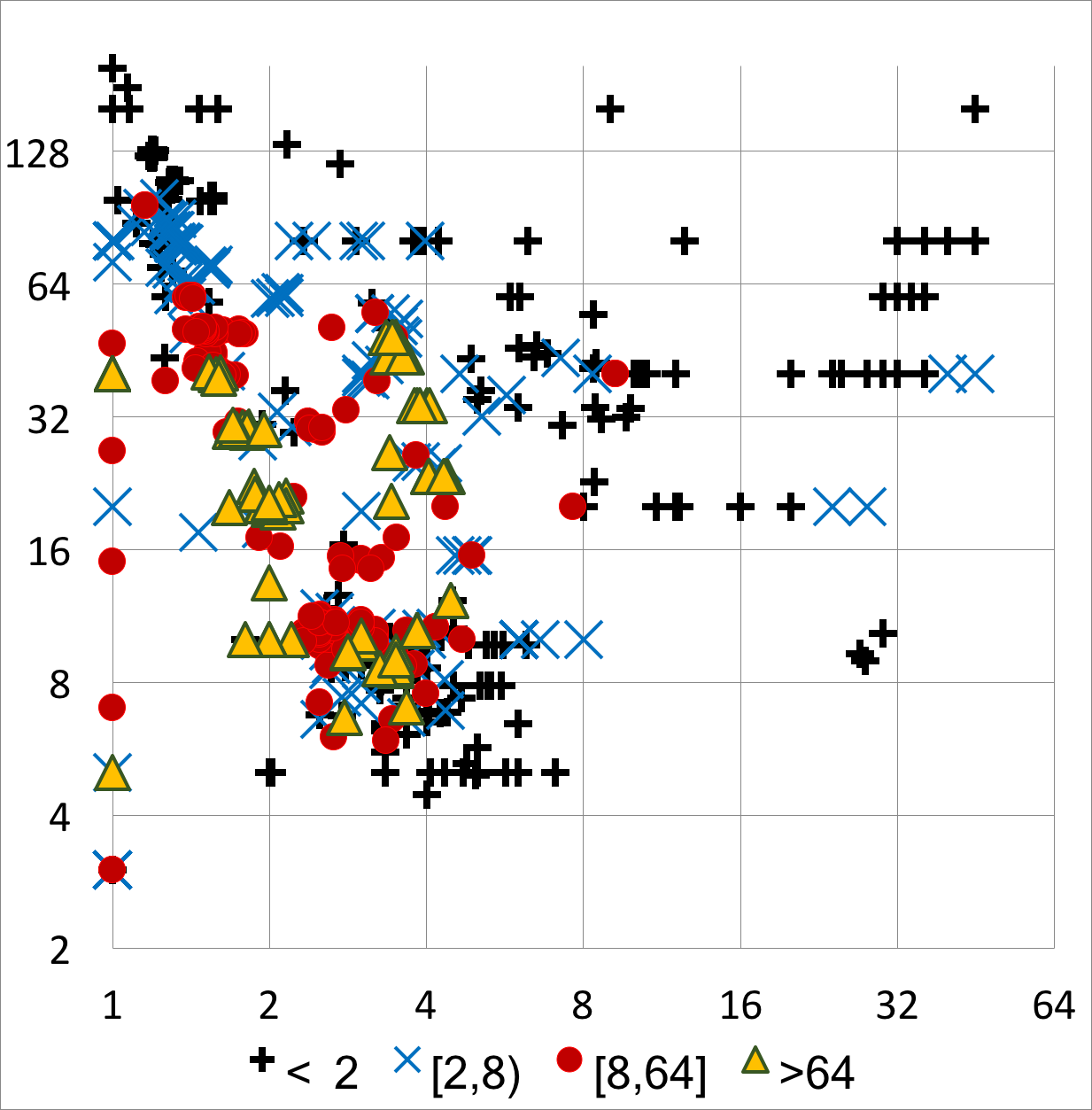}

  
  \caption{Minimum runtimes (sec) for instances with less than 300 vertices.}
  \label{fig:hardness}
\end{figure}


%% file: Y-pace_2019.tex
\begin{figure}
  \centering
  \includegraphics[width=\columnwidth]{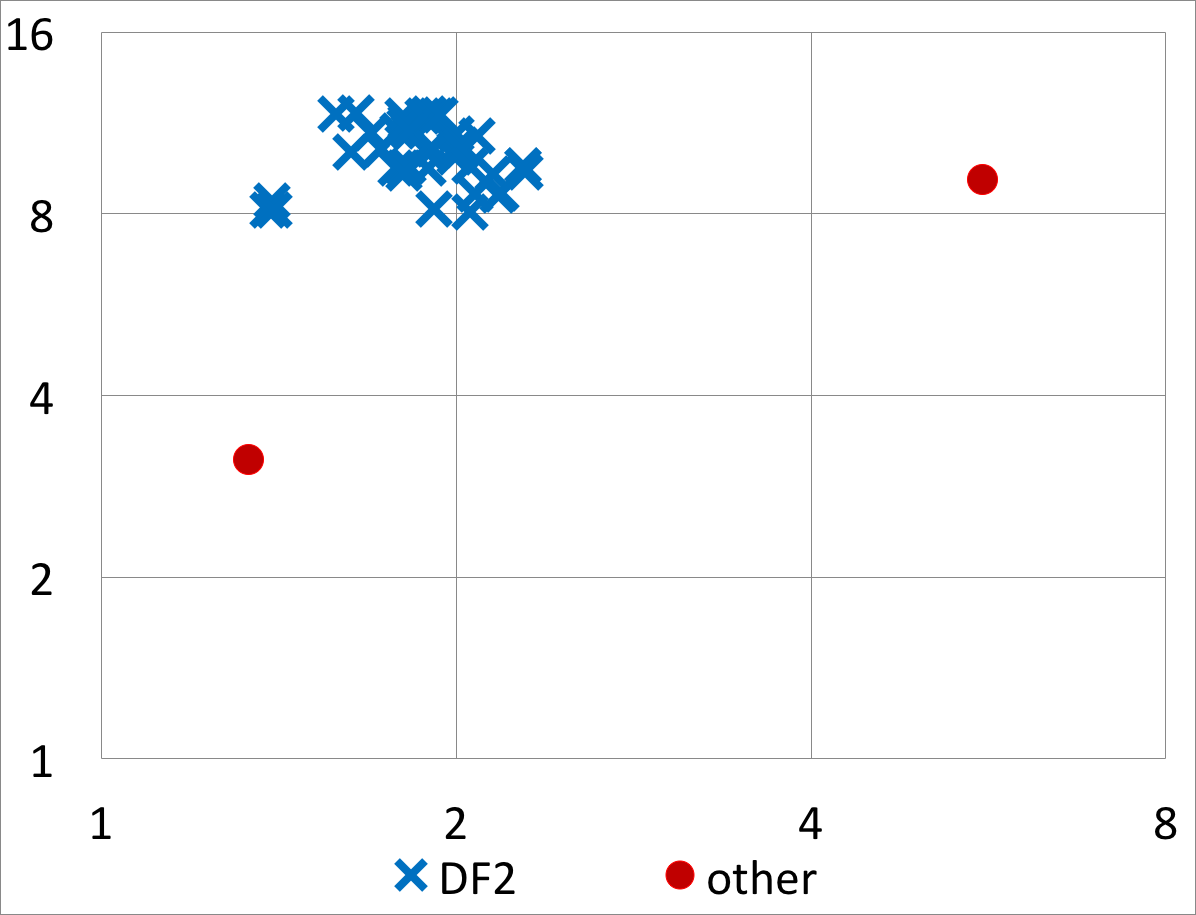}

  \medskip
  
  \caption{Runtime landscape for selected 2019 PACE challenge instances.}
  \label{fig:pace_2019}
\end{figure}


%% file: 4-special_cases.tex
We turn now to instances where our hypotheses do not apply, analyzing these
in detail.
The main takeaways are (i)~large, sparse graphs, if amenable to branch and
reduce at all, require a broader suite of reductions -- but these are still
somewhat predictable; and (ii)~graphs with
special structure, e.g., geometric and planar graphs, benefit from customized \configs.

\subsection{Large sparse networks} \label{sec:real_world}

\input{Y-real_world}

Akiba and Iwata~\cite{BB:Akiba16} report results for a corpus of sparse
real-world networks (Table~1 in their paper). Most of these are either
trivial (the instance is reduced at the root with runtimes less than two seconds) or unsuitable for a \br implementation. In the latter
category are the road networks and meshes
motivated by graphics, which have average degree between 3 and 6 and \spread
$<2$, i.e., they confirm Hypothesis~\ref{hyp:hardness}.
Table~\ref{tab:real_world}
shows runtime results and degree statistics for five moderately difficult
instances and one that is out of reach for all but the full blown \rthreelf config.
To accommodate instances of this size we used the more powerful
server and a 24-hour timeout.

Large instances make it extremely important to reduce as many vertices as
possible at the root, thus decreasing the number of branches (exponential in
the number of undecided vertices).
Two reductions that are particularly good at this are (i)~\unconfinedRed, as observed
both experimentally -- Table~\ref{tab:reduction_effectiveness_efficiency}(a),
and theoretically -- see Xiao and Nagamochi~\cite{BB:Xiao13,BB:Xiao17}; and
(ii)~\lpRed -- these are linear time after the $O(m\sqrt{n})$ preprocessing
and are unique in their ability to reduce a large number of vertices at once.

It also appears that instances with moderate density and degree spread
  (\textsf{web-NotreDame} and \textsf{baidu-relatedpages}) favor
  \unconfinedRed reductions, while those that are very sparse with large degree
  spread (\textsf{libimseti} and \textsf{petster-friendships-dog}) favor
  \lpRed reductions.
  The reason for this is unclear. In any case the favored reductions lead to only
  a small amount of branching.
  If both \unconfinedRed and \lpRed reductions are added to \cheap, we get
  competitive runtimes for all but the \textsf{web-Stanford} instance.
  
  OCT percentages, below 20\% for all of these instances, are
  misleading -- what matters is the OCT value \emph{after} low degree
  vertices have been removed by \degRed{} and \foldRed{} reductions.
  We have not performed this measurement, but the low density and large
  spread of the two instances that favor \lpRed reductions
  (\textsf{libimseti} and \textsf{petster-friendships-dog}) suggest that the
  simple reductions prune a large percentage of vertices.

The first instance, \textsf{as-skitter} (a social network) is hard primarily
because of its size -- runtimes are in the same two-hour ballpark for all but
one config, and for CPLEX.
The last instance, \textsf{web-Stanford}, has the \nad and \spread
similar to those of \textsf{web-NotreDame} and \textsf{baidu-relatedpages},
but a
more detailed look at its profile in relation to the others reveals some important characteristics:
(i)~less
than 5\% of the vertices are degree-1; and
(ii)~its maximum degree is considerably less, proportionally, than that of
\textsf{baidu-relatedpages} (so clique lower bounds are less
likely).\footnote{Statistics reported by \VCSP show that \emph{all} lower
  bounds for \textsf{baidu-relatedpages} are clique lower bounds, while the
  \rthreelf config relies heavily on LP lower bounds when solving
  \textsf{web-Stanford}.}

The \textsf{Cheap+LPU}, r2\_l4, and r3\_l4 configs solve
\textsf{petster-friendships-dog} without branching and
\textsf{baidu-relatedpages} with less than ten branches. Runtimes for all
three of these configs are roughly the same on these two instances.
This supports our conjecture that eliminating many vertices early is
important.

Of special note is the \textsf{libimseti} instance (another social network).
It has by far the largest spread and it appears that the presence of many
vertices of moderately large degree causes reductions such as dominance and
unconfined to be inefficient. The most time consuming reductions for r2\_l4
and r3\_l4 on this instance are \textsf{funnel} reductions.
These are the likely reasons for \cheapLP to be more than a factor of
\emph{five} faster than \rthreelf.

\input{Y-real_world-effectiveness_efficiency}

Finally,
note that the \config
that includes everything \emph{except dominance} (and packing) is competitive
in all but the web-Stanford instance.
Table~\ref{tab:real_world-effectiveness_efficiency} has more detailed
information about relative efficiency and effectiveness of reductions on
these instances.

\subsection{Geometric and Planar Graphs}

Two other graph categories that are not amenable to simple \configs such as
\dft, \rOlI, or \cheap are the geometric and planar graphs we generated.

\input{Y-geo_drawings}

\input{Y-hard_geo}

\medskip\noindent
\textbf{Geometric graphs.} Table~\ref{tab:hard_geo}
shows data for the harder 512-vertex geometric graphs.
Any \config that did not include \domRed reductions timed out on all
instances
and the full suite provided by \rthreelf gave the best runtimes.

For the sparsest geometric graphs, not shown in the table (runtimes for competitive
\configs were less than $1/100$ second),
\domRed is effective on its own because there are many
dominated degree-3 vertices -- see Fig.~\ref{fig:geo_drawings}(a).
 In the mid-range, Fig.~\ref{fig:geo_drawings}(b),
CPLEX is the better
choice; the neighborhood of most vertices does
not yield opportunities for dominance or even unconfined reductions; and
the induced cliques overlap.
At the highest densities, also not shown, there are enough large
cliques so that \domRed{} can reduce the graph at the root.

All of these instances are easy for CPLEX -- the only reason it does not
always have the best runtimes is because of preprocessing at
the root.

Table~\ref{tab:reduction_eff_geo} shows that the profile of reduction
effectiveness is radically different for geometric graphs than for the
general population. The \domRed and \unconfinedRed reductions play a much
more pronounced role. Profiles of easier geometric instances are also
shown -- there, \degRed, \domRed, and \foldRed reductions, the
ones composing \rOlI, do almost all the work.

\input{Y-planar_instances}

\medskip\noindent
\textbf{Planar graphs.}
While complete Delaunay triangulations are not necessarily representative of
maximal planar graphs, nor are their duals representative of 3-regular planar
graphs, these two classes do illustrate another situation necessitating more
complex \configs.

The full triangulations -- Table~\ref{tab:planar_instances}(a), are
remarkable in their wide range of difficulty.
All have the same number of edges, 2994, and almost exactly the same degree spread,~2.
We ran \rthreelf on~30 different permutations (vertices renumbered, edges
reordered) of the most difficult instance, \textsf{tri\_inf-1000\_9}, and
found (only) a factor of two difference between minimum and maximum runtime
with a small standard deviation (roughly~3).
There was even less variance for the easiest instance,
\textsf{tri\_inf-1000\_7}.
So the difference must lie in subtle structural properties, such as higher
degree vertices with degree-3 neighbors, leading to degree-2 reductions
(\foldRed or \domRed on degree-2 vertices).

The duals -- Table~\ref{tab:planar_instances}(b), are less prone to varying runtimes, but still more so than can be
accounted for by input permutation. Statistics for permuted runs are almost
identical to those of the full triangulations.
And nothing stands out when looking at differences in efficiency and
effectiveness of reductions.
Table~\ref{tab:planar_instances}(c) shows runtimes for 300-vertex 3-regular
graphs, which have comparable runtimes but are much smaller.
Simple reductions suffice here and runtimes vary much less.

\input{Y-reduction_eff_geo}

Table~\ref{tab:reduction_eff_geo} shows relative efficiency and effectiveness
of reductions for geometric, planar, and 3-regular graphs in comparison with
the main goldilocks corpus.
We have already discussed geometric graphs.
Duals of triangulations have profiles very similar to 3-regular graphs,
except for the importance of desk reductions in the former: a vertex of
degree four in the original triangulation leads to a chordless 4-cycle in the
dual.
And 3-regular graphs differ from the general population in that \lpRed
reductions are neither efficient nor effective in the former. Nor are they
efficient/effective in any of the geometric or planar classes.
Finally, the full triangulations differ mostly from the 3-regular ones in the
prominence of dominance reductions.
The triangulations, like the sparse geometric graphs, are likely to have
$K_4$'s with an 'unconnected' middle vertex -- see Fig.~\ref{fig:geo_drawings}(a).


%% file: Y-real_world.tex
\begin{table*}
  \caption{Large sparse real-world instances in the goldilocks zone.}
  \label{tab:real_world}

  \medskip
  \begin{minipage}{\textwidth}
    \begin{center}
    (a) Runtimes for a selection of competitive
      configurations.

      \smallskip
    \begin{tabular}{|l||r|r|r|r|r|r|}
      \hline
      Instance & Cheap+U & Cheap+LP & Cheap+LPU & r2\_l4 & min & CPLEX \\ \hline
      as-skitter & \emph{7019.8} & 17,245.3 & \emph{8595.3} & \emph{7922.7}
      & \textbf{\emph{5548.5}}\footnote{Achieved by DF2+U} & \emph{6968.1} \\ \hline\hline
      web-NotreDame & \textbf{33.4} & t/o & \textbf{31.6} & \textbf{33.2} &
      31.6 & num\footnote{CPLEX terminated before proving optimality due to
        reaching numercial tolerance limit. See~\cite{BB:Akiba16}.} \\ \hline
baidu-relatedpages & \textbf{2.9} & t/o & \textbf{2.7} & \textbf{2.9} & 2.7 & 856.3
\\ \hline
\hline
\textbf{\emph{libimseti}} & t/o & \textbf{\emph{468.9}} & \emph{668.1} &
\textsf{1651.7}\footnote{The r3\_l4 config used by Akiba and
  Iwata~\cite{BB:Akiba16} took \textsf{2025.1} seconds, almost a factor of
  five worse than \textsf{Cheap+LP}.} & 468.9 &
mem\footnote{Ran out of memory.} \\ \hline
petster-friendships-dog & t/o & \emph{50.3} & \emph{66.3} & \emph{59.1} &
\textbf{\emph{39.1}}\footnote{Achieved by DF2+LPU}
 & 1487.3 \\ \hline\hline
web-Stanford & t/o & t/o & t/o & t/o & 38,960.0\footnote{Achieved by r3\_l4.}
& num
\\ \hline
    \end{tabular}

    \medskip

    (b) Degree statistics for the instances.

    \smallskip
    \small
    \begin{tabular}{|l||r|r|r|r|r|r|r|r|r|r|r|}
      \hline
Instance & n & m & min & b & med & t & max & nad & spread \\ \hline
as-skitter & 1,696,415 & 11,095,298 & 1 & 1 & 5 & 37 & 35,455 & 13.08 & 37 \\ \hline
\hline
web-NotreDame & 325,729 & 1,103,836 & 1 & 1 & 2 & 24 & 10,721 & 6.78 & 24 \\ \hline
baidu-relatedpages & 415,641 & 2,374,053 & 1 & 1 & 8 & 23 & 127,066 & 11.42 & 23 \\ \hline
\hline
libimseti & 220,970 & 17,233,144 & 1 & 1 & 57 & 542 & 33,389 & 0.14 & 542 \\ \hline
petster-friendships-dog & 426,820 & 8,545,065 & 1 & 1 & 12 & 105 & 46,504 &
0.02 & 105 \\ \hline
\hline
web-Stanford & 281,903 & 1,992,636 & 1 & 2 & 6 & 38 & 38,625 & 14.14 & 19
\\ \hline
    \end{tabular}
    
    \end{center}
    
    \medskip
  \end{minipage}

  \fbox{
    \begin{minipage}{0.9\textwidth}
      Time limit was 24 hours or 86,400 seconds.
      Runtimes are in italics if competitive, bold if within~1.1 of minimum
      and bold-italic if minimum (usually by a large margin).

      The \textsf{libimsetti} instance stands out: \cheapLP
      outperforms \rthreelf by a factor of more than five.

      In two of the instances, as-skitter and petster-friendships-dog, a
      \textsf{DF2+$x$} \config significantly outperformed the corresponding
      \textsf{Cheap+$x$} \config. These minima are highlighted in the
      \textsf{min} column of the table.
      In all but one other case, the difference was slight.
      However,in the case of libimsetti, the \textsf{DF2+$x$} \configs took almost twice
      as long as the corresponding \textsf{Cheap+$x$} \configs.
    \end{minipage}
  }

\end{table*}


%% file: Y-real_world-effectiveness_efficiency.tex
\begin{table*}
  \caption{Efficiency and effectiveness of reductions on large sparse
    real-world instances.
  }
  \label{tab:real_world-effectiveness_efficiency}

  \medskip

  \begin{minipage}{\textwidth}
  (a) Efficiency -- microseconds per vertex reduced. The \emph{gold} column
  gives the geometric mean for the goldilocks instances in the general
  corpus. Reductions are sorted by increasing efficiency of the goldilocks
  instances.

  \begin{center}
  \medskip
  \begin{tabular}{|l||r||r||r|r||r|r||r||}
    \hline
    reduction & gold. & ~~~as-k & web-N & ~~baidu & ~~libim & ~~~pets & web-S \\
    \hline\hline
    fold2 & 1.3 & 1.1 & 2.4 & 5.1 & 161.6 & 152.0 & 1.2 
    \\
    \hline
    deg1 & 8.1 & 7.3 & 2.3 & 0.8 & 5.8 & 1.4 & 7.3
    \\
    \hline
    unconf. & 21.6 & 39.2 & 52.7 & 10.9 & 45,880.2 & 71.4 & 24.9
    \\
    \hline
    lp & 70.0 & 1,124.8 & 19.9 & 5.7 & 182.2 & 6.4 & 112.7
    \\
    \hline
    pack. & 69.4 & 13.2 & 9.2 & \textbf{\emph{136,113.0}}\footnote[1]{No vertices were
      reduced. The time reported is the total time spent attempting the reductions.}
    & 25,462.9 & \textbf{\emph{663,132.0}} & 36.5
    \\
    \hline
    fun. & 136.3 & 32.5 & 647.3 & \textbf{\emph{179,117.0}}
    & 95,881.2
    & \textbf{\emph{24.7}} & 32.1
    \\
    \hline
    twin & 117.2 & 237.6 & 3.7 & 52.5 & 5,135.2 & 184.1 & 67.6
    \\
    \hline
    desk & 129.7 & 355.4 & 3.0 & 301.1 & \textbf{\emph{701,832.4}} & 397.6 & 71.1
    \\
    \hline
    dom. & 331.7 & 54.7 & 37.7 & 37.2 & \textbf{\emph{0.0}} & \textbf{\emph{37.7}} & 69.2 
    \\
    \hline
  \end{tabular}
  \end{center}

  \medskip
  (b) Effectiveness, the percent of vertices reduced by each
  reduction. Reductions are sorted by decreasing effectiveness on the
  goldilocks instances.

 \begin{center}
  \medskip
  \begin{tabular}{|l||r||r||r|r||r|r||r||}
    \hline
    reduction & gold. & ~~~as-k & web-N & ~~baidu & ~~libim & ~~~pets & web-S \\
    \hline\hline
    fold2 & 62.4 & 74.5 & 26.9 & 21.9 & 21.7 & 46.1 & 75.9
    \\
    \hline
    unconf. & 9.9 & 6.2 & 2.2 & 7.0 & 0.3 & 3.4 & 7.1
    \\
    \hline
    lp & 3.5 & 0.1 & 3.1 & 5.3 & 74.4 & 20.4 & 1.2
    \\
    \hline
    pack. & 2.4 & 5.9 & 1.4 & \textbf{\emph{0.0}} & 0.1 & \textbf{\emph{0.0}} & 0.4
    \\
    \hline
    deg1 & 1.9 & 7.5 & 20.9 & 64.7 & 2.6 & 19.2 & 6.8
    \\
    \hline
    fun. & 1.2 & 3.3 & 0.1 & \textbf{\emph{0.0}} & 0.2 & \textbf{\emph{0.0}} & 2.2
    \\
    \hline
    desk & 0.8 & 0.3 & 25.0 & 0.1 & \textbf{\emph{0.0}} & 0.5 & 3.6 
    \\
    \hline
    twin & 0.5 & 0.2 & 18.0 & 0.5 & 0.7 & 10.4 & 1.6
    \\
    \hline
    dom. & 0.3 & 2.0 & 2.3 & 0.7 & \textbf{\emph{0.0}} & \textbf{\emph{0.0}} & 1.4
    \\
    \hline
  \end{tabular}
  \end{center}
  \end{minipage}

  \medskip
  \fbox{
    \begin{minipage}{0.9\textwidth}
      With the exception of the as-skitter and web-Stanford instances, the reduction
      effeciency/effectiveness profiles of these large sparse instances
      differs radically from those of the general population.
      There are more \degRed reductions -- not surprising,
      but there are \emph{significantly fewer} \foldRed reductions.

      The tables clearly point out why \cheapLP is so effective for
      libimsetti.
      \lpRed reductions do most of the work when using the \rthreelf \config, on
      which the data are based, but a lot of time is wasted on unsuccessful
      \packingRed, \unconfinedRed, \funnelRed, and \deskRed reductions. 
    \end{minipage}
  }
\end{table*}


%% file: Y-geo_drawings.tex
\begin{figure*}
  \begin{center}
    \begin{tabular}{l @{~~~~~~~~~} r}
      \includegraphics[width=0.35\textwidth]{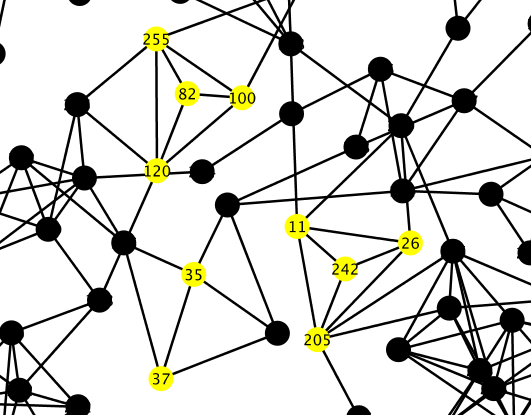}
      &
      \includegraphics[width=0.28\textwidth]{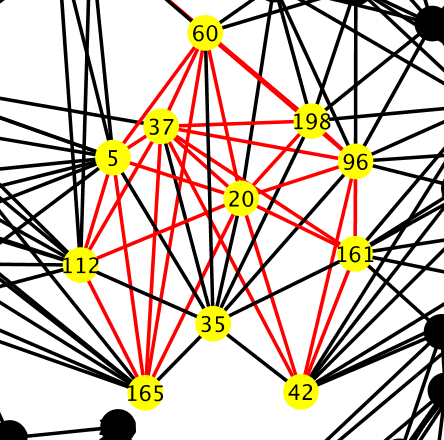}

      \\ \\
      (a)~256 vertices, 512 edges
      &
      (b)~256 vertices, 2048 edges
    \end{tabular}

    \medskip
    \fbox{
      \begin{minipage}{0.9\textwidth}
        (a)~In the sparser geometric graphs there are many $K_4$'s with a
        middle vertex not connected to the rest of the graph or vertices at
        the junction of two triangles, such as vertex~37.

        (b)~The highlighted vertices and red edges form (at least) three overlapping
        $K_5$'s: vertex sets $\{5,20,37,112,165\}$, $\{20,37,60,96,198\}$
        (the edge between 60 and 96 is obscured), and $\{20,37,42,96,161\}$.
      \end{minipage}
    }
    
  \end{center}
  \caption{Closeup view of two geometric graphs.}
  \label{fig:geo_drawings}
\end{figure*}


%% file: Y-hard_geo.tex
\begin{table*}
  \centering
  \caption{Runtime and branching data for harder 512-vertex geometric instances.}
  \label{tab:hard_geo}

  \medskip
  \small
  \begin{tabular}{|l||r|r||r|r|r|r||r|r|r|r|}
    \hline
    Instance & edges & spread & \multicolumn{4}{c||}{runtime} & \multicolumn{4}{c|}{branches}
    \\
    \cline{4-11}
 &  & & DD & r0\_l1+U & r3\_l4 & CPLEX & DD & r0\_l1+U & r3\_l4 & CPLEX \\ \hline\hline
W-16384-0 & 15,968 & 1.4 & 900.0 & 900.0 & \textbf{\emph{707.3}} & \textbf{\emph{4.5}} & 2,723,618 & 927,514 & 280,594 & 80 \\ \hline
W-16384-1 & 15,868 & 1.4 & 900.0 & 900 & \textbf{\emph{567.8}} & \textbf{\emph{3.4}} & 2,790,936 & 950,663 & 237,319 & 27 \\ \hline
W-16384-4 & 15,920 & 1.4 & 900.0 & 900 & \textbf{\emph{485.2}} & \textbf{\emph{3.0}} & 3,064,004 & 1,040,497 & 198,080 & 51 \\ \hline
W-16384-3 & 16,118 & 1.6 & 900.0 & 900.0 & \textbf{\emph{420.8}} & \textbf{\emph{4.0}} & 3,125,229 & 1,028,770 & 176,676 & 85 \\ \hline
W-16384-2 & 16,020 & 1.7 & 900 & 900 & \textbf{\emph{372.7}} & \textbf{\emph{4.8}} & 3,396,657 & 1,042,980 & 161,528 & 185 \\ \hline
W-08192-0 & 7,026 & 2.5 & 22.0 & \textbf{\emph{1.2}} & \textbf{\emph{1.3}} & \textbf{\emph{0.5}} & 272,410 & 868 & 328 & 7 \\ \hline
G-08192-0 & 7,026 & 2.5 & 22.4 & \textbf{\emph{1.1}} & \textbf{\emph{1.2}} & \textbf{\emph{0.5}} & 272,410 & 868 & 328 & 7 \\ \hline
W-04096-2 & 3,954 & 2.6 & 225.7 & 1.3 & \textbf{\emph{1.1}} & \textbf{\emph{0.4}} & 2,497,458 & 1,234 & 400 & 0 \\ \hline
G-08192-1 & 7,046 & 2.2 & 22.2 & 1.1 & \textbf{\emph{1.0}} & \textbf{\emph{0.5}} & 279,029 & 1,088 & 290 & 5 \\ \hline
W-08192-1 & 7,046 & 2.2 & 21.9 & 1.1 & \textbf{\emph{1.0}} & \textbf{\emph{0.5}} & 279,029 & 1,088 & 290 & 5 \\ \hline
W-04096-3 & 3,933 & 2.6 & 900 & 1.9 & \textbf{\emph{0.9}} & \textbf{\emph{0.0}} & 11,460,458 & 1,262 & 228 & 417 \\ \hline
W-08192-3 & 7,188 & 2.6 & 10.8 & \textbf{\emph{0.7}} & \textbf{\emph{0.8}} & \textbf{\emph{0.4}} & 94,997 & 378 & 134 & 0 \\ \hline
G-16384-4 & 13,623 & 2.2 & 2.1 & 0.8 & \textbf{\emph{0.8}} & \textbf{\emph{0.3}} & 11,661 & 440 & 136 & 0 \\ \hline
G-08192-3 & 7,188 & 2.6 & 10.7 & 1.7 & \textbf{\emph{0.7}} & \textbf{\emph{0.4}} & 94,997 & 378 & 134 & 0 \\ \hline
G-08192-2 & 7,161 & 2.7 & 13.0 & \textbf{\emph{0.6}} & \textbf{0.7} & \textbf{\emph{0.6}} & 179,228 & 430 & 150 & 9 \\ \hline
W-08192-2 & 7,161 & 2.7 & 12.9 & \textbf{\emph{0.5}} & \textbf{\emph{0.6}} & \textbf{\emph{0.6}} & 179,228 & 430 & 150 & 9 \\ \hline
G-16384-3 & 13,547 & 2.7 & 0.7 & \textbf{\emph{0.4}} & \textbf{0.6} & \textbf{\emph{0.5}} & 2,806 & 80 & 72 & 0 \\ \hline
W-04096-4 & 3,944 & 2.6 & 209.0 & \textbf{0.6} & \textbf{\emph{0.5}} & \textbf{\emph{0.2}} & 1,686,029 & 274 & 60 & 0 \\ \hline
G-16384-2 & 13,528 & 2.6 & \emph{0.7} & \textbf{\emph{0.4}} & \textbf{0.4} & \textbf{0.6} & 4,791 & 62 & 60 & 0 \\ \hline
G-16384-0 & 13,346 & 2.3 & 1.1 & \textbf{\emph{0.2}} & 0.5 & \textbf{0.3} & 4,266 & 62 & 54 & 0 \\ \hline
W-04096-0 & 3,771 & 2.2 & 24.8 & \textbf{\emph{0.1}} & \textbf{0.2} & \emph{0.2} & 267,644 & 84 & 42 & 0 \\ \hline
G-08192-4 & 7,205 & 2.6 & 15.2 & \textbf{\emph{0.1}} & \textbf{0.1} & 0.9 & 145,475 & 50 & 30 & 158 \\ \hline
W-08192-4 & 7,205 & 2.6 & 15.4 & \textbf{\emph{0.1}} & \textbf{0.1} & 0.9 & 145,475 & 50 & 30 & 158 \\ \hline
G-04096-0 & 3,611 & 2.8 & 1.0 & \textbf{\emph{0.1}} & \textbf{\emph{0.1}} & 0.3 & 12,868 & 26 & 23 & 0 \\ \hline
G-04096-3 & 3,747 & 3.1 & 7.8 & \textbf{\emph{0.1}} & \textbf{\emph{0.1}} & \textbf{0.1} & 129,550 & 28 & 20 & 0 \\ \hline
W-04096-1 & 3,922 & 2.4 & 124.1 & \textbf{\emph{0.1}} & \textbf{0.1} & 0.4 & 1,752,487 & 32 & 14 & 3 \\ \hline
G-16384-1 & 13,291 & 2.2 & 0.4 & \textbf{0.1} & \textbf{\emph{0.1}} & \emph{0.2} & 1,576 & 10 & 10 & 0 \\ \hline
  \end{tabular}

  \fbox{
    \begin{minipage}{\textwidth}
      The 'W' instances are geometric graphs with wraparound, e.g., points
      at distance $d$ from the left edge of the unit square are treated as if
      they were distance $d$ to the right of the right edge.
      The 'G' instances have no wraparound.
      Instance numbers reflect the desired number of edges -- actual number
      of edges differ from these.
      
      The \configs that do not include dominance timed out on these instances
      and \rthreelf always performed at least as well as \rtlf.
      Except for the easier instances, CPLEX has the best runtimes, but we
      indicate the best runtimes among the \VCS configs in bold italics nonetheless.
    \end{minipage}
  }

\end{table*}


%% file: Y-planar_instances.tex
\begin{table*}
  \centering
  \caption{Runtime and branching data for planar and 3-regular graphs.}
  \label{tab:planar_instances}

  \medskip
  \small
  \begin{tabular}{|l||r|r|r|r||r|r|r|r|}
    \hline
    & \multicolumn{4}{c||}{runtime} & \multicolumn{4}{c|}{branches}
    \\
    \hline
Instance & r0\_l1+FU & r2\_l4 & r3\_l4 & CPLEX & r0\_l1+FU & r2\_l4 & r3\_l4 & CPLEX \\ \hline
tri\_inf-1000\_9 & 169.3 & 207.8 & \textbf{\emph{24.4}} & \textbf{\emph{5.9}} & 100,179 & 200,689 & 16,353 & 383 \\ \hline
tri\_inf-1000\_4 & \textbf{6.2} & 11.0 & \textbf{\emph{4.7}} & \textbf{\emph{2.2}} & 3,757 & 7,500 & 2,137 & 0 \\ \hline
tri\_inf-1000\_2 & \textbf{\emph{0.4}} & \textbf{0.6} & \textbf{0.5} & 3.9 & 107 & 211 & 151 & 170 \\ \hline
tri\_inf-1000\_6 & \textbf{0.6} & 1.0 & \textbf{\emph{0.4}} & 3.3 & 122 & 245 & 73 & 3 \\ \hline
tri\_inf-1000\_3 & \textbf{1.0} & 1.4 & \textbf{\emph{0.7}} & 9.2 & 172 & 345 & 117 & 1,712 \\ \hline
tri\_inf-1000\_0 & \textbf{\emph{0.4}} & \textbf{0.5} & \textbf{0.4} & 2.9 & 68 & 137 & 63 & 0 \\ \hline
tri\_inf-1000\_8 & \textbf{\emph{0.3}} & \emph{0.5} & \textbf{0.5} & 2.6 & 48 & 97 & 83 & 0 \\ \hline
tri\_inf-1000\_1 & \textbf{\emph{0.1}} & \emph{0.2} & \emph{0.2} & 1.9 & 22 & 45 & 43 & 0 \\ \hline
tri\_inf-1000\_5 & \textbf{\emph{0.2}} & \textbf{0.3} & \emph{0.4} & 3.4 & 42 & 85 & 75 & 158 \\ \hline
tri\_inf-1000\_7 & \textbf{\emph{0.1}} & \textbf{0.2} & \textbf{\emph{0.1}} & 29.9 & 25 & 51 & 35 & 4,261 \\ \hline
  \end{tabular}

  \smallskip
  (a)~Full triangulations, including infinite face. 
  
  \medskip
  \begin{tabular}{|l||r|r|r|r||r|r|r|r|}
    \hline
    & \multicolumn{4}{c||}{runtime} & \multicolumn{4}{c|}{branches}
    \\
    \hline
Instance & Cheap+FU & r2\_l4 & r3\_l4 & CPLEX & Cheap+FU & r2\_l4 & r3\_l4 & CPLEX \\ \hline
dual\_1024-4 & \textbf{15.0} & \emph{17.7} & \textbf{\emph{9.8}} & \textbf{\emph{1.8}} & 19,951 & 8,565 & 4,205 & 6 \\ \hline
dual\_1024-6 & \textbf{13.3} & \textbf{15.4} & \textbf{\emph{9.2}} & \textbf{\emph{2.5}} & 16,107 & 7,113 & 3,695 & 0 \\ \hline
dual\_1024-1 & \emph{14.5} & \emph{14.3} & \textbf{\emph{8.2}} & \textbf{\emph{2.7}} & 16,076 & 6,891 & 3,665 & 139 \\ \hline
dual\_1024-9 & \emph{13.9} & \emph{17.0} & \textbf{\emph{6.8}} & \textbf{\emph{2.0}} & 15,119 & 6,761 & 2,667 & 50 \\ \hline
dual\_1024-5 & \emph{8.0} & 9.8 & \textbf{\emph{4.0}} & \textbf{\emph{2.4}} & 8,559 & 3,559 & 1,333 & 0 \\ \hline
dual\_1024-2 & 9.2 & 10.1 & \textbf{\emph{3.3}} & \textbf{\emph{2.5}} & 10,169 & 4,378 & 1,227 & 29 \\ \hline
dual\_1024-8 & \textbf{5.6} & \textbf{6.1} & \textbf{\emph{4.9}} & \textbf{\emph{3.0}} & 6,227 & 2,071 & 1,128 & 331 \\ \hline
dual\_1024-3 & 10.5 & 11.7 & \textbf{\emph{3.4}} & \textbf{\emph{1.5}} & 12,665 & 5,272 & 1,077 & 0 \\ \hline
dual\_1024-7 & 3.8 & 3.8 & \textbf{\emph{1.7}} & \emph{3.1} & 2,036 & 838 & 285 & 77 \\ \hline
dual\_1024-0 & \textbf{1.6} & \emph{2.5} & \textbf{\emph{1.4}} & \textbf{2.1} & 921 & 407 & 192 & 21 \\ \hline
  \end{tabular}

  \smallskip
  (b)~Duals of full triangulations.

  \medskip
  \begin{tabular}{|l||r|r|r|r||r|r|r|r|}
    \hline
    & \multicolumn{4}{c||}{runtime} & \multicolumn{4}{c|}{branches}
    \\
    \hline
00-Instance & DF2 & r2\_l4 & r3\_l4 & CPLEX & DF2 & r2\_l4 & r3\_l4 & CPLEX \\ \hline
reg3-300\_4 & \textbf{\emph{6.8}} & \emph{12.2} & \emph{12.2} & \textbf{\emph{4.2}} & 112,353 & 32,701 & 24,548 & 3,446 \\ \hline
reg3-300\_2 & \textbf{\emph{6.7}} & \textbf{10.0} & \textbf{10.1} & \textbf{7.3} & 103,040 & 24,301 & 19,696 & 7,542 \\ \hline
reg3-300\_9 & \textbf{\emph{6.6}} & \textbf{9.8} & \emph{10.4} & \emph{11.3} & 84,795 & 22,361 & 17,263 & 10,516 \\ \hline
reg3-300\_7 & \textbf{\emph{6.3}} & \emph{10.0} & \emph{10.0} & 13.0 & 101,756 & 24,074 & 19,184 & 11,697 \\ \hline
reg3-300\_5 & \textbf{\emph{6.0}} & \emph{10.8} & \emph{10.0} & 20.1 & 86,157 & 24,505 & 18,507 & 22,569 \\ \hline
reg3-300\_3 & \textbf{\emph{5.6}} & \textbf{\emph{5.7}} & \textbf{7.1} & \textbf{\emph{5.9}} & 67,051 & 12,998 & 10,831 & 5,678 \\ \hline
reg3-300\_8 & \textbf{\emph{5.3}} & \emph{8.2} & \textbf{7.1} & \emph{8.6} & 71,221 & 19,058 & 14,043 & 7,127 \\ \hline
reg3-300\_0 & \textbf{\emph{5.2}} & \textbf{6.8} & \textbf{6.3} & 15.0 & 74,005 & 13,164 & 9,999 & 10,739 \\ \hline
reg3-300\_1 & \textbf{\emph{5.0}} & \textbf{7.8} & \textbf{8.0} & \emph{9.0} & 72,186 & 18,631 & 15,087 & 5,569 \\ \hline
reg3-300\_6 & \textbf{\emph{4.4}} & \textbf{6.5} & \textbf{6.8} & 9.2 & 55,902 & 12,234 & 9,455 & 8,417 \\ \hline
  \end{tabular}

  \smallskip
  (c)~3-regular graphs with 300 vertices.

  \medskip
  \fbox{
    \begin{minipage}{0.9\textwidth}
      For the two types of planar instances \unconfinedRed reductions play a
      key role; performance is even better when \funnelRed reductions are
      included.
      For the harder instances, \packingRed appears to be a major factor.
      The \deskRed reduction plays an important role in the duals -- any
      degree-4 vertex in the original graph leads to a chordless 4-cycle in
      the dual. CPLEX also does well on these instances.
    \end{minipage}
  }
  
\end{table*}


%% file: Y-reduction_eff_geo.tex
\begin{table*}
  \caption{The median effectiveness and efficiency of various reductions on
    geometric and planar graphs versus comparable other graphs.}
  \label{tab:reduction_eff_geo}

  \medskip
  \begin{minipage}{\textwidth}
    \centering
    \begin{tabular}{c c}
      \begin{tabular}{|l||r|r||r|r|}
        \hline
        reduction
        & \multicolumn{2}{c||}{$\mu$sec/vertex}
        & \multicolumn{2}{c|}{\% reduced} \\
        \cline{2-5}
        & \multicolumn{1}{c|}{med} & \multicolumn{1}{c||}{geo}
        & \multicolumn{1}{c|}{med} & \multicolumn{1}{c|}{geo}
        \\
        \hline\hline
        fold2 & 5.0 & 4.6 & 49.9 & 45.9 \\ \hline
        dom. & 36.0 & 36.4 & 19.2 & 21.2 \\ \hline
        unconf. & 105.9 & 107.5 & 20.9 & 19.9 \\ \hline
        deg1 & 16.6 & 16.0 & 6.1 & 6.0 \\ \hline
        pack. & 260.6 & 265.2 & 1.1 & 1.0 \\ \hline
        fun. & 382.3 & 463.7 & 0.8 & 0.6 \\ \hline
        desk & 448.5 & 395.7 & 0.7 & 0.8 \\ \hline
        twin & 423.3 & 419.5 & 0.3 & 0.2 \\ \hline
        lp & 1,234.5 & 1,379.4 & 0.3 & 0.3 \\ \hline
      \end{tabular}

      &
      \begin{tabular}{|l||r|r||r|r|}
        \hline
        reduction
        & \multicolumn{2}{c||}{$\mu$sec/vertex}
        & \multicolumn{2}{c|}{\% reduced} \\
        \cline{2-5}
        & \multicolumn{1}{c|}{med} & \multicolumn{1}{c||}{geo}
        & \multicolumn{1}{c|}{med} & \multicolumn{1}{c|}{geo}
        \\
        \hline\hline
        dom. & 7.4 & 11.6 & 42.2 & 47.5 \\ \hline
        fold2 & 6.3 & 11.7 & 39.5 & 18.7 \\ \hline
        deg1 & 19.3 & 22.8 & 14.5 & 9.8 \\ \hline
        \textbf{\emph{fun.}}\footnote{In most instances no vertices were
          reduced.} & 3.0 & 3.8 & $<0.1$ & $<0.1$ \\ \hline
        \textbf{\emph{unconf.}} & 3.0 & 5.4 & $<0.1$ & $<0.1$ \\ \hline
        \textbf{\emph{pack.}} & 11.0 & 18.4 & $<0.1$ & $<0.1$ \\ \hline
        \textbf{\emph{lp}} & 36.0 & 43.1 & $<0.1$ & $<0.1$  \\ \hline
        \textbf{\emph{twin}} & 249.0 & 352.6 & $<0.1$ & $<0.1$ \\ \hline
        \textbf{\emph{desk}} & 412.0 & 385.3 & $<0.1$ & $<0.1$ \\ \hline
      \end{tabular}

      \\ \\
      \shortstack[l]{
        (a) Harder geometric instances.\\
        512 vertices, average degree 16 and 32.
      }

      &

      \shortstack[l]{
        (b) Easier geometric instances.\\
        512 vertices, average degree 4, 8 and 64.
      }
      
      \\ \\

      \begin{tabular}{|l||r|r||r|r|}
        \hline
        reduction
        & \multicolumn{2}{c||}{$\mu$sec/vertex}
        & \multicolumn{2}{c|}{\% reduced} \\
        \cline{2-5}
        & \multicolumn{1}{c|}{med} & \multicolumn{1}{c||}{geo}
        & \multicolumn{1}{c|}{med} & \multicolumn{1}{c|}{geo}
        \\
        \hline\hline
        fold2 & 2.5 & 2.3  & 69.4 & 69.1     \\ \hline
        unconf. & 84.0 & 80.8 & 13.5 & 13.7   \\ \hline
        dom. & 47.3 & 37.5 & 7.7 & 8.0     \\ \hline
        deg1 & 18.1 & 15.1 & 6.2 & 6.1     \\ \hline
        fun. & 138.3 & 117.6 & 1.7 & 1.6   \\ \hline
        desk & 357.2 & 599.2 & 0.3 & 0.3   \\ \hline
        pack. & 209.7 & 240.4 & 0.3 & 0.2  \\ \hline
        twin & 540.8 & 975.7 & 0.2 & 0.1   \\ \hline
        lp & 1,746.5 & 2,431.3 & 0.1 & 0.1 \\ \hline
      \end{tabular}

      &

      \begin{tabular}{|l||r|r||r|r|}
        \hline
        reduction
        & \multicolumn{2}{c||}{$\mu$sec/vertex}
        & \multicolumn{2}{c|}{\% reduced} \\
        \cline{2-5}
        & \multicolumn{1}{c|}{med} & \multicolumn{1}{c||}{geo}
        & \multicolumn{1}{c|}{med} & \multicolumn{1}{c|}{geo}
        \\
        \hline\hline
        fold2 & 0.7 & 0.9 & 75.1 & 75.2     \\ \hline
        desk & 29.7 & 33.1 & 7.9 & 8.2    \\ \hline
        unconf. & 25.3 & 28.2 & 7.0 & 6.9 \\ \hline
        deg1 & 4.2 & 4.7 & 6.0 & 6.0      \\ \hline
        fun. & 27.5 & 26.3 & 2.4 & 2.3    \\ \hline
        dom. & 61.4 & 72.8 & 1.0 & 1.0    \\ \hline
        lp & 2,479.7 & 2,484.2 & 0.1 & 0.1 \\ \hline
        twin & 756.0 & 698.6 & 0.1 & 0.1  \\ \hline
        pack. & 148.6 & 213.5 & 0.1 & 0.1 \\ \hline
      \end{tabular}

      \\ \\

      \shortstack[l]{
        (c) Planar: Full triangulations.\\
        1000 vertices, 2994 edges.
      }
      
      &

      \shortstack[l]{(d) Duals of full triangulations.\\
        1024 vertices, 1536 edges, 3-regular.}

      \\ \\

      \begin{tabular}{|l||r|r||r|r|}
        \hline
        reduction
        & \multicolumn{2}{c||}{$\mu$sec/vertex}
        & \multicolumn{2}{c|}{\% reduced} \\
        \cline{2-5}
        & \multicolumn{1}{c|}{med} & \multicolumn{1}{c||}{geo}
        & \multicolumn{1}{c|}{med} & \multicolumn{1}{c|}{geo}
        \\
        \hline\hline
        fold2 & 1.2 & 1.3 & 71.2 & 62.4 \\ \hline
        unconf. & 24.6 & 21.6 & 9.1 & 9.9 \\ \hline
        lp & 51.5 & 70.0 & 4.5 & 3.5 \\ \hline
        pack. & 59.4 & 69.4 & 3.5 & 2.4 \\ \hline
        deg1 & 6.1 & 8.1 & 2.8 & 1.9 \\ \hline
        fun. & 81.8 & 136.3 & 2.0 & 1.2 \\ \hline
        desk & 108.5 & 129.7 & 0.8 & 0.8 \\ \hline
        twin & 96.3 & 117.2 & 0.5 & 0.5 \\ \hline
        dom. & 236.8 & 331.7 & 0.3 & 0.3 \\ \hline
      \end{tabular}

      &

      \begin{tabular}{|l||r|r||r|r|}
        \hline
        reduction
        & \multicolumn{2}{c||}{$\mu$sec/vertex}
        & \multicolumn{2}{c|}{\% reduced} \\
        \cline{2-5}
        & \multicolumn{1}{c|}{med} & \multicolumn{1}{c||}{geo}
        & \multicolumn{1}{c|}{med} & \multicolumn{1}{c|}{geo}
        \\
        \hline\hline
        fold2 &  0.3   & 0.3   & 89.7 & 89.6      \\ \hline
        deg1 &   3.6   & 3.7   &  3.0 &  3.0      \\ \hline
        fun. & 16.0    & 16.3 &   2.6 &  2.6   \\ \hline
        unconf. & 23.3 & 23.3 &   2.0 &  2.0   \\ \hline
        desk &   73.8  & 72.8 &   1.0 &  1.0   \\ \hline
        pack. & 11.2   & 11.8 &   0.9 &  0.9   \\ \hline
        lp &     347.5 & 292.2 &  0.4 &  0.4    \\ \hline
        dom. &   107.4 & 125.8 &  0.2 &  0.2    \\ \hline
        twin &   167.5 & 207.9 &  0.2 &  0.2   \\ \hline
      \end{tabular}

      \\ \\

      \shortstack[l]{
        (e)~General corpus -- see Table~\ref{tab:reduction_effectiveness_efficiency}:\\
        626~goldilocks instances.
      }

      &

      \shortstack[l]{
        (f)~3-regular graphs:\\
        300~vertices, 450 edges.
      }
    \end{tabular}

    \medskip
    In all tables, reductions are sorted by decreasing frequency (geometric means).
  \end{minipage}

\end{table*}


%% file: 5-conclusions.tex
Using a large corpus of problem instances from multiple sources we have
validated hypotheses that allow a \br solver (automated or with human guidance) to
conclude when to (i)~use no reductions at all; (ii)~use the simplest
reductions (\degRed, \foldRed) only; and (iii)~use more sophisticated
reductions such as \domRed, \unconfinedRed, and \lpRed.
We have formulated five hypotheses and tested them on our randomly generated
instances, several benchmark collections, and instances from the recent PACE challenge. 

\balance

Our study raises many questions and offers opportunity for new avenues of
investigation.
A C++ solver based on \VCSP was used in our early experiments -- we are now
poised to add sophistication to it.
Here are some ideas for future work.

\begin{compactitems}
  \item Solutions produced early by \VCS are usually close to optimal; future
    experiments can address how it fares as an 'anytime algorithm' (use the best
    solution produced within a given time limit), compared with
    metaheuristics (see, e.g., Andrade et al.~\cite{LS:Andrade12}), particularly on large
    instances where optima are not known.
  \item \VCSP does not currently provide a mechanism for changing the order
    of the reductions; our experiments suggest that, for most instances,
    \foldRed should be early
    in the sequence and
    \domRed later (see Algorithm~\ref{alg:reductions} in
    Appendix~\ref{app:branch_reduce}
    for the \VCS sequence).
  \item Automation appears to be a promising prospect; given what we know
    about both the larger and the small/medium instances, it makes sense to
    apply the \cheapLPU config at the root and then measure the instance (or
    instances if there are multiple components) to
    determine how to proceed.
  \item
    Runtime for dominance depends on the degrees of the vertices under
    consideration. It may make sense, for some graphs (characteristics need
    to be determined experimentally), to set a threshold for the degree of
    vertices considered as dominated. In other words, restrict the search to
    vertices $v$ of degree $\leq d$ and, for each such $v$, check
    whether it is dominated by any of its neighbors.
\end{compactitems}


%% file: A-branch_reduce.tex
Here we show the details of \VCS in three main parts.
The first, Algorithm~\ref{alg:BBGeneric} is the basic branching strategy.
Variable $I$ represents a (sub)instance and $|I|$ is the number of vertices
in $I$.
When a branching vertex $v$ is chosen, one subinstance
includes $v$ in the cover and the other omits $v$, but includes all its
neighbors.
We use max-degree branching -- always choose a vertex of maximum degree.
\VCS does not create new instances: \emph{only one copy} of the graph is
maintained and vertices are marked
in a \emph{status vector} as \textsf{in}, \textsf{out}, or
\textsf{undecided} depending on whether, based on decisions leading to the
current branch, they are known to be in the cover,
not in the cover or still part of the current instance.
When branching creates smaller instances, a stack keeps track of \emph{only
  the vertices whose status has changed}.
Thus there is also \emph{only one copy of the status vector}.
Small additional status vectors keep track of vertices introduced during folding.

Algorithm~\ref{alg:process_node} is responsible for applying reductions
(using Algorithm~\ref{alg:reductions}) and deciding what to do with the current
\emph{node}, which is either the root (original instance) or a branch.
Both of these functions have an option for brute force solution when the
instance is sufficiently small.
\textsc{ProcessNode} is also alerted when the graph becomes disconnected.
The procedure \textsc{ComponentSolve} hides these details.

\textsc{Reduce} applies reductions in a fixed order (our \VCSP does not
change this) and, if any reduction reduces at least one vertex, the sequence
starts at the beginning. For example, if \textsc{degree-One},
\textsc{dominance}, and \textsc{unconfined} fail, and \textsc{lp} succeeds,
then \degRed{} reductions are applied again, etc.
A reduction is applied only if selected by its runtime option.

\begin{algorithm}[t]
  \caption{A branching algorithm for \mvc{}.}
    \label{alg:BBGeneric}

  \begin{algorithmic}
    \Function{Solve}{$I,C$} 
    \State $\Var{status} \gets \Call{ProcessNode}{I}$
    \If {$\Var{status}$ = \emph{solved}}
    \State{$C \gets \min(|I|,C)$}
    \EndIf
    \If {$\Var{status}$ = \emph{alive}}
    \State save vertex status vector on stack
    \State $x \gets \Call{Select-Branching-Candidate}{I}$
    \State $C_l \gets \Call{Solve}{I \setminus \set{x}, C - 1}$
    \State restore status vector and save it again
    \State $C_r \gets \Call{Solve}{I \setminus N[x], C}$
    \State restore status vector
    \State $C \gets \min\set{C, C_l + 1, C_r}$
    \ElsIf {$\Var{status}$ = \emph{solved}}
    \State $C \gets \min(|I|,C)$
    \EndIf
    \State \Return $C$
    \EndFunction
  \end{algorithmic}
\end{algorithm}

\begin{algorithm}
  \caption{Processing the root/branch.}
    \label{alg:process_node}

  \begin{algorithmic}
    \Function{ProcessNode}{$I$}
    \State $\Var{n} \gets \text{\# of vertices in the graph}$
    \State $\Var{undecided} \gets \text{\# of undecided vertices}$
    \State \Var{status} $\gets$ \Call{Reduce}{\Var{I}}
    \IfThen {$\Var{status} =$ reduction\_cut}{\Return \emph{solved}}
    \If{$\Var{undecided} = 0$}
    \State \Return \emph{solved}
    \EndIf
    \If{$\Var{I} ~ \text{\textbf{not} connected}$}
    \State solve components separately
    \State \Return \emph{solved}
    \ElsIf {\Var{undecided} is small}
    \State solve by brute force
    \State \Return \emph{solved}
    \EndIf
    \State \Return \emph{alive}
    \EndFunction
  \end{algorithmic}
\end{algorithm}

\begin{algorithm}[b]
  \caption{Applying reductions.}
  \label{alg:reductions}

  \begin{algorithmic}
    \Function{Reduce}{$I$}
    \State $\rhd$ Each reduction function returns
    \State $\rhd$ \True{} if at least one vertex is reduced,
    \State $\rhd$ \False{} otherwise
    \State $\Var{n} \gets \text{\# of vertices in the graph}$
    \State $\Var{undecided} \gets \text{\# of undecided vertices}$
    \While{$\Var{undecided} > 0$}
    \IfThen{\Call{degree-one}{\Var{I}}}{\continue}
    \State $\rhd$ if \# undecided vertices below threshold,
    \State $\rhd$ solve by brute force
    \If{$\Var{n} \cdot \Var{SHRINK} \ge \Var{undecided}$}
    \State $\Call{Component-Solve}{\Var{I}}$ 
    \State \Return \emph{reduction\_cut}
    \EndIf
    \IfThen{\Call{dominance}{\Var{I}}}{\continue}
    \IfThen{\Call{unconfined}{\Var{I}}}{\continue}
    \IfThen{\Call{lp}{\Var{I}}}{\continue}
    \IfThen{\Call{packing}{\Var{I}}}{\continue}
    \IfThen{\Call{fold2}{\Var{I}}}{\continue}
    \IfThen{\Call{twin}{\Var{I}}}{\continue}
    \IfThen{\Call{funnel}{\Var{I}}}{\continue}
    \IfThen{\Call{desk}{\Var{I}}}{\continue}
    \State \Break
    \EndWhile
    \IfThen {$|I| = 0$}{\Return \emph{reduction\_cut}}
    \State \Return \emph{alive}
    \EndFunction
  \end{algorithmic}
\end{algorithm}


%% file: A-vcsp.tex
\input{Y-vcsp_options}

\input{Y-vcsp_stats}

Table~\ref{tab:vcsp_options} shows the options provided by \VCSP. Almost all
of these are related to selecting specific reductions and lower bounds.
The order of application for the reductions is not affected, i.e., the order
of the relevant options on the command line makes no difference.

Table~\ref{tab:vcsp_stats} shows the important part of the output of \VCSP
(header information giving version, date, input file name, options, etc., is
omitted).
The most useful data are the runtimes and number of vertices reduced by each
reduction. Also provided are runtimes for lower bounds and number of times
each lower bound was effective in cutting off a branch
and information about how many times the procedure invoking each
reduction was called and how many times it succeeded in reducing at least one vertex.

At the end of the output is a string of 0's and 1's representing the solution
found by \VCSP: a 1 in position~$i$ if vertex $i$ is included, a 0 if not.
Indexing is 0-based (some benchmark instances have vertices numbered 0) and an
underscore (\verb+_+) is used for missing vertices (aside from a missing 0-vertex some
benchmarks have non-contiguous numbering).
We provide a script for verifying solutions in this format.


%% file: Y-vcsp_options.tex
\begin{table*}
  \caption{Options provided by \VCSP}
  \label{tab:vcsp_options}

  \small
  \medskip
\begin{verbatim}
  -b, --branching <int>(2)	0: random, 1: mindeg, 2: maxdeg
  -d, --debug <int>(0)	0: no debug output, 1: basic branching and decompose,
                        2: detailed branching and decompose and basic reduction,
                        3: detailed reduction
  -t, --timeout <int>(3600)	timeout in seconds
      --trace <int>(0)	0: no trace, 1: short version without solution vectors,
                        2: full trace with solution vectors
      --quiet <boolean>(false) Don't print progress messages
      --root <boolean>(false)	Only process root node -- no branching
      --show_solution <boolean>(false)	Enable printing of solution vector
      --clique_lb <boolean>(false)	Enable clique lower bound
      --lp_lb <boolean>(false)	Enable lp lower bound
      --cycle_lb <boolean>(false)	Enable cycle lower bound
      --deg1 <boolean>(false)	Enable degree1 reduction
      --dom <boolean>(false)	Enable dominance reduction
      --fold2 <boolean>(false)	Enable fold2 reduction
      --LP <boolean>(false)	Enable LP reduction
      --unconfined <boolean>(false)	Enable unconfined reduction
      --twin <boolean>(false)	Enable twin reduction
      --funnel <boolean>(false)	Enable funnel reduction
      --desk <boolean>(false)	Enable desk reduction
      --packing <boolean>(false)	Enable packing reduction
      --all_red <boolean>(false)	Enable all reductions except packing,
                                    equivalent to old '-r2 -l3'
      --help	Show this message
  \end{verbatim}
\end{table*}


%% file: Y-vcsp_stats.tex
\begin{table}
  \caption{Statistics printed by \VCSP}
  \label{tab:vcsp_stats}

  \small
  \medskip
\begin{verbatim}
num_vertices        	             510
num_edges           	            4032
value               	             412
runtime             	     3.595
num_branches        	            5682
Reduction Times (ms):
deg1Time            	    41.760
domTime             	   329.665
fold2Time           	   144.886
lpTime              	   248.026
twinTime            	    80.137
deskTime            	   260.051
unconfinedTime      	  1401.675
funnelTime          	   303.038
packingTime         	     0.000
Vertices Reduced:
deg1Count           	           10956
domCount            	            8321
fold2Count          	          161319
lpCount             	           18254
twinCount           	             843
deskCount           	            2793
unconfinedCount     	           84497
funnelCount         	            7833
packingCount        	               0
Effective Reduction Calls:
deg1Calls           	            3239
domCalls            	            4010
fold2Calls          	           16402
lpCalls             	            3531
twinCalls           	             167
deskCalls           	             287
unconfinedCalls     	           15179
funnelCalls         	            2437
packingCalls        	               0
Total Reduction Calls:
deg1AllCalls        	           30097
domAllCalls         	           19311
fold2AllCalls       	           53388
lpAllCalls          	           17343
twinAllCalls        	           36986
deskAllCalls        	           36819
unconfinedAllCalls  	           32522
funnelAllCalls      	           13812
packingAllCalls     	               0
Effective Lower Bounds:
trivialLBCount      	            4933
cliqueLBCount       	             736
lpLBCount           	              11
cycleLBCount        	               0
Lower Bound Times (ms):
cliqueLBTime        	   220.685
cycleLBTime         	     0.000
num_leftcuts        	               2
root_lb             	             380
  \end{verbatim}
\end{table}


%% file: A-dimacs.tex
\providecommand{\phat}{\textsf{p\_hat}\xspace}

In this appendix, we describe the provenance of the DIMACS instances.
Although these instances were originally used for the minimum coloring problem, Akiba and Iwata~\cite{BB:Akiba16} use the \emph{complement} graphs of these instances in their experiments.
To maintain consistency between our results and theirs, we also use the complement graphs.\footnote{The complement versions can be found at \textsf{https://turing.cs.hbg.psu.edu/txn131/vertex\_cover.html}}
There are three types of DIMACS instances: (i)~random graph generators based on the Erdos-Renyi models, (ii)~random graph generators that embed a clique, and (iii)~graphs taken from applications.

\textbf{Random.}
The \emph{C} and \emph{DSJC} instances are based on the $G(n,p)$ Erdos-Renyi model -- a graph of $n$ vertices where each edge has probability $p$ being added independent from every other edge.
The \emph{C} instances were created by Michael Trick using a graph generator
written by Craig Morgenstern.\footnote{For more information visit
  \textsf{http://mat.gsia.cmu.edu/COLOR02/clq.html}}
The \emph{DSJC} instances were used by Johnson~\etal{}~\cite{Source:DSJC} in simulated annealing experiments for graph coloring and number partitioning.
The naming convention for these two types are \Verb+C+$n.p$ and \Verb+DSJC+$n.p$ where $n$ and $p$ are the parameters of the $G(n,p)$ model.

\begin{table}
  \centering
  \begin{tabular}{l@{\hskip 0.5in}rrr}
    \toprule
    $X$ & $a = $ & $b = $ & Expected Density\\
    \midrule
    1 & 0.00 & 0.50 & 0.25\\
    2 & 0.00 & 1.00 & 0.50\\
    3 & 0.50 & 1.00 & 0.75\\
    \bottomrule
  \end{tabular}
  \caption{Parameter values for \phat{}$n$-$x$ instances}
  \label{table:phat_params}
\end{table}

The \emph{p\_hat} instances use the \phat{} generator
introduced by Gendreau et al.~\cite{Source:phat}, using
a generalization of the $G(n,p)$ model that accepts three parameters: $n$, the number of vertices; $a$ and $b$ which are both real numbers such that $0 \le a \le b \le 1$.
Each vertex $x$ is assigned a value $a \le p[x] \le b$ -- the probability that an edge $uv$ is added is $\frac{1}{2}(p[u] + p[v])$.
When $a = b$, the \phat{} model is equivalent to the $G(n,p)$ model.
The \phat{} instances use the following naming convention: \Verb+p_hat+$n$-$x$ where $n$ is the number of vertices, and $x$ denotes a combination of values for $a$ and $b$ (see Table~\ref{table:phat_params} for specific parameter values).

The \emph{sanr} instances
are random graphs that generated using the procedure introduced by Sanchis and Jagota~\cite{Source:san}.

\textbf{Embedded clique.}
The \emph{brock} instances use the DELTA generator written by Mark Brockington and Joe Culberson.\footnote{For details about the DELTA generator, visit: \textsf{http://webdocs.cs.ualberta.ca/7Ejoe/Coloring/Generators/brock.html}}
The Brock instances use the following naming convention:
\Verb+brock+$n$\_$x$ where $n$ is the number of vertices and $x$ is a
distinguishing tag.

The \emph{san} generator, created by Sanchis and Jagota~\cite{Source:san},
accepts three parameters: $n$, the number of vertices; $m$, the number of edges; and $c$, the size of the embedded clique.
The following naming convention is used: \Verb+san+$n$\_$d$\_$r$ where
$n$ is the number of vertices, $d$ is the density, and $r$ is an integer
used to distinguish instances with the same number of vertices and edges
but different embedded clique sizes.

\textbf{Other.}
The \emph{c-fat} instances are taken from Berman and Pelc's~\cite{Source:cfat} work on fault diagnosis for multiprocessor systems.
For a given parameter $c$, a \emph{$c$-fat} ring is a graph, $G = (V,E)$, constructed as follows:
Let $k = \floor*{\frac{|V|}{c\mathrm{log}|V|}}$ and $W_0, ..., W_{k-1}$ be a partition of $V$ such that $c\mathrm{log}|V| \le |W_i| \le 1 + \ceil*{c\mathrm{log}|V|}$ for all $i = 0, ..., k-1$.
For each $u \in W_i$ and $v \in W_j$, add the edge $uv$ if $u \not = v$ and $|i - j| \in \set{0,1,k-1}$.
Naming convention is: \Verb+c-fat+$n$-$c$ where $n$ is the number of
vertices and $c$ is the construction parameter.

The \emph{\mann{}}
instances are based on Steiner triples.
For a set $S = \set{1,\ldots,n}$, a \emph{Steiner triple system} of $S$ is a set $F_S = \set{T_1, ..., T_m}$ where $T_i \subset S$ and $|T_i| = 3$ (we call $T_i$ a Steiner triple), such that for every $u,v \in S$, $u$ and $v$ are contained in exactly one $T_i$.
It is known that a Steiner triple system exists if and only if $n \ge 3$ and $n \equiv 1,3 $ (mod 6)~\cite{steinerTriple}.
To construct a \mann{} graph from a Steiner triple system,
add a vertex $v_s$ for each $s \in S$.
Then for each Steiner triple $T = \set{i,j,k}$, add a clique of size 3 and let $t_i$, $t_j$, $t_k$ be the associated vertices.
For each $s$, add an edge between $v_s$ and $T_s$ for all Steiner triples that contain $s$.
The resulting graph is a set of triangles connected by high degree ($> 3$) vertices.
We use \mann{}\_{a$n$} to denote a \mann{} graph constructed from a
Steiner triple system of $\set{1,\ldots,n}$.


%% file: A-oct.tex
\input{Y-oct_landscape}

Fig.~\ref{fig:oct_landscape} shows the distribution of the OCT instances and
their synthetic derivatives on our landscape.
None of these instances have both large degree and wide spread. The
synthetic versions, as expected, track the originals pretty well -- they
are designed to have roughly the same average degree. Barabasi-Albert
graphs, while guaranteed to be connected, have less control over degree spread.


%% file: Y-oct_landscape.tex
\begin{figure*}
  \centering

  \begin{tabular}{c c}
  \includegraphics[width=0.45\textwidth]{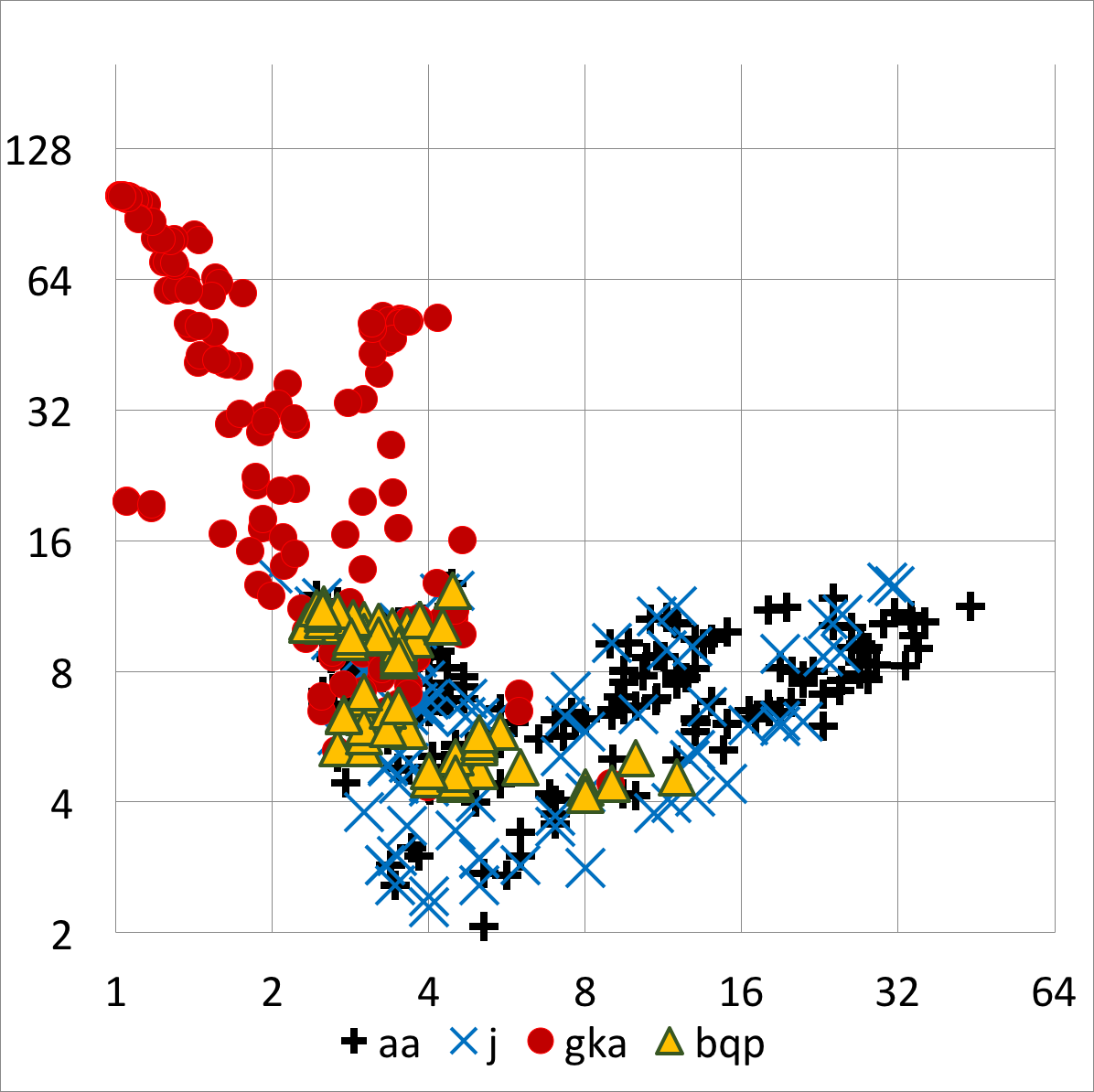}

  &
  \includegraphics[width=0.45\textwidth]{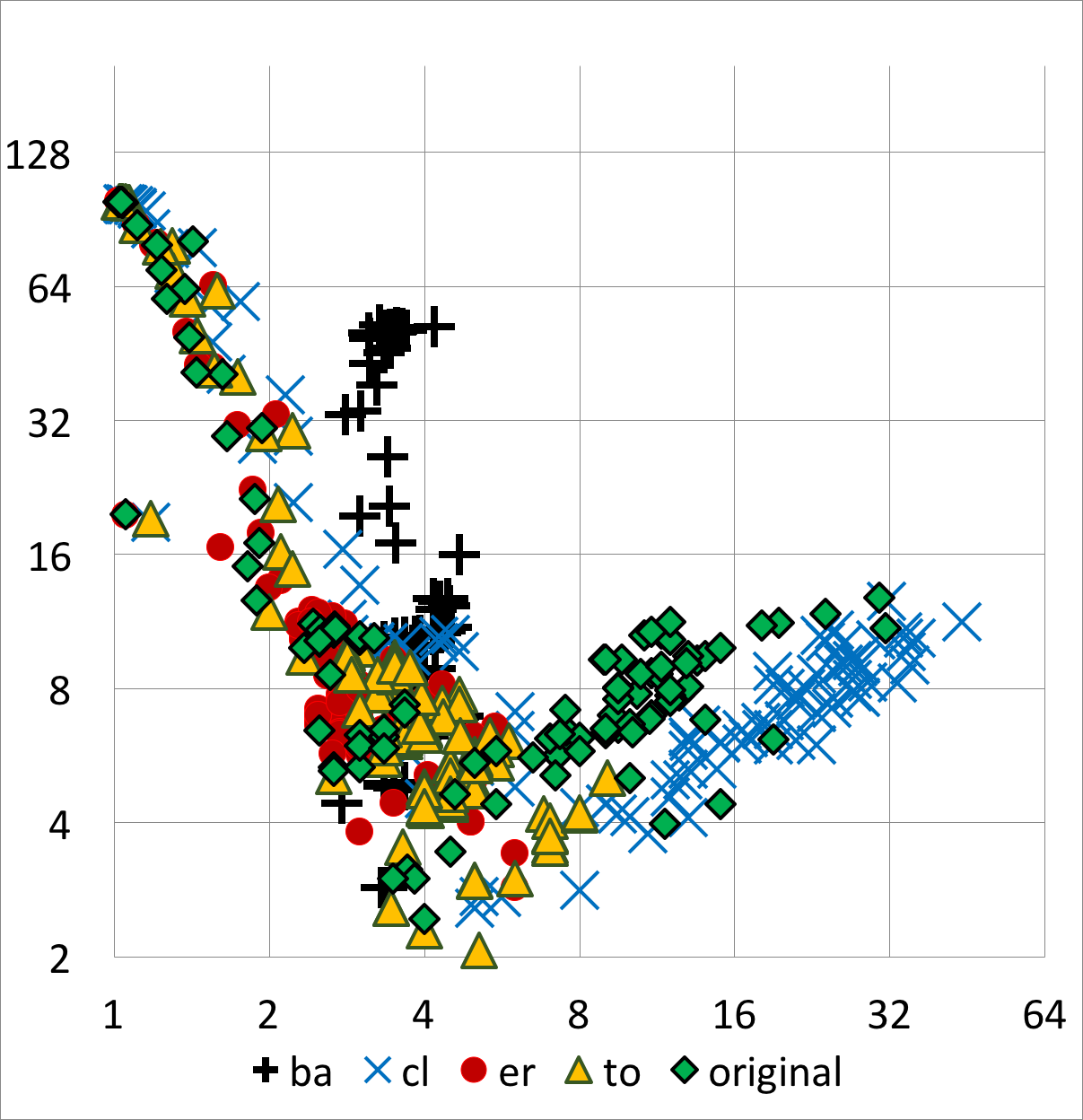}

  \\ \\
  \begin{minipage}{0.45\textwidth}
    (a)~Oct instance categories on our landscape:
    the \textsf{aa} and \textsf{j} instances come from the genetic database
    published by Wernicke~\cite{wernicke2014algorithmic}, the
    \textsf{gka} from the dataset used by Glover et
    al.~\cite{glover1998adaptive}, and the \textsf{bqp} instances are from from Beasley's OR library.
  \end{minipage}
  
  &
    
  \begin{minipage}{0.45\textwidth}
    (b)~Synthetic oct instances on our landscape, generated to have roughly
    the same size, density, and oct percentages as the originals.
    The generators used are Barabasi-Albert~(\textsf{ba}),
    Chung-Lu~(\textsf{cl}), Erd\"os-Renyi~(\textsf(er)), and \emph{tunable
      oct}~(\textsf{to}). See Goodrich et
al.~\cite{goodrich2018practical}.
    
  \end{minipage}

  \end{tabular}

  
  \caption{Oct and synthetic versions on our landscape.}
  \label{fig:oct_landscape}
\end{figure*}


%% file: A-exceptions.tex
\input{Y-exceptions}

In Fig.~\ref{fig:none_df2_r0_l1}
there are three red dots in the region where
\none should be competitive and nine scattered instances where the simplest
configs are not competitive.
Table~\ref{tab:exceptions} gives runtime and degree data for
all~12 exceptions.

The first three \textsf{ba} instances are easy for all configs except \none,
and \dft is competitive when measured against \rtlf.
The last \blg instances, where dominance is a major factor (bold),
are unusual in that at least 5\% of vertices also
have the maximum degree, close to the maximum possible.
Unconfined reductions are the major factor in all but one of the remaining
exceptions -- in that instance (DSJR500.1) \lpRed{} plays a key role.
The runtimes for all of these instances are small, but they hint at behavior
in similar, larger graphs.
We see, in Section~\ref{sec:real_world}, the effectiveness of the \cheapLPU \config.


%% file: Y-exceptions.tex
\begin{table*}
  \caption{Exceptions to our hypotheses.}
  \label{tab:exceptions}

  \centering
  
  \medskip
  \small
  \begin{tabular}{|l||r|r|r|r|r|r|r|r|r|r|r||r|r|}
    \hline
Instance & None & DF2 & r0\_l1 & r0\_l1+U & r2\_l4 & min & cpx & config &
\textsf{br-conf} &
\textsf{br-cpx} \\ \hline
\hline
ba\_512\_008\_3 & t/o & 0.21 & 0.16 & \textbf{\emph{0.07}} & \textbf{0.11} &
0.07 & 2.10 & r0\_l1+U & 4 & 0\\ \hline
ba\_512\_008\_5 & t/o & 0.19 & 0.15 & \textbf{\emph{0.06}} & \textbf{0.10} &
0.06 & 0.54 & r0\_l1+U & 4 & 0\\ \hline
ba\_999\_008\_5 & t/o & 1.43 & 1.23 & \textbf{\emph{0.47}} & \textbf{0.90} &
0.47 & 5.97 & r0\_l1+U & 55 & 0 \\ \hline
\hline
blg-200\_020\_16\_05d020 & 9.88 & 11.23 & 5.05 & \textbf{\emph{2.33}} &
\textbf{3.42} & 2.33 & \textbf{\emph{0.02}} & r0\_l1+U & 1,516 & 0\\ \hline
blg-200\_040\_16\_05d060 & 3.56 & 0.42 & 0.39 & \textbf{\emph{0.05}} &
\textbf{\emph{0.05}} & 0.05 & 0.05 & r0\_l1+U & 0 & 0\\ \hline
blg-200\_120\_20\_07d199 & 2.66 & 0.93 & \textbf{\emph{0.02}} &
\textbf{\emph{0.02}} & \textbf{0.03} & 0.02 & 0.17 & r0\_l1 & 48 & 78 \\ \hline
blg-200\_160\_01\_03d199 & 1.25 & 1.06 & \textbf{\emph{0.50}} & \textbf{0.76}
& 1.50 & 0.49 & 10.85 & r0\_l1 & 1,441 & 405\\ \hline
blg-250\_050\_16\_05d100 & 2.17 & 0.36 & 0.17 & \textbf{\emph{0.05}} &
\textbf{\emph{0.05}} & 0.05 & 0.05 & r0\_l1+U & 0 & 0\\ \hline
blg-250\_200\_16\_05d225 & 3.01 & 2.35 & \textbf{0.03} & \textbf{0.03} &
\textbf{0.03} & 0.02 & 0.11 & DD & 0 & 0\\ \hline
\hline
DSJR500.1 & t/o & t/o & 6.38 & 0.82 & \textbf{\emph{0.38}} & 0.38 & 0.36 &
r2\_l4 & 125 & 0\\ \hline
DSJR500.5 & 1.78 & 2.03 & \textbf{0.14} & \textbf{\emph{0.10}} &
\textbf{0.13} & 0.10 & 0.52 & r0\_l1+U & 0 & 0\\ \hline
\hline
gka\_16 & 2.17 & 1.04 & 0.35 & \textbf{\emph{0.04}} & \textbf{0.05} & 0.04 &
0.07 & r0\_l1+U & 0 & 6 \\ \hline
  \end{tabular}

  
  \medskip
  \begin{minipage}{\textwidth}
  (a)~The config column shows the config with the fewest reductions that
  comes within 1.1 of the minimum. Runtimes for all such configurations are
  shown in bold italics. Bold runtimes are competitive. The cpx column shows
  CPLEX runtimes.
  The \textsf{br-conf} and \textsf{br-cpx} columns show the number of
  branches for the minimum runtime configuration and CPLEX, respectively.
  \end{minipage}
  
  \medskip
  \begin{tabular}{|l||r|r||r|r|r|r|r|r|r||r|r|}
    \hline
Instance & n & m & min & bottom & med & mean & top & max & stdev & spread & nad \\ \hline
\hline
ba\_512\_008\_3 & 512 & 2,032 & 4 & 4 & 5 & 7.9 & 22 & 79 & 7.7 & 5.5 & 7.9 \\ \hline
ba\_512\_008\_5 & 512 & 2,032 & 1 & 4 & 5 & 7.9 & 21 & 85 & 8.1 & 5.4 & 7.9 \\ \hline
ba\_999\_008\_5 & 999 & 3,980 & 1 & 4 & 5 & 8.0 & 21 & 125 & 8.8 & 5.3 & 8.0 \\ \hline
\hline
blg-200\_020\_16\_05d020 & 200 & 2,000 & 10 & 20 & 20 & 20.0 & 20 & 25 & 1.2 & 1.0 & 20.0 \\ \hline
blg-200\_040\_16\_05d060 & 200 & 4,000 & 5 & 5 & 60 & 40.0 & 60 & 60 & 25.0 & 12.0 & 40.0 \\ \hline
\textbf{blg-200\_120\_20\_07d199} & 200 & 12,000 & 71 & 73 & 76 & 120.0 & 199 & 199 & 59.3 & 2.7 & 120.0 \\ \hline
\textbf{blg-200\_160\_01\_03d199} & 200 & 16,000 & 121 & 125 & 157 & 160.0 & 199 & 199 & 26.5 & 1.6 & 160.0 \\ \hline
blg-250\_050\_16\_05d100 & 250 & 6,250 & 5 & 5 & 18 & 50.0 & 100 & 100 & 44.0 & 20.0 & 40.0 \\ \hline
\textbf{blg-250\_200\_16\_05d225} & 250 & 25,000 & 5 & 5 & 225 & 200.0 & 225 & 225 & 64.7 & 45.0 & 160.0 \\ \hline
\hline
\emph{DSJR500.1} & 500 & 3,555 & 4 & 7 & 14 & 14.2 & 21 & 25 & 4.4 & 3.0 & 14.2 \\ \hline
\textbf{DSJR500.5} & 500 & 58,862 & 103 & 142 & 231 & 235.4 & 363 & 388 & 64.7 & 2.6 & 94.2 \\ \hline
\hline
gka\_16 & 180 & 8,016 & 87 & 88 & 89 & 89.1 & 90 & 90 & 0.9 & 1.0 & 99.0 \\ \hline
  \end{tabular}

  \medskip
  \begin{minipage}{\textwidth}
    (b)~Degree statistics for the exceptions. The instances where adding
    \domRed{} is important are in bold. The one where the full suite of
    reductions appears to be necessary, DSJR500.1, is in italics. The rest rely primarily
    on \unconfinedRed{} reductions.
  \end{minipage}
\end{table*}


%% file: A-runtime_tables.tex
Tables~\ref{tab:blg_instances}--\ref{tab:pace_instances} show runtimes for various configs, grouped by
sets of instances.
In each table, competitive runtimes are bold and ones within
1.1 of minimum (roughly) are also italic -- the corresponding configs are
listed in the rightmost column (the table may not show runtimes for all of these, given
the large number of \configs in our trials).
Rows are sorted by decreasing minimum runtime.
To avoid boring the reader with a plethora of large numbers we omit
information on branches except to note that
major decreases in branching with added reductions
are also accompanied by major decreases in runtime.
There were a few cases listed in the tables where CPLEX did no branching and a collection of
PACE instances
where neither CPLEX nor \VCS (best config) did branching.

Except for Table~\ref{tab:pace_instances}, the PACE instances, the data are
based on the nine configs (\none, \degOne, \foldTwo, \dft, \dd, \rOlI,
\rOlIU, \rIlf, and \rtlf -- we also ran \rthreelf on all instances later to
confirm that, in almost all cases, it performed no better than \rtlf) using a timeout of 900~seconds on our server.

For the PACE instances we used the more powerful
server\footnote{Dual Intel E5-2695v2 (2.4GHz, 30MB cache) processors and
  128GB DDR3 RAM, running Ubuntu Server 18.10.} with a four-hour timeout and a
different set of configs:
 \none, \dft, \rOlI, \cheap,
\cheapU, \cheapLP, \cheapLPU, \rOlIU, \rtlf, \rthreelf, \all.
Here, the bold runtimes are within 1.5 of minimum and those in italics only
are within a factor of two.
Minimum runtimes -- \dft in all but the three highlighted cases --
are bold-italic.
The table shows instances with minimum runtimes $>5$ seconds.

Table~\ref{tab:pace_2019-degree_stats} gives degree statistics for the same instances.
Here the ones where \rthreelf runtimes were within 1.5 or 1.1 of minimum are highlighted
in bold or bold-italic, respectively.

\input{Y-blg_instances}

\input{Y-ba_instances}

\input{Y-cl_instances}

\input{Y-oct_instances}

\input{Y-benchmark_instances}

\input{Y-pace_table}

\input{Y-pace_2019-degree_stats}


%% file: Y-blg_instances.tex
\begin{table*}
  \caption{Runtime data for selected Bucket List instances}
  \label{tab:blg_instances}
  \small

  \medskip
  \centering
  \begin{tabular}{|l||r|r||r|r|r|r|r|r||l||}
    \hline
Instance & spread & nad & None & DF2 & r0\_l1 & r2\_l4 & min & CPLEX & configs \\ \hline
blg-200\_005\_00\_05d005 & 1 & 5 & t/o & \textbf{\emph{257.91}} & \textbf{305.6} & \textbf{456.31} & 244.27 & 254.05 & Fold2;DF2 \\ \hline
blg-200\_040\_00\_05d060 & 1 & 40 & \textbf{\emph{156.55}} & \textbf{173.65} & \textbf{181.82} & 768.50 & 156.55 & t/o & None;Deg1;Fold2 \\ \hline
blg-150\_007\_00\_07d007 & 1 & 7 & 252.69 & \textbf{\emph{47.89}} & \textbf{56.11} & 121.83 & 44.19 & 86.68 & Fold2;DF2 \\ \hline
blg-150\_020\_00\_20d040 & 1 & 26.7 & \textbf{\emph{32.04}} & \textbf{\emph{33.22}} & \textbf{39.07} & 136.00 & 31.35 & 128.15 & None;Deg1;Fold2;DF2 \\ \hline
blg-250\_050\_40\_05d249 & 9.2 & 40 & t/o & \textbf{22.38} & \textbf{24.62} & \textbf{37.32} & 19.19 & 119.39 & Fold2;r1\_l4 \\ \hline
blg-200\_010\_16\_05d060 & 4.7 & 10 & t/o & \textbf{\emph{19.64}} & \textbf{21.96} & \textbf{28.55} & 18.85 & 29.25 & Fold2;DF2 \\ \hline
blg-150\_005\_00\_05d005 & 1 & 5 & 119.97 & \textbf{\emph{7.01}} & \textbf{\emph{6.99}} & 17.31 & 6.47 & 10.36 & Fold2;DF2;r0\_l1 \\ \hline
blg-250\_050\_16\_05d225 & 45 & 40 & t/o & \textbf{7.09} & \textbf{7.71} & \textbf{10.52} & 6.21 & 17.41 & Fold2 \\ \hline
blg-250\_100\_01\_05d225 & 4 & 80 & \textbf{6.77} & \textbf{6.77} & \textbf{8.24} & 20.17 & 6.07 & t/o & Fold2 \\ \hline
blg-200\_080\_00\_05d080 & 1 & 80 & \textbf{\emph{4.72}} & \textbf{5.24} & \textbf{6.43} & 19.81 & 4.72 & 234.70 & None;Deg1 \\ \hline
blg-200\_010\_01\_05d020 & 4 & 10 & 94.89 & \textbf{4.34} & \textbf{4.34} & 10.20 & 3.42 & 18.79 & Fold2 \\ \hline
\emph{blg-200\_020\_16\_05d020} & 1 & 20 & 9.88 & 11.23 & 5.05 & \textbf{3.42} & 2.33 & 0.02 & r0\_l1+U \\ \hline
blg-200\_040\_01\_05d080 & 10 & 40 & 4.25 & \textbf{\emph{1.78}} & \textbf{2.06} & 5.82 & 1.78 & 94.69 & DF2 \\ \hline
blg-250\_100\_16\_20d125 & 6.3 & 80 & \textbf{2.93} & \textbf{\emph{1.75}} & \textbf{2.15} & 3.82 & 1.75 & 0.53 & DF2 \\ \hline
blg-200\_080\_01\_05d100 & 2.3 & 80 & \textbf{1.46} & \textbf{2.35} & 2.89 & 6.47 & 1.26 & 116.43 & Deg1 \\ \hline
blg-200\_160\_01\_05d180 & 1.5 & 160 & \textbf{1.99} & \textbf{1.89} & 2.59 & 3.52 & 1.07 & 17.30 & Fold2 \\ \hline
blg-200\_003\_00\_03d003 & 1 & 3 & t/o & \textbf{\emph{0.29}} & \textbf{0.36} & \textbf{0.48} & 0.29 & 1.03 & DF2 \\ \hline
blg-200\_080\_16\_05d180 & 36 & 80 & 4.18 & 0.30 & \textbf{\emph{0.02}} & \textbf{0.03} & 0.02 & 0.23 & r0\_l1;r0\_l1+U;r1\_l4 \\ \hline

  \end{tabular}
\end{table*}


%% file: Y-ba_instances.tex
\begin{table*}
  \caption{Runtime data for selected Barabasi-Albert instances}
  \label{tab:ba_instances}
  \small

  \medskip
  \centering
  \begin{tabular}{|l||r|r||r|r|r|r|r|r||l||}
    \hline
Instance & spread & nad & None & DF2 & r0\_l1 & r2\_l4 & min & CPLEX & configs \\ \hline
ba\_256\_064\_5 & 3.60 & 43.80 & \textbf{240.99} & \textbf{\emph{196.03}} & \textbf{260.32} & 684.79 & 180.83 & 900.4 & Fold2;DF2 \\ \hline
ba\_256\_048\_2 & 4.1 & 34.0 & \textbf{253.52} & \textbf{\emph{168.68}} & \textbf{225.01} & 566.95 & 156.96 & t/o & Fold2;DF2 \\ \hline
ba\_256\_032\_5 & 4.3 & 23.4 & 350 & \textbf{\emph{122.14}} & \textbf{157.61} & 399.68 & 113.52 & t/o & Fold2;DF2 \\ \hline
ba\_256\_072\_5 & 3.4 & 48.3 & \textbf{102.87} & \textbf{\emph{92.04}} & \textbf{123.65} & 338.24 & 85.22 & t/o & Fold2;DF2 \\ \hline
ba\_256\_016\_4 & 4.9 & 15.5 & 484.46 & \textbf{\emph{10.43}} & \textbf{11.61} & 24.74 & 9.52 & 515.12 & Fold2;DF2 \\ \hline
ba\_256\_084\_4 & 3.4 & 54.9 & \textbf{6.88} & \textbf{\emph{5.64}} & \textbf{7.69} & 21.57 & 5.64 & 43.61 & Fold2;DF2 \\ \hline
ba\_999\_008\_3 & 4.8 & 8 & t/o & \textbf{4.38} & \textbf{3.07} & \textbf{2.78} & 2.22 & 12.6 & r0\_l1+U \\ \hline
ba\_512\_008\_1 & 4.8 & 7.9 & t/o & \textbf{0.56} & \textbf{0.65} & \textbf{0.7} & 0.48 & 5.33 & r0\_l1+U \\ \hline
ba\_200\_010\_1 & 5.6 & 9.8 & 5.32 & \textbf{0.17} & \textbf{0.18} & \textbf{0.19} & 0.11 & 1.57 & r0\_l1+U \\ \hline
  \end{tabular}
\end{table*}


%% file: Y-cl_instances.tex
\begin{table*}
  \caption{Runtime data for selected Chung-Lu instances}
  \label{tab:cl_instances}
  \small

  \medskip
  \centering
  \begin{tabular}{|l||r|r||r|r|r|r|r|r||l||}
    \hline
Instance & spread & nad & None & DF2 & r0\_l1 & r2\_l4 & min & CPLEX & configs \\ \hline
cl\_200\_020\_020\_3 & 2.0 & 20.0 & \textbf{298.57} & \textbf{\emph{217.46}} & \textbf{271.34} & 831.80 & 201.52 & t/o & Fold2;DF2 \\ \hline
cl\_200\_010\_010\_2 & 3.0 & 10.2 & t/o & \textbf{\emph{83.14}} & \textbf{100.87} & 230.93 & 82.22 & 645.53 & Fold2;DF2 \\ \hline
cl\_200\_040\_040\_6 & 1.6 & 39.2 & \textbf{\emph{81.22}} & \textbf{\emph{84.53}} & \textbf{\emph{88.34}} & 363.67 & 81.22 & t/o & None;Fold2;DF2;r0\_l1 \\ \hline
cl\_200\_020\_040\_4 & 2.5 & 29.5 & \textbf{50.62} & \textbf{41.64} & \textbf{46.53} & 161.62 & 37.71 & t/o & Fold2 \\ \hline
cl\_200\_010\_020\_2 & 3.0 & 15.2 & 102.16 & \textbf{\emph{24.38}} & \textbf{29.04} & 76.38 & 23.46 & 163.48 & Fold2;DF2 \\ \hline
cl\_200\_060\_060\_3 & 1.4 & 60.1 & \textbf{\emph{8.5}} & \textbf{9.93} & \textbf{11.19} & 43.80 & 8.50 & 581.79 & None \\ \hline
cl\_200\_060\_080\_2 & 1.5 & 69.8 & \textbf{\emph{5.83}} & \textbf{6.92} & \textbf{7.06} & 24.95 & 5.83 & 290.82 & None;Fold2 \\ \hline
cl\_200\_040\_080\_2 & 2.1 & 59.5 & \textbf{\emph{5.5}} & \textbf{6.81} & \textbf{6.99} & 26.18 & 5.50 & 266.19 & None;Fold2 \\ \hline
cl\_200\_010\_040\_6 & 3.9 & 25.3 & \textbf{9.7} & \textbf{\emph{5.27}} & \textbf{\emph{5.73}} & 14.69 & 5.27 & 88.74 & DF2;r0\_l1 \\ \hline
cl\_200\_010\_040\_5 & 3.7 & 25.2 & \textbf{7.29} & \textbf{5.18} & \textbf{6.49} & 15.89 & 4.45 & 40.43 & Fold2 \\ \hline
cl\_200\_080\_080\_3 & 1.4 & 79.9 & \textbf{\emph{3.46}} & \textbf{\emph{3.57}} & \textbf{4.41} & 14.64 & 3.46 & 160.38 & None;DF2 \\ \hline
cl\_200\_020\_080\_2 & 3.6 & 53.4 & \textbf{\emph{2.5}} & \textbf{3.44} & \textbf{3.76} & 9.05 & 2.50 & 87.31 & None \\ \hline
cl\_200\_020\_080\_3 & 3.5 & 51.3 & \textbf{2.92} & \textbf{\emph{2.66}} & \textbf{3.44} & 10.29 & 2.46 & 145.23 & Fold2;DF2 \\ \hline
cl\_200\_100\_100\_4 & 1.3 & 99.6 & \textbf{\emph{1.79}} & \textbf{\emph{1.82}} & \textbf{\emph{1.89}} & 6.05 & 1.79 & 144.64 & None;DF2;DD;r0\_l1 \\ \hline
cl\_200\_080\_120\_5 & 1.6 & 98.7 & \textbf{1.47} & \textbf{1.44} & \textbf{1.52} & 5.50 & 1.26 & 119.20 & Fold2 \\ \hline
cl\_200\_010\_080\_4 & 4.9 & 43.3 & \textbf{2.32} & \textbf{1.41} & \textbf{1.53} & 5.86 & 1.21 & 45.44 & Fold2 \\ \hline
cl\_200\_100\_120\_6 & 1.3 & 110.5 & \textbf{\emph{1.02}} & \textbf{1.63} & \textbf{1.23} & 4.37 & 1.02 & 93.45 & None \\ \hline
cl\_200\_005\_040\_1 & 8.4 & 22.8 & 3.73 & \textbf{\emph{0.99}} & \textbf{1.09} & 2.84 & 0.99 & 17.23 & Fold2;DF2 \\ \hline
\multicolumn{10}{|c|}{}\\ \hline
clspk\_200\_05\_040\_020\_1 & 2.0 & 20.0 & \textbf{460.63} & \textbf{\emph{301.94}} & \textbf{369.1} & t/o & 282.59 & t/o & Fold2;DF2 \\ \hline
clspk\_200\_05\_040\_030\_3 & 1.7 & 29.9 & \textbf{213.26} & \textbf{\emph{192.09}} & \textbf{210.85} & 796.56 & 188.95 & t/o & Fold2;DF2 \\ \hline
clspk\_200\_05\_080\_030\_4 & 1.7 & 29.9 & \textbf{167.05} & \textbf{\emph{143.87}} & \textbf{163.27} & 616.88 & 139.13 & t/o & Fold2;DF2 \\ \hline
clspk\_200\_05\_040\_040\_4 & 1.6 & 39.5 & \textbf{\emph{60.82}} & \textbf{\emph{62.49}} & \textbf{66.42} & 266.88 & 60.08 & t/o & None;Fold2;DF2 \\ \hline
clspk\_200\_05\_199\_010\_4 & 2.5 & 11.3 & 880.61 & \textbf{\emph{57.12}} & \textbf{68.74} & 161.93 & 55.30 & 360.33 & Fold2;DF2 \\ \hline
clspk\_200\_05\_080\_045\_4 & 1.6 & 44.6 & \textbf{\emph{27.39}} & \textbf{30.92} & \textbf{35.47} & 137.21 & 27.39 & t/o & None \\ \hline
clspk\_200\_05\_160\_010\_5 & 2.5 & 10.4 & 403.30 & \textbf{\emph{23.73}} & \textbf{28.45} & 56.14 & 23.27 & 87.88 & Fold2;DF2 \\ \hline
clspk\_200\_05\_080\_010\_4 & 3.0 & 10.3 & 257.52 & \textbf{\emph{19.96}} & \textbf{24.43} & 47.35 & 19.15 & 84.90 & Fold2;DF2 \\ \hline
clspk\_200\_05\_199\_050\_4 & 1.5 & 50.6 & \textbf{\emph{17.26}} & \textbf{19.64} & \textbf{21.96} & 84.01 & 17.26 & 625.85 & None;Fold2 \\ \hline
clspk\_200\_05\_160\_050\_5 & 1.5 & 51.1 & \textbf{\emph{15.78}} & \textbf{\emph{16.17}} & \textbf{18.59} & 69.11 & 15.78 & 615.77 & None;Fold2;DF2 \\ \hline
clspk\_200\_05\_040\_010\_1 & 3.2 & 10.2 & 226.49 & \textbf{16.08} & \textbf{17.29} & 39.05 & 14.24 & 69.32 & Fold2 \\ \hline
clspk\_200\_05\_080\_065\_3 & 1.4 & 64.9 & \textbf{\emph{7.32}} & \textbf{\emph{7.39}} & \textbf{8.78} & 32.77 & 7.32 & 356.67 & None;Fold2;DF2 \\ \hline
clspk\_200\_05\_199\_010\_5 & 2.8 & 10.3 & 150.83 & \textbf{\emph{6.13}} & \textbf{\emph{6.02}} & 14.82 & 6.02 & 29.74 & Fold2;DF2;r0\_l1 \\ \hline
clspk\_200\_05\_080\_080\_5 & 1.4 & 79.1 & \textbf{\emph{3.67}} & \textbf{4.59} & \textbf{4.79} & 15.64 & 3.67 & 224.79 & None \\ \hline
clspk\_200\_05\_160\_085\_4 & 1.3 & 83.9 & \textbf{\emph{3.19}} & \textbf{\emph{3.45}} & \textbf{3.8} & 9.57 & 3.19 & 199.05 & None;DF2 \\ \hline
clspk\_200\_05\_199\_090\_5 & 1.3 & 89.9 & \textbf{2.93} & \textbf{3.2} & \textbf{3.71} & 7.92 & 2.30 & 223.58 & Fold2 \\ \hline
clspk\_200\_05\_160\_125\_4 & 1.2 & 124.3 & \textbf{1.13} & \textbf{1.09} & \textbf{1.15} & 3.52 & 0.87 & 132.35 & Fold2 \\ \hline
clspk\_200\_05\_199\_130\_4 & 1.2 & 128.2 & \textbf{\emph{0.69}} & \textbf{0.87} & \textbf{1.32} & 3.13 & 0.69 & 138.07 & None \\ \hline
  \end{tabular}
\end{table*}


%% file: Y-oct_instances.tex
\begin{table*}
  \caption{Runtime data for selected OCT instances: original}
  \label{tab:oct_instances_original}
  \small

  \medskip
  \centering
  \begin{tabular}{|l||r|r||r|r|r|r|r|r||l||}
    \hline
Instance & spread & nad & None & DF2 & r0\_l1 & r2\_l4 & min & CPLEX & configs \\ \hline
bqp100\_10 & 3.8 & 10.5 & t/o & \textbf{\emph{196.43}} & \textbf{221.59} & 545.76 & 196.43 & 455.14 & Fold2;DF2 \\ \hline
bqp100\_1 & 3.8 & 10.5 & 779.88 & \textbf{\emph{96.52}} & \textbf{114.34} & 270.61 & 96.52 & 66.15 & Fold2;DF2 \\ \hline
gka\_28 & 3.4 & 26.5 & \textbf{\emph{84.45}} & \textbf{\emph{89.16}} & \textbf{114.04} & 403.02 & 84.45 & 611.78 & None;Deg1;Fold2;DF2 \\ \hline
gka\_24 & 3.7 & 10.4 & 378.98 & \textbf{\emph{60.01}} & \textbf{71.65} & 160.26 & 60.01 & 64.50 & Fold2;DF2 \\ \hline
gka\_33 & 1.3 & 77.8 & \textbf{\emph{1.58}} & \textbf{2.91} & 3.88 & 7.80 & 1.58 & 12.58 & None \\ \hline
gka\_21 & 2.1 & 36.6 & \textbf{\emph{0.59}} & \textbf{\emph{0.62}} & \textbf{0.91} & 2.63 & 0.59 & 0.72 & None;DF2 \\ \hline
aa41 & 44 & 11.3 & 2.87 & \textbf{\emph{0.58}} & \textbf{0.68} & \textbf{1.12} & 0.58 & 2.88 & DF2 \\ \hline
  \end{tabular}
\end{table*}

\begin{table*}
  \caption{Runtime data for selected OCT instances: synthetic Barabasi-Albert}
  \label{tab:oct_instances_ba}
  \small

  \medskip
  \centering
  \begin{tabular}{|l||r|r||r|r|r|r|r|r||l||}
    \hline
Instance & spread & nad & None & DF2 & r0\_l1 & r2\_l4 & min & CPLEX & configs \\ \hline
aa38-ba & 4.3 & 12.6 & t/o & \textbf{\emph{375.22}} & \textbf{451.1} & \textbf{595.42} & 347.08 & 900.04 & DF2;r1\_l4 \\ \hline
j20-ba & 4.8 & 6.9 & t/o & \textbf{\emph{40.73}} & \textbf{44.41} & \textbf{47.99} & 38.07 & 685.62 & DF2;r1\_l4 \\ \hline
aa17-ba & 4.3 & 10.7 & t/o & \textbf{\emph{20.76}} & \textbf{23.82} & \textbf{41.46} & 20.76 & 205.87 & DF2 \\ \hline
gka\_27-ba & 2.2 & 21 & 56.60 & \textbf{\emph{21.11}} & \textbf{23.41} & 65.42 & 19.27 & 58.81 & Fold2;DF2 \\ \hline
aa32-ba & 4.2 & 10.7 & t/o & \textbf{\emph{18.31}} & \textbf{20.9} & \textbf{36.49} & 18.31 & 118.07 & DF2 \\ \hline
aa40-ba & 4.2 & 10.6 & 664.12 & \textbf{\emph{9.61}} & \textbf{\emph{9.99}} & 21.30 & 9.61 & 30.58 & Fold2;DF2;r0\_l1 \\ \hline
aa33-ba & 4.7 & 6.9 & t/o & \textbf{10.18} & \textbf{\emph{8.89}} & \textbf{15.76} & 8.89 & 102.04 & r0\_l1 \\ \hline
gka\_28-ba & 1.9 & 28.2 & \textbf{12.0} & \textbf{8.45} & \textbf{9.25} & 26.25 & 6.80 & 35.91 & Fold2 \\ \hline
aa50-ba & 3.9 & 10.6 & 50.93 & \textbf{3.54} & \textbf{4.45} & 9.36 & 2.96 & 7.61 & Fold2 \\ \hline
j24-ba & 4 & 6.9 & t/o & \textbf{\emph{2.08}} & \textbf{\emph{1.9}} & 4.90 & 1.90 & 56.51 & DF2;r0\_l1 \\ \hline
bqp100\_10-ba & 3.6 & 10.1 & 10.73 & \textbf{1.76} & \textbf{\emph{1.36}} & 3.88 & 1.36 & 1.73 & Fold2;r0\_l1 \\ \hline
gka\_25-ba & 4.3 & 10.5 & 11.42 & \textbf{\emph{1.34}} & \textbf{\emph{1.4}} & 4.42 & 1.34 & 1.74 & Fold2;DF2;r0\_l1 \\ \hline
bqp100\_1-ba & 3.2 & 9.6 & 10.55 & \textbf{\emph{1.32}} & \textbf{1.57} & 4.04 & 1.32 & 1.74 & DF2 \\ \hline
gka\_26-ba & 4.5 & 10.6 & 10.84 & \textbf{2.0} & \textbf{1.37} & 4.30 & 1.23 & 1.74 & Fold2 \\ \hline
gka\_8-ba & 6 & 6.4 & 18.77 & \textbf{1.17} & \textbf{\emph{1.05}} & 2.41 & 1.05 & 2.36 & r0\_l1 \\ \hline
gka\_22-ba & 2.2 & 29.5 & 3.00 & \textbf{1.28} & \textbf{1.31} & 4.99 & 0.95 & 1.77 & Fold2 \\ \hline
gka\_31-ba & 1.5 & 58.2 & \textbf{\emph{0.78}} & \textbf{\emph{0.74}} & \textbf{1.13} & 4.26 & 0.74 & 4.54 & None;Deg1;DF2 \\ \hline
aa27-ba & 4 & 6.9 & 45.50 & \textbf{0.74} & \textbf{0.77} & \textbf{0.91} & 0.60 & 2.24 & r0\_l1+U \\ \hline
  \end{tabular}
\end{table*}

\begin{table*}
  \caption{Runtime data for selected OCT instances: synthetic Chung-Lu}
  \label{tab:oct_instances_cl}
  \small

  \medskip
  \centering
  \begin{tabular}{|l||r|r||r|r|r|r|r|r||l||}
    \hline
Instance & spread & nad & None & DF2 & r0\_l1 & r2\_l4 & min & CPLEX & configs \\ \hline
gka\_27-cl & 1.9 & 22.3 & \textbf{186.82} & \textbf{\emph{130.43}} & \textbf{159.04} & 499.65 & 119.92 & 551.86 & Fold2;DF2 \\ \hline
gka\_28-cl & 1.7 & 31.1 & \textbf{\emph{50.27}} & \textbf{\emph{49.93}} & \textbf{67.41} & 239.43 & 48.16 & 153.92 & None;Deg1;Fold2;DF2 \\ \hline
bqp100\_9-cl & 2.8 & 9.6 & 471.76 & \textbf{\emph{37.2}} & \textbf{44.02} & 109.98 & 37.20 & 125.66 & DF2 \\ \hline
gka\_25-cl & 2.5 & 10.9 & 170.27 & \textbf{\emph{25.69}} & \textbf{29.51} & 79.72 & 25.69 & 74.67 & Fold2;DF2 \\ \hline
bqp100\_2-cl & 2.5 & 10.7 & 205.82 & \textbf{\emph{24.91}} & \textbf{28.93} & 70.06 & 24.91 & 19.63 & Fold2;DF2 \\ \hline
gka\_29-cl & 1.5 & 42.5 & \textbf{\emph{12.66}} & \textbf{14.16} & \textbf{19.47} & 64.60 & 12.66 & 232.08 & None;Deg1 \\ \hline
gka\_30-cl & 1.4 & 50.3 & \textbf{\emph{8.07}} & \textbf{9.07} & \textbf{12.49} & 41.40 & 8.07 & 119.05 & None;Fold2 \\ \hline
aa41-cl & 3.3 & 7.9 & 72.57 & \textbf{\emph{5.15}} & \textbf{\emph{5.61}} & 15.08 & 5.15 & 88.42 & Fold2;DF2;r0\_l1 \\ \hline
gka\_31-cl & 1.3 & 60.5 & \textbf{\emph{4.74}} & \textbf{5.05} & \textbf{6.25} & 17.95 & 4.53 & 80.52 & None;Deg1;Fold2 \\ \hline
gka\_4-cl & 2.6 & 8.8 & 17.84 & \textbf{3.72} & \textbf{\emph{2.89}} & 10.06 & 2.89 & 3.59 & r0\_l1 \\ \hline
gka\_32-cl & 1.3 & 69.6 & \textbf{4.14} & \textbf{3.98} & \textbf{4.01} & 9.16 & 2.83 & 90.11 & Deg1 \\ \hline
aa42-cl & 34.4 & 9.7 & 15.41 & \textbf{\emph{2.28}} & \textbf{2.52} & 6.15 & 2.28 & 20.12 & Fold2;DF2 \\ \hline
gka\_33-cl & 1.2 & 79 & \textbf{\emph{1.9}} & \textbf{2.51} & \textbf{3.18} & 8.07 & 1.90 & 11.89 & None \\ \hline
aa24-cl & 35 & 9 & 41.97 & \textbf{\emph{1.63}} & \textbf{2.2} & 7.60 & 1.63 & 25.80 & DF2 \\ \hline
gka\_21-cl & 1.6 & 42.2 & \textbf{\emph{0.9}} & \textbf{\emph{0.95}} & \textbf{1.32} & 3.30 & 0.90 & 1.61 & None;DF2 \\ \hline
j20-cl & 21 & 6.1 & 47.44 & \textbf{\emph{0.85}} & \textbf{1.03} & 2.67 & 0.85 & 6.49 & DF2 \\ \hline
aa32-cl & 33.8 & 10.7 & 3.35 & \textbf{0.82} & \textbf{0.84} & 1.85 & 0.72 & 7.91 & Fold2 \\ \hline
  \end{tabular}
\end{table*}

\begin{table*}
  \caption{Runtime data for selected OCT instances: synthetic Erdos-Renyi}
  \label{tab:oct_instances_er}
  \small

  \medskip
  \centering
  \begin{tabular}{|l||r|r||r|r|r|r|r|r||l||}
    \hline
Instance & spread & nad & None & DF2 & r0\_l1 & r2\_l4 & min & CPLEX & configs \\ \hline
bqp100\_2-er & 3.5 & 8.8 & t/o & \textbf{\emph{362.37}} & \textbf{432.31} & t/o & 362.37 & 486.10 & Fold2;DF2 \\ \hline
gka\_26-er & 2.8 & 9.4 & t/o & \textbf{\emph{330.76}} & \textbf{399.71} & t/o & 330.76 & 384.69 & Fold2;DF2 \\ \hline
j28-er & 2 & 13.4 & 712.32 & \textbf{\emph{291.93}} & \textbf{363.92} & t/o & 282.88 & 383.09 & Fold2;DF2 \\ \hline
aa27-er & 2.8 & 6.7 & t/o & \textbf{\emph{84.77}} & \textbf{98.46} & \textbf{166.64} & 84.77 & 263.67 & DF2 \\ \hline
gka\_23-er & 2.1 & 16.2 & \textbf{23.43} & \textbf{\emph{18.44}} & \textbf{22.89} & 65.25 & 18.00 & 41.33 & Fold2;DF2 \\ \hline
aa26-er & 4 & 7.5 & 126.77 & \textbf{\emph{10.61}} & \textbf{\emph{10.74}} & 21.82 & 10.61 & 14.00 & DF2;r0\_l1 \\ \hline
j17-er & 2.8 & 9.1 & 35.36 & \textbf{7.94} & \textbf{8.35} & 18.91 & 6.63 & 7.71 & Fold2 \\ \hline
gka\_31-er & 1.4 & 59.9 & \textbf{\emph{3.73}} & \textbf{4.66} & \textbf{5.18} & 13.95 & 3.73 & 80.65 & None;Deg1 \\ \hline
gka\_4-er & 3 & 7.2 & 14.38 & \textbf{5.44} & \textbf{5.11} & 12.78 & 3.68 & 4.24 & Fold2 \\ \hline
aa51-er & 4.3 & 8.1 & 21.10 & \textbf{\emph{3.36}} & \textbf{\emph{3.49}} & 8.49 & 3.36 & 9.56 & Fold2;DF2;r0\_l1 \\ \hline
aa43-er & 2.7 & 11.6 & \textbf{3.7} & \textbf{\emph{2.76}} & \textbf{\emph{2.75}} & 6.38 & 2.63 & 4.82 & Deg1;Fold2;DF2;r0\_l1 \\ \hline
gka\_34-er & 1.1 & 87.7 & \textbf{\emph{2.23}} & \textbf{2.9} & \textbf{3.62} & 7.41 & 2.17 & 2.55 & None;Deg1;Fold2;DD \\ \hline
gka\_33-er & 1.2 & 78.6 & \textbf{\emph{1.82}} & \textbf{3.58} & 3.80 & 7.86 & 1.82 & 17.22 & None \\ \hline
gka\_3-er & 3.3 & 6.5 & 6.31 & \textbf{1.53} & \textbf{\emph{1.15}} & 5.12 & 1.13 & 1.58 & Fold2;r0\_l1 \\ \hline
gka\_21-er & 1.6 & 40.6 & \textbf{0.69} & \textbf{\emph{0.64}} & \textbf{0.82} & 2.52 & 0.61 & 0.98 & Deg1;Fold2;DF2 \\ \hline
  \end{tabular}
\end{table*}


%% file: Y-benchmark_instances.tex
\begin{table*}
  \caption{Runtime data for selected benchmark instances: DIMACS and SAT}
  \label{tab:benchmark_instances}
  \small

  \medskip
  \centering
  \begin{tabular}{|l||r|r||r|r|r|r|r|r||l||}
    \hline
Instance & spread & nad & None & DF2 & r0\_l1 & r2\_l4 & min & CPLEX & configs \\ \hline
p\_hat1500-1 & 1.4 & 149.2 & \textbf{\emph{430.79}} & \textbf{\emph{465.1}} & t/o & t/o & 430.79 & t/o & None;Deg1;Fold2;DF2 \\ \hline
san400\_0.7\_3 & 1.3 & 59.9 & \textbf{\emph{271.06}} & \textbf{300.5} & \textbf{437.49} & t/o & 271.06 & t/o & None;Deg1;Fold2 \\ \hline
sanr200\_0.9 & 2 & 20.4 & \textbf{276.44} & \textbf{\emph{205.02}} & \textbf{261.34} & 785.22 & 191.06 & t/o & Fold2;DF2 \\ \hline
DSJC500.5 & 1.2 & 100.2 & \textbf{\emph{115.74}} & \textbf{\emph{127.01}} & \textbf{205.24} & 752.15 & 115.74 & t/o & None;Deg1;Fold2;DF2 \\ \hline
p\_hat700-2 & 2.7 & 100.3 & \textbf{\emph{87.16}} & \textbf{\emph{89.33}} & \textbf{146.42} & 519.28 & 84.96 & 934.44 & None;Fold2;DF2 \\ \hline
san200\_0.9\_3 & 1.7 & 19.9 & \textbf{\emph{71.2}} & \textbf{\emph{74.54}} & \textbf{101.92} & 358.31 & 68.37 & 3.32 & None;Deg1;Fold2;DF2 \\ \hline
san400\_0.7\_2 & 1.2 & 59.9 & \textbf{\emph{50.93}} & \textbf{56.28} & \textbf{89.06} & 340.32 & 50.93 & 727.01 & None;Deg1;Fold2 \\ \hline
p\_hat1000-1 & 1.4 & 150.9 & \textbf{\emph{48.49}} & \textbf{\emph{52.19}} & 135.30 & 436.41 & 48.49 & 924.23 & None;Fold2;DF2 \\ \hline
sanr400\_0.5 & 1.2 & 99.5 & \textbf{\emph{34.11}} & \textbf{38.09} & \textbf{59.12} & 210.13 & 34.11 & 904.47 & None;Fold2 \\ \hline
san1000 & 1.2 & 99.6 & \textbf{\emph{30.93}} & \textbf{34.5} & 87.99 & 305.46 & 30.93 & t/o & None;Fold2 \\ \hline
p\_hat300-3 & 2.6 & 50.9 & \textbf{31.26} & \textbf{\emph{30.97}} & \textbf{43.32} & 143.69 & 28.38 & t/o & Fold2;DF2 \\ \hline
DSJC1000.9 & 1 & 179.8 & \textbf{\emph{27.9}} & \textbf{\emph{28.23}} & 91.33 & 232.12 & 26.90 & t/o & None;Deg1;Fold2;DF2 \\ \hline
\emph{san400\_0.7\_1} & 1.2 & 59.9 & \textbf{\emph{19.83}} & \textbf{\emph{21.08}} & \textbf{32.47} & 122.46 & 19.83 & 7.96 & None;Fold2;DF2 \\ \hline
\emph{johnson16-2-4} & 1 & 46.7 & \textbf{\emph{19.21}} & \textbf{\emph{20.72}} & \textbf{21.32} & 42.72 & 19.21 & 0.01 & None;Fold2;DF2;DD \\ \hline
p\_hat700-1 & 1.4 & 149.9 & \textbf{\emph{12.47}} & \textbf{\emph{12.9}} & 32.99 & 97.47 & 12.47 & t/o & None;Fold2;DF2 \\ \hline
queen16\_16 & 1.3 & 38.6 & \textbf{\emph{11.66}} & \textbf{13.49} & \textbf{15.36} & 46.98 & 11.66 & 0.05 & None \\ \hline
p\_hat500-2 & 2.6 & 98.9 & \textbf{\emph{10.53}} & \textbf{\emph{11.42}} & \textbf{19.11} & 63.64 & 10.53 & t/o & None;Fold2;DF2 \\ \hline
flat300\_28\_0 & 1.2 & 96.4 & \textbf{\emph{10.47}} & \textbf{11.62} & \textbf{17.34} & 55.34 & 10.47 & t/o & None;Fold2 \\ \hline
flat300\_20\_0 & 1.1 & 95 & \textbf{\emph{6.27}} & \textbf{7.28} & \textbf{10.53} & 33.37 & 6.27 & t/o & None \\ \hline
p\_hat500-1 & 1.4 & 149.1 & \textbf{\emph{5.12}} & \textbf{5.65} & 10.75 & 28.57 & 5.12 & t/o & None \\ \hline
DSJC500.9 & 1 & 179.9 & \textbf{\emph{3.6}} & \textbf{4.15} & 9.10 & 22.40 & 3.60 & t/o & None;Deg1;Fold2 \\ \hline
DSJC250.5 & 1.2 & 100.3 & \textbf{\emph{2.5}} & \textbf{3.25} & \textbf{3.49} & 11.20 & 2.50 & 392.63 & None;Fold2 \\ \hline
\emph{le450\_25a} & 19.5 & 16.3 & 41.75 & 5.45 & \textbf{\emph{2.48}} & \textbf{3.37} & 2.48 & 0.03 & r0\_l1 \\ \hline
school1\_nsh & 18.2 & 47.2 & 4.46 & 4.54 & \textbf{\emph{2.06}} & \textbf{3.96} & 2.06 & 5.15 & r0\_l1;r0\_l1+U \\ \hline
\emph{queen14\_14} & 1.3 & 43.6 & \textbf{2.42} & \textbf{2.07} & \textbf{2.07} & 6.24 & 1.88 & 0.04 & DD \\ \hline
DSJC125.1 & 2.6 & 11.8 & 3.74 & \textbf{\emph{1.6}} & \textbf{1.94} & 4.03 & 1.60 & 2.43 & Fold2;DF2 \\ \hline
DSJC250.9 & 1.1 & 178.5 & \textbf{1.26} & \textbf{\emph{0.91}} & 2.67 & 3.00 & 0.91 & 19.40 & Deg1;DF2 \\ \hline
  \end{tabular}
\end{table*}


%% file: Y-pace_table.tex


\begin{table*}
  \caption{Runtime data for selected PACE-2019 instances}
  \label{tab:pace_instances}
  \small

  \medskip
  \centering
  \begin{tabular}{|l||r|r||r|r|r|r|r|r||r||r||}
    \hline
Instance & spread & nad & DF2 & r0\_l1 & Cheap & Ch+LP & r2\_l4 & r3\_l4 & min & CPLEX \\ \hline
\textbf{vc-exact\_040} & 1.0 & 5.0 & t/o & t/o & t/o & t/o & t/o & \textbf{\emph{8,420.8}} & 8,420.8 & \textbf{\emph{5.8}} \\ \hline
vc-exact\_091 & 1.8 & 11.6 & \textbf{\emph{195.4}} & \textbf{223.4} & \textbf{288.1} & \emph{384.6} & 534.4 & \emph{358.6} & 195.4 & 113.0 \\ \hline
vc-exact\_043 & 1.4 & 8.4 & \textbf{\emph{186.8}} & \textbf{222.4} & \emph{287.1} & 399.2 & 475.5 & \textbf{276.6} & 186.8 & \textbf{\emph{18.0}} \\ \hline
\textbf{vc-exact\_039} & 1.3 & 3.1 & t/o & t/o & t/o & t/o & \emph{256.8} & \textbf{\emph{162.6}} & 162.6 & \textbf{\emph{30.2}} \\ \hline
vc-exact\_083 & 1.6 & 11.7 & \textbf{\emph{123.3}} & \textbf{148.1} & \textbf{179.1} & \emph{220.3} & 348.2 & \emph{264.1} & 123.3 & \textbf{\emph{60.4}} \\ \hline
vc-exact\_046 & 1.4 & 8.1 & \textbf{\emph{120.3}} & \textbf{141.9} & \emph{186.1} & 249.4 & 331.6 & \textbf{176.0} & 120.3 & \textbf{\emph{12.2}} \\ \hline
vc-exact\_031 & 1.4 & 8.1 & \textbf{\emph{104.2}} & \textbf{120.4} & \emph{161.3} & 221.4 & 225.3 & \textbf{126.6} & 104.2 & \textbf{\emph{8.0}} \\ \hline
vc-exact\_081 & 1.8 & 10.0 & \textbf{\emph{91.5}} & \textbf{110.7} & \textbf{131.5} & \emph{166.7} & 254.7 & 201.6 & 91.5 & 76.0 \\ \hline
vc-exact\_056 & 1.7 & 10.9 & \textbf{\emph{82.3}} & \textbf{98.1} & \textbf{116.1} & \emph{150.8} & 210.5 & \emph{142.7} & 82.3 & \textbf{\emph{33.2}} \\ \hline
vc-exact\_067 & 1.6 & 11.7 & \textbf{\emph{76.6}} & \textbf{92.9} & \textbf{107.4} & \emph{137.9} & 224 & 181.3 & 76.6 & 38.6 \\ \hline
vc-exact\_093 & 1.9 & 11.6 & \textbf{\emph{69.2}} & \textbf{83.2} & \textbf{100.3} & \emph{129.6} & 207.7 & 155.5 & 69.2 & 76.4 \\ \hline
vc-exact\_044 & 1.9 & 11.5 & \textbf{\emph{65.0}} & \textbf{79.7} & \textbf{92.3} & \emph{116.8} & 195.9 & 142.9 & 65.0 & 38.6 \\ \hline
vc-exact\_082 & 1.9 & 9.5 & \textbf{\emph{63.9}} & \textbf{75.1} & \emph{96.3} & \emph{118.2} & 159.2 & \emph{101.1} & 63.9 & 38.9 \\ \hline
vc-exact\_060 & 1.8 & 11.2 & \textbf{\emph{53.2}} & \textbf{65.2} & \textbf{75.8} & \emph{96.3} & 146.2 & 116.4 & 53.2 & 32.4 \\ \hline
vc-exact\_063 & 1.6 & 10.1 & \textbf{\emph{51.8}} & \textbf{63.2} & \textbf{75.2} & \emph{91.9} & 140.5 & \emph{99.6} & 51.8 & 25.2 \\ \hline
vc-exact\_062 & 1.8 & 11.3 & \textbf{\emph{50.2}} & \textbf{57.0} & \textbf{71.7} & \emph{95.6} & 148.6 & 102.4 & 50.2 & 34.8 \\ \hline
vc-exact\_058 & 1.9 & 11.7 & \textbf{\emph{49.1}} & \textbf{56.5} & \textbf{68.9} & \emph{87.5} & 118.7 & \textbf{71.6} & 49.1 & 35.3 \\ \hline
vc-exact\_057 & 1.9 & 11.6 & \textbf{\emph{48.3}} & \textbf{56.4} & \textbf{68.7} & \emph{90.6} & 134.6 & 99.2 & 48.3 & 28.9 \\ \hline
vc-exact\_047 & 1.8 & 10.9 & \textbf{\emph{46.4}} & \textbf{58.1} & \textbf{67.4} & \emph{81.1} & 118.2 & \emph{85.3} & 46.4 & 33.0 \\ \hline
vc-exact\_053 & 1.0 & 10.3 & \textbf{\emph{45.8}} & \textbf{56.8} & \emph{69.0} & \emph{83.2} & 127.5 & \emph{91.7} & 45.8 & 34.6 \\ \hline
vc-exact\_050 & 1.7 & 10.2 & \textbf{\emph{45.0}} & \textbf{53.1} & \textbf{65.6} & \emph{84.0} & 126.7 & \emph{74.0} & 45.0 & 48.4 \\ \hline
vc-exact\_041 & 2.0 & 10.2 & \textbf{\emph{43.8}} & \textbf{51.6} & \textbf{63.9} & \emph{81.5} & 110.1 & \emph{73.9} & 43.8 & 23.6 \\ \hline
vc-exact\_051 & 1.8 & 10.0 & \textbf{\emph{42.2}} & \textbf{50.4} & \textbf{58.5} & \emph{71.9} & 107.4 & 85.8 & 42.2 & 25.4 \\ \hline
vc-exact\_073 & 2.1 & 10.8 & \textbf{\emph{34.2}} & \textbf{41.9} & \textbf{49.5} & \emph{59.5} & 94.8 & \emph{64.7} & 34.2 & 30.9 \\ \hline
vc-exact\_069 & 1.0 & 10.8 & \textbf{\emph{33.4}} & \textbf{38.6} & \textbf{47.4} & \emph{62.6} & 95.9 & \emph{67.9} & 33.4 & 22.7 \\ \hline
vc-exact\_071 & 1.8 & 9.5 & \textbf{\emph{32.5}} & \textbf{37.1} & \textbf{44.6} & \emph{52.0} & 72.2 & \emph{57.5} & 32.5 & 16.7 \\ \hline
vc-exact\_065 & 1.9 & 10.1 & \textbf{\emph{31.7}} & \textbf{39.1} & \textbf{43.8} & \emph{54.9} & 86.2 & \emph{57.0} & 31.7 & 32.8 \\ \hline
vc-exact\_042 & 1.8 & 9.5 & \textbf{\emph{30.1}} & \textbf{35.8} & \textbf{43.8} & \emph{53.1} & 80.1 & \emph{59.2} & 30.1 & \textbf{\emph{14.1}} \\ \hline
vc-exact\_072 & 1.9 & 11.7 & \textbf{\emph{28.9}} & \textbf{35.2} & \textbf{40.8} & \emph{51.6} & 79.5 & 67.8 & 28.9 & 18.1 \\ \hline
vc-exact\_054 & 1.8 & 9.6 & \textbf{\emph{25.9}} & \textbf{30.9} & \textbf{37.8} & \emph{47.3} & 70.9 & 54.8 & 25.9 & 20.6 \\ \hline
vc-exact\_052 & 1.0 & 9.9 & \textbf{\emph{25.7}} & \textbf{30.9} & \textbf{37.5} & \emph{50.1} & 66.0 & \textbf{36.9} & 25.7 & 27.1 \\ \hline
vc-exact\_048 & 1.9 & 10.2 & \textbf{\emph{25.4}} & \textbf{30.2} & \textbf{36.5} & \emph{45.4} & 60.9 & \emph{39.6} & 25.4 & 28.9 \\ \hline
vc-exact\_064 & 2.0 & 10.4 & \textbf{\emph{24.3}} & \textbf{28.7} & \textbf{34.0} & \emph{42.7} & 62.0 & \emph{44.8} & 24.3 & 26.9 \\ \hline
vc-exact\_045 & 2.0 & 10.2 & \textbf{\emph{19.9}} & \textbf{22.8} & \textbf{27} & \emph{32.4} & 50.5 & \emph{38.5} & 19.9 & 22.7 \\ \hline
vc-exact\_061 & 2.2 & 9.3 & \textbf{\emph{18.0}} & \textbf{23.4} & \textbf{28.3} & \emph{35.7} & 51.3 & \emph{36.2} & 18.0 & 24.7 \\ \hline
vc-exact\_049 & 1.8 & 9.3 & \textbf{\emph{18.8}} & \textbf{21.0} & \textbf{26.9} & \emph{32.3} & 44.9 & \emph{31.1} & 18.8 & 17.3 \\ \hline
\textbf{vc-exact\_038} & 5.6 & 9.1 & t/o & 477.8 & t/o & t/o & 116.3 & \textbf{\emph{16.4}} & 16.4 & 10.1 \\ \hline
vc-exact\_077 & 2.3 & 9.6 & \textbf{\emph{14.6}} & \textbf{14.8} & \textbf{21.2} & \emph{26.9} & 34.1 & \emph{26} & 14.6 & 17.2 \\ \hline
vc-exact\_059 & 2.3 & 9.6 & \textbf{\emph{14.1}} & \textbf{14.0} & \textbf{20.1} & \emph{26.8} & 33.9 & \emph{22.1} & 14.1 & 17.1 \\ \hline
vc-exact\_070 & 2.2 & 8.6 & \textbf{\emph{11.7}} & \textbf{13.1} & \textbf{15.8} & \emph{19.9} & 28.0 & \emph{21.1} & 11.7 & 17.7 \\ \hline
vc-exact\_068 & 2.1 & 9.6 & \textbf{\emph{10.5}} & \textbf{12.2} & \textbf{14.2} & \emph{17.3} & 22.9 & \textbf{15.8} & 10.5 & 12.6 \\ \hline
vc-exact\_037 & 1.9 & 8.2 & \textbf{\emph{10.5}} & \textbf{11.9} & \textbf{14.4} & \emph{17.9} & 24.4 & \emph{17.9} & 10.5 & 11.9 \\ \hline
vc-exact\_066 & 2.2 & 8.7 & \textbf{\emph{10.5}} & \textbf{12.7} & \emph{16.2} & \emph{19.4} & 26.6 & \textbf{14.7} & 10.5 & 16.9 \\ \hline
vc-exact\_074 & 2.1 & 8.1 & \textbf{\emph{9.6}} & \textbf{10.7} & \textbf{13.0} & \emph{18.4} & 22.1 & \textbf{14.2} & 9.6 & 11.3 \\ \hline
vc-exact\_035 & 2.1 & 8.6 & \textbf{\emph{9.1}} & \textbf{10.8} & \textbf{12.0} & \emph{15.6} & \emph{18.2} & \emph{15.1} & 9.1 & 15.1 \\ \hline
vc-exact\_055 & 2.3 & 9.4 & \textbf{\emph{7.6}} & \textbf{8.7} & \textbf{10.3} & \emph{12.9} & 17.4 & \emph{14.2} & 7.6 & 16.1 \\ \hline
  \end{tabular}
  \end{table*}


%% file: Y-pace_2019-degree_stats.tex
\begin{table*}
  \caption{Degree statistics for selected PACE-2019 instances}
  \label{tab:pace_2019-degree_stats}

  \medskip
  \centering
  \small
    \begin{tabular}{|l||r|r|r|r|r|r|r|r|r|r|r|}
      \hline
00-Instance & n & m & min & bottom & med & mean & top & max & stdev & spread & nad \\ \hline
\textbf{\emph{vc-exact\_040}} & 210 & 625 & 5 & 6 & 6 & 6.0 & 6 & 6 & 0.2 & 0.0 & 6.0 \\ \hline
vc-exact\_091 & 200 & 1,163 & 4 & 6 & 11 & 11.6 & 17 & 22 & 3.3 & 1.8 & 11.6 \\ \hline
\textbf{vc-exact\_043} & 200 & 841 & 3 & 5 & 8 & 8.4 & 11 & 15 & 2.0 & 1.4 & 8.4 \\ \hline
\textbf{\emph{vc-exact\_039}} & 6,795 & 10,620 & 2 & 2 & 3 & 3.1 & 3 & 21 & 2.3 & 1.3 & 3.1 \\ \hline
vc-exact\_083 & 200 & 1,172 & 3 & 7 & 11 & 11.7 & 17 & 20 & 3.3 & 1.6 & 11.7 \\ \hline
\textbf{vc-exact\_046} & 200 & 812 & 4 & 5 & 8 & 8.1 & 11 & 15 & 2.0 & 1.4 & 8.1 \\ \hline
\textbf{vc-exact\_031} & 200 & 813 & 3 & 5 & 8 & 8.1 & 11 & 13 & 2.1 & 1.4 & 8.1 \\ \hline
vc-exact\_081 & 199 & 1,091 & 2 & 6 & 11 & 11.0 & 17 & 20 & 3.1 & 1.8 & 11.0 \\ \hline
vc-exact\_056 & 200 & 1,089 & 4 & 6 & 11 & 10.9 & 16 & 20 & 3.1 & 1.7 & 10.9 \\ \hline
vc-exact\_067 & 200 & 1,174 & 4 & 7 & 12 & 11.7 & 18 & 23 & 3.4 & 1.6 & 11.7 \\ \hline
vc-exact\_093 & 200 & 1,162 & 3 & 6 & 11 & 11.6 & 18 & 24 & 3.7 & 1.9 & 11.6 \\ \hline
vc-exact\_044 & 200 & 1,147 & 3 & 6 & 11 & 11.5 & 18 & 23 & 3.5 & 1.9 & 11.5 \\ \hline
vc-exact\_082 & 200 & 954 & 3 & 5 & 9.5 & 9.5 & 15 & 18 & 2.9 & 1.9 & 9.5 \\ \hline
vc-exact\_060 & 200 & 1,118 & 3 & 6 & 11 & 11.2 & 17 & 22 & 3.4 & 1.8 & 11.2 \\ \hline
vc-exact\_063 & 200 & 1,011 & 4 & 6 & 10 & 10.1 & 15 & 20 & 3.1 & 1.6 & 10.1 \\ \hline
vc-exact\_062 & 199 & 1,128 & 3 & 6 & 11 & 11.3 & 17 & 24 & 3.4 & 1.8 & 11.3 \\ \hline
\textbf{vc-exact\_058} & 200 & 1,171 & 4 & 6 & 12 & 11.7 & 18 & 24 & 3.7 & 1.9 & 11.7 \\ \hline
vc-exact\_057 & 200 & 1,160 & 3 & 6 & 12 & 11.6 & 18 & 20 & 3.4 & 1.9 & 11.6 \\ \hline
vc-exact\_047 & 200 & 1,093 & 1 & 6 & 11 & 10.9 & 17 & 21 & 3.4 & 1.8 & 10.9 \\ \hline
vc-exact\_053 & 200 & 1,026 & 4 & 5 & 10 & 10.3 & 16 & 21 & 3.2 & 2.0 & 10.3 \\ \hline
vc-exact\_050 & 200 & 1,025 & 4 & 6 & 10 & 10.3 & 16 & 21 & 3.1 & 1.7 & 10.3 \\ \hline
vc-exact\_041 & 200 & 1,023 & 3 & 5 & 10 & 10.2 & 16 & 22 & 3.2 & 2.0 & 10.2 \\ \hline
vc-exact\_051 & 200 & 1,098 & 4 & 6 & 11 & 11.0 & 17 & 24 & 3.5 & 1.8 & 11.0 \\ \hline
vc-exact\_073 & 200 & 1,078 & 3 & 5 & 11 & 10.8 & 17 & 21 & 3.5 & 2.1 & 10.8 \\ \hline
vc-exact\_069 & 200 & 1,083 & 2 & 5 & 11 & 10.8 & 16 & 19 & 3.5 & 2.0 & 10.8 \\ \hline
vc-exact\_071 & 200 & 952 & 4 & 5 & 9.5 & 9.5 & 14 & 18 & 2.8 & 1.8 & 9.5 \\ \hline
vc-exact\_065 & 200 & 1,011 & 4 & 5 & 10 & 10.1 & 15 & 18 & 3.0 & 1.9 & 10.1 \\ \hline
vc-exact\_042 & 200 & 952 & 3 & 5 & 9 & 9.5 & 14 & 19 & 2.8 & 1.8 & 9.5 \\ \hline
vc-exact\_072 & 200 & 1,167 & 3 & 6 & 11 & 11.7 & 18 & 25 & 3.8 & 1.9 & 11.7 \\ \hline
vc-exact\_054 & 200 & 961 & 1 & 5 & 9 & 9.6 & 14 & 18 & 3.0 & 1.8 & 9.6 \\ \hline
\textbf{vc-exact\_052} & 200 & 992 & 3 & 5 & 10 & 9.9 & 16 & 18 & 3.2 & 2.0 & 9.9 \\ \hline
vc-exact\_048 & 200 & 1,025 & 2 & 5 & 10 & 10.3 & 15 & 21 & 3.2 & 1.9 & 10.3 \\ \hline
vc-exact\_064 & 200 & 1,042 & 2 & 5 & 10 & 10.4 & 16 & 19 & 3.4 & 2.0 & 10.4 \\ \hline
vc-exact\_045 & 200 & 1,020 & 3 & 5 & 10 & 10.2 & 16 & 24 & 3.4 & 2.0 & 10.2 \\ \hline
vc-exact\_061 & 200 & 931 & 1 & 4 & 9 & 9.3 & 14 & 17 & 3.1 & 2.2 & 9.3 \\ \hline
vc-exact\_049 & 200 & 933 & 2 & 5 & 9 & 9.3 & 14 & 19 & 2.9 & 1.8 & 9.3 \\ \hline
\textbf{\emph{vc-exact\_038}} & 786 & 14,024 & 1 & 4 & 26 & 35.7 & 91 & 136 & 28.5 & 5.6 & 9.1 \\ \hline
vc-exact\_077 & 200 & 961 & 2 & 4 & 9 & 9.6 & 15 & 18 & 3.3 & 2.3 & 9.6 \\ \hline
vc-exact\_059 & 200 & 961 & 2 & 4 & 9 & 9.6 & 15 & 18 & 3.3 & 2.3 & 9.6 \\ \hline
vc-exact\_070 & 200 & 860 & 2 & 4 & 8 & 8.6 & 14 & 19 & 2.9 & 2.2 & 8.6 \\ \hline
\textbf{vc-exact\_068} & 200 & 961 & 3 & 5 & 9 & 9.6 & 16 & 20 & 3.4 & 2.1 & 9.6 \\ \hline
vc-exact\_037 & 198 & 808 & 2 & 4 & 8 & 8.2 & 12 & 16 & 2.6 & 1.9 & 8.2 \\ \hline
\textbf{vc-exact\_066} & 200 & 866 & 3 & 4 & 8 & 8.7 & 14 & 19 & 3.0 & 2.2 & 8.7 \\ \hline
\textbf{vc-exact\_074} & 200 & 805 & 1 & 4 & 8 & 8.1 & 13 & 16 & 2.8 & 2.1 & 8.1 \\ \hline
vc-exact\_035 & 200 & 864 & 2 & 4 & 8 & 8.6 & 13 & 18 & 2.9 & 2.1 & 8.6 \\ \hline
vc-exact\_055 & 200 & 938 & 1 & 4 & 9 & 9.4 & 15 & 21 & 3.4 & 2.3 & 9.4 \\ \hline
    \end{tabular}
\end{table*}
